\documentclass[11pt, notitlepage, nofootinbib,
onecolumn,preprintnumbers,amsmath,amssymb,prx,superscriptaddress,noshowpacs]{revtex4-1} 

\usepackage[normalem]{ulem}

\usepackage{amsthm}
\theoremstyle{theorem}

\newcommand{\SO}{{\rm SO}}
\newcommand{\Spin}{{\rm Spin}}
\newcommand{\rO}{{\rm O}}
\newcommand{\U}{{\rm U}}

\newcommand{\Pin}{{\rm Pin}}

\newcommand{\bea}{\begin{eqnarray}}
\newcommand{\eea}{\end{eqnarray}}

\usepackage{graphicx}
\usepackage{amsmath}
\usepackage{amsthm}
\usepackage{amstext}
\usepackage{amssymb}
\usepackage{mathrsfs}
\usepackage{amsfonts}
\usepackage{amsbsy} 
\usepackage{dcolumn}
\usepackage{bm}
\usepackage{multirow}

\usepackage[usenames, dvipsnames]{color}
\usepackage[svgnames]{xcolor}
\usepackage[colorlinks,citecolor=RoyalBlue, urlcolor=RoyalBlue, linkcolor=RoyalBlue ]{hyperref} 

\definecolor{red}{rgb}{1,0,0}
\definecolor{blue}{rgb}{0,0,1}
\definecolor{dblue}{rgb}{0,0,0.4}
\definecolor{green}{rgb}{0,1,0}
\definecolor{black}{rgb}{0,0,0}
\definecolor{white}{rgb}{1,1,1}

\definecolor{brn}{rgb}{.8,.4,.0}
\definecolor{redo}{rgb}{1,.5,.0}
\definecolor{ddgrn}{rgb}{0,0.4,0}
\definecolor{dgrn}{rgb}{0,0.55,0}
\definecolor{dbl}{rgb}{0,0,0.5}

\usepackage[bbgreekl]{mathbbol}
\usepackage{amscd}

\newcommand{\Z}{\mathbb{Z}}

\renewcommand{\t}[1]{\tilde{#1}} 
\newcommand{\ii}{\hspace{1pt}\mathrm{i}\hspace{1pt}}

\newcommand{\dd}{\hspace{1pt}\mathrm{d}}

\newcommand{\Refe}[1]{Ref.~\cite{#1}}
\newcommand{\Eq}[1]{Eq.\,(\ref{#1})} 
\newcommand{\eq}[1]{(\ref{#1})} 
\newcommand{\eqn}[1]{eqn.~(\ref{#1})}

\newcommand{\bpm}{\begin{pmatrix}}
\newcommand{\epm}{\end{pmatrix}}
\newcommand{\bmm}{\begin{matrix}}
\newcommand{\emm}{\end{matrix}}

\newcommand{\cA}{ {\cal A} } 
\newcommand{\cB}{ {\cal B} }
 
\newcommand{\cD}{ {\cal D} }

\newcommand{\cH}{ {\cal H} }

\newcommand{\cN}{ {\cal N} }

\newcommand{\cS}{ {\cal S} } 
\newcommand{\cT}{ {\cal T} } 
 
\newcommand{\cX}{ {\cal X} } 
\newcommand{\cY}{ {\cal Y} }

\newcommand{\al}{\alpha}

\newcommand{\ga}{\gamma}

\newcommand{\nn}{{\nonumber}}

\def\B{\mathrm{B}}

\def\Sq{\mathrm{Sq}}

\newcommand{\sfU}{\mathsf{U}}

\newcommand{\tm}{{\rm m}}

\usepackage{sseq}
\usepackage[all,cmtip]{xy}
\usepackage{tikz-cd}
\usepackage{tikz}
\usetikzlibrary{matrix}
\usetikzlibrary{decorations.markings}
\usetikzlibrary{positioning}
\usetikzlibrary{shadings}

\usepackage{datetime}

\usepackage{enumitem} 
\usepackage{moreenum}

\usepackage{forest}

\def \Hom{\operatorname{Hom}}

\newcommand{\Sec}[1]{Sec.~\ref{#1}} 
\newcommand{\Fig}[1]{Fig.~\ref{#1}}

\newcommand\cW{{\cal W}}

\newcommand{\diag}{\mathop{\mathrm{diag}}}

\newcommand{\re}{\mathrm{e}}

\newcommand{\ri}{\mathrm{i}}
\newcommand{\rii}{\mathrm{ii}}
\newcommand{\riii}{\mathrm{iii}}

\newcommand{\I}{\text{I}}
\newcommand{\II}{\text{II}}
\newcommand{\III}{\text{III}}
\newcommand{\IV}{\text{IV}}
\newcommand{\V}{\text{V}}
\newcommand{\VI}{\text{VI}}
\newcommand{\VII}{\text{VII}}
\newcommand{\VIII}{\text{VIII}}
\newcommand{\IX}{\text{IX}}
\newcommand{\X}{\text{X}}

\def\CO{{\cal O}}
\newcommand{\GSD}{\text{GSD}}

\numberwithin{equation}{section}

\def\ra{{\mathrm{a}}}
\def\rb{{\mathrm{b}}}

\def\bZ{{\mathbf{Z}}}

\newcommand{\mi}[1]{\mathit{#1}} 

\usepackage{comment} 
\newcommand\hcup[1]{\underset{{\scriptscriptstyle #1}}{\cup}}

\begin{document}


\fontsize{12}{12}\selectfont
\title{
\LARGE{{
Non-Liquid Cellular States}}:\\[2mm] 
\large{Gluing 
Gauge-(Higher)-Symmetry-Breaking vs 
 -Extension Interfacial Defects 
 }
}

\author{Juven Wang}
\affiliation{Center of Mathematical Sciences and Applications, Harvard University, MA 02138, USA}
\email{jw@cmsa.fas.harvard.edu}


\begin{abstract} 


The existence of quantum non-liquid states and fracton orders, both gapped and gapless states, challenges our understanding of phases of entangled matter. We generalize the cellular topological states to liquid or non-liquid cellular states. We propose a mechanism to construct more general non-abelian states by gluing gauge-symmetry-breaking vs gauge-symmetry-extension interfaces as extended defects in a cellular network, including defects of higher-symmetries, in any dimension. Our approach also naturally incorporates the anyonic particle/string condensations and composite string (related to particle-string or p-string)/membrane condensations. This approach shows gluing the familiar extended topological quantum field theory or conformal field theory data via topology, geometry, and renormalization consistency criteria (via certain modified group cohomology or cobordism theory data) in a tensor network can still guide us to analyze the non-liquid states. (Part of the abelian construction can be understood from the K-matrix Chern-Simons theory approach and the coupled-layer-by-junction constructions.) This approach may also lead us toward a unifying framework for quantum systems of both
\emph{higher-symmetries} and 
\emph{sub-system/sub-dimensional symmetries}.


\end{abstract}


\maketitle

%

\fontsize{10}{10}\selectfont
\tableofcontents


\fontsize{10}{10}\selectfont

\section{Introduction 
and Summary
}

The paradigm of statistical physics and condensed matter theory in the 20th century is  largely governed by the Landau-Ginzburg theory \cite{Landau1958} and Wilsonian renormalization group (RG) \cite{Wilson1974mbRevModPhys47773}.
However, it gradually becomes clear to us 
that many topological phases of quantum matter cannot be characterized or classified by the old paradigm alone
 \cite{Wen2016ddy1610.03911}.
For example, there are short-range entangled gapped phases with the same global symmetry (or the same global symmetry-breaking),
but with different short-range entanglement patterns protected by the global symmetry --- known as the 
symmetry-protected topological states (SPTs)  \cite{Wen2016ddy1610.03911}.
There are also long-range entangled gapped phases with the same global symmetry (or the same global symmetry-breaking, 
or even without any global symmetry),
but with different long-range entanglement patterns with emergent gauge fields 
--- known as the intrinsic topologically ordered states (in short, as topological order or TO)  \cite{Wen2016ddy1610.03911}.
The bulk gapped phase of topological order \emph{cannot} be directly detected through 
Ginzburg-Landau symmetry breaking order parameter
\emph{nor} any local operator  $\CO(x)$, \emph{nor} the gapless long-range order like Nambu-Goldstone modes. 
Since the late 1980 and early 1990, researchers learn to relate the low energy physics of topologically ordered states \cite{Wen1990tm}
to the Schwarz type of Topological Quantum Field Theories (TQFTs) \cite{Schwarz1978cnLMP, Witten1988hfTQFTJones}.

However, it is important to emphasize the following, the starting definitions of
 topological order and TQFT are rather different:
\begin{itemize}[leftmargin=-2mm] 
\item 
Topological order is realized as a long-range entangled state 
(normally defined as gapped liquids \cite{ZengWen20151406.5090, Swingle2014qpaMcGreevy1407.8203} and usually the ground state) 
of quantum many-body systems
as a Hamiltonian quantum theory
with a UV completion (usually by a high-energy cutoff, on a lattice or a simplicial complex).
\item TQFTs are defined via a path integral formulation (e.g. Lagrangian) with a partition function ${\bf Z}$ obeying Atiyah's axiom \cite{Atiyah1989vu}.
\end{itemize}
Luckily, it happened that many SPTs and topological order can be characterized by Atiyah's axiomatic TQFTs at IR low-energy.
Those TQFTs assign a  manifold with a boundary to a state-vector living in a finite-dimensional Hilbert space.  
They have the properties:
\begin{enumerate}[label=\textcolor{blue}{(\arabic*).}, ref={(\arabic*)},leftmargin=*]
\item \label{TQFT1} Allow a path integral formulation and a lattice or simplicial-complex path integral formulation, as a well-behaved quantum theory.

\item \label{TQFT2} Have a finite-dimensional Hilbert space (the finite ground state degeneracy [GSD]; or a finite number of 
ground state zero-energy modes) on a closed spatial manifold.
So the path integral on the manifold $M^{D-1} \times S^1$ is
${\bf Z}(M^{D-1} \times S^1)= \dim {\cal{H}}_{M^{D-1}}\equiv \GSD$.
At the large spacetime volume limit, the GSD can only be sensitive to the spacetime topology, 
but independent of the geometry or the details of the lattice cutoff.
\end{enumerate}
For example, the low energy physics of $G$-SPT with a $G$ symmetry can be characterized by 
invertible Topological Quantum Field Theories (iTQFTs) of a flat-connection $G$-background field associated with a classifying space $\B G$
that has $|{\bf Z}|=1$ on any closed manifold \cite{Kapustin2014tfa1403.1467, 1405.7689, Kapustin2014dxa1406.7329, Wen2014zga1410.8477, Freed2016rqq1604.06527}.
On the other hand,
the low energy physics of topological order
can be characterized by various types of continuum TQFT formulations in any dimension \cite{Putrov2016qdo, Kapustin2017jrc1701.08264, Wang2018edf1801.05416, Guo2018vij1812.11959} of 
bosonic or fermionic TQFTs of
Chern-Simons-Witten gauge theories  \cite{Witten1988hfTQFTJones},  Dijkgraaf-Witten twisted discrete gauge theories \cite{Dijkgraaf1989pzCMP}, and category theories \cite{rowell2009classification0712.1377ZhenghanWang, Lan2017iqpLKW1602.05946},
whose $|{\bf Z}| \neq 1$ is generically non-invertible on a closed manifold.

In recent years, the discovery of fracton order (see the recent review \cite{RahulNandkishore2018sel1803.11196, PretkoReview2020cko2001.01722})
indicates that the gapped fracton orders 
\cite{2005PRL0404182Chamon, 1108.2051CastelnovoChamon, 2011PhRvAHaah1101.1962, Vijay2015mka1505.02576VijayHaahFu, Vijay2016phm1603.04442},\footnote{In some sense, the gapped fracton orders should be regarded as an exotic 
subclass of gapped entangled states.}
even at the low energy, may \emph{not} have a complete data formulation merely based on TQFT of Atiyah's axiom.
So the certain fracton orders\footnote{We shall
call them gapped fracton orders without referring to topological orders, 
since fracton orders are not gapped quantum liquids \cite{ZengWen20151406.5090, Swingle2014qpaMcGreevy1407.8203}, 
thus they are not the conventional topological order.
} may not have a correspondence to TQFT of Atiyah's axiom even at the low energy!
The reasons are that\\[-6mm] 
\begin{enumerate}[label=\textcolor{blue}{(\arabic*').}, ref={},leftmargin=*]
\item 
Many fracton orders are only formulated on lattice Hamiltonian models so far, which may not obviously or necessarily have a 
path integral formulation, thus which may violate \ref{TQFT1}.
\item
Gapped fracton orders have extensive ground state degeneracy depending on the system size and the details of lattice sites and cutoffs, thus which violate \ref{TQFT2}.
In addition, gapped fracton orders can also have excitations, either being immobile in isolation or being mobile along in lower subdimensions.
\end{enumerate}
Another way to illustrate the above conflicts, is that 
familiar topological orders of Atiyah's axiomatic TQFT types are
{{\bf\emph{gapped liquid states}}} (which includes the fractional quantum Hall states and gapped spin quids and discrete gauge theories),
while the fracton orders are
{{\bf\emph{gapped non-liquid states}}} \cite{ZengWen20151406.5090, Swingle2014qpaMcGreevy1407.8203}.\footnote{By
 \cite{ZengWen20151406.5090},
 gapped liquid states can \emph{dissolve} any tensor product states on additional sites (with \emph{appropriate} distribution).
 Such that the liquid takes no shapes to increase its size by dissolving the tensor product state on the additional sites.}

\noindent
{{\bf \emph{Question}}:} This raises the puzzle that how can the familiar TQFT data of Atiyah's axiomatic TQFT (and the familiar topological orders) guide
us to analyze the non-liquid states, especially the fracton orders?

In this work, we aim to use the familiar data of TQFT in $D$-dimensions and the data of
their associated various lower-dimensional defects in $(D-1)$, $(D-2)$, $(D-3)$, etc., -dimensions\footnote{The
$(D-d)$-dimension is also the co-dimension-$d$.}
to 
the building blocks for both liquid phases
and non-liquid phases.\footnote{For the dimensional notations, we write
the $n+1$D, 
as the $n$ dimensional space and 1 dimensional time.
We write $n'$d as $n'$-dimensional spacetime where $n'=n+1$.}

In fact, such an idea is recently also pursued and initiated by \cite{Wen2020pri2002.02433, Aasen2020zru2002.05166}.
Our construction in certain cases is somehow more similar to \Refe{Wen2020pri2002.02433}.
\Refe{Wen2020pri2002.02433} proposes the new terminology
{{\bf\emph{cellular topological states}}}:
\Refe{Wen2020pri2002.02433} divides the 2D space into several cells, while the 2+1D
topological order (the familiar gapped liquid states) live on 2D cell surfaces.
The 1D edges sit between 2+1D cell surfaces host 1+1D 
anomalous gapped topological orders.\footnote{Here the
anomalous gapped topological orders mean that the gapped states are the boundary only phenomena.
In this case, the ``anomalous'' can require no global symmetry thus beyond the $G$-'t Hooft anomalies associated with $G$-global symmetries.} 
\Refe{Wen2020pri2002.02433} then uses the classification of gapped topological interface formulas 
derived in \cite{Lan2014uaa1408.6514}, applicable to both 2+1D abelian and non-abelian TQFTs, given by integer-valued tunneling matrix $\cW$, 
then solve the consistency gluing conditions and lattice renormalization formulas.

However, the 1+1D 
anomalous gapped topological orders employed in \cite{Wen2020pri2002.02433}
are only limited to
those that can be realized as certain {{\bf\emph{anyon-condensation}}}-induced gapped interfaces
(see a recent review and Reference therein \cite{Burnell2017otf1706.04940},
the set of condensed anyons associated with the Lagrangian subgroup).\footnote{In this work, 
for {anyon condensation} induced gapped boundaries,
we particularly follow the approach in \cite{Wang2012am1212.4863, Levin2013gaa1301.7355, Lan2014uaa1408.6514}.
Meanwhile we may supplement various different types of interpretations of {anyon condensation} on the lattice Hamiltonian \cite{9811052BravyiKitaev, 1104.5047KitaevKong, Cong2017ffh1707.04564}, category \cite{1104.5047KitaevKong, Kong2013aya1307.8244}, Chern-Simons gauge theory \cite{1008.0654KapustinSaulina}, the defect approaches \cite{Barkeshli2013jaa1304.7579, Barkeshli2013yta1305.7203, Barkeshli2014cna1410.4540},
't Hooft anomaly perspectives \cite{Wang2013yta1307.7480, Wang2018ugf1807.05998},
or the quantum group breaking \cite{Bais2008ni0808.0627, Hung2013qpa1308.4673, Hung2014tba1408.0014}.
The Lagrangian subgroup
is in analogy with
the Lagrangian subspace in symplectic geometry \cite{1008.0654KapustinSaulina}.} 

\Refe{Wang2017loc1705.06728} points out that 
for many cases of topological orders with a 2+1D (3d) discrete gauge theory (can be twisted by group cohomology cocycle as \cite{Dijkgraaf1989pzCMP}), the {anyon condensation} types of gapped boundaries
in fact have a one-to-one correspondence to the {{\bf\emph{gauge-(symmetry) breaking}}} type of gapped boundary construction.\footnote{Since gauge-symmetry is only a gauge description redundancy, but not a global symmetry, from now we will refer the gauge-symmetry breaking simply as {{\bf\emph{gauge-breaking}}}. See the details in Appendix E and F of \Refe{Wang2017loc1705.06728} on this mechanism,
see also Sec.~7 of \Refe{Wang2018edf1801.05416}.} 
While the {gauge-(symmetry)}-breaking gapped boundaries can also be systematically obtained from gauging the 
 {{\bf\emph{global-symmetry breaking}}} gapped boundaries, we have the following schematic relation for 1+1D (2d) topological 
 gapped boundaries or gapped interfaces:\footnote{The gapped interface between two gapped topological states $\bf{Z_A}$ and $\bf{Z_B}$ has a one-to-one correspondence to a gapped boundary of the folded gapped topological states $\bf{Z_A} \otimes \bf{\overline{Z}_B}$ by the folding trick {(here ${\bf{\overline{Z}_B}}$ 
 takes the opposite orientation into account)}.
 Thus, we can always interchange the gapped boundary or the gapped interface construction to each other.
 In \Eq{eq:breaking-1} and others, when we use the notation ``$\Longleftrightarrow$,'' this means there is a way to go between the
 left and right hand sides. When we use the notation ``$\Rightarrow$,'' 
 this means there is a way to understand the
 left hand side approach in terms of the right hand side approach. 
 By ``$\stackrel{\text{un/gauging}}{\Longleftrightarrow}$,'' we mean that
 going to the right we have ``$\stackrel{\text{gauging}}{\Longrightarrow}$,''
 while going to the left we have 
 ``$\stackrel{\text{ungauging}}{\Longleftarrow}$.''
 }
 \bea \label{eq:breaking-1}
 \text{global-symmetry-breaking} \stackrel{\text{un/gauging}}{\Longleftrightarrow} \text{gauge-(symmetry)-breaking} \Rightarrow
 \text{anyon condensation}.
\eea
In addition, \Refe{Wang2017loc1705.06728} and \Refe{Wang2018edf1801.05416}'s 
systematic construction of {global-symmetry-breaking} and {gauge-(symmetry)-breaking}
 gapped boundaries can be applied to 
 gapped bulk phases  of any physical dimension and higher dimensions (2+1D, 3+1D, 4+1D bulk, etc.). 
\newpage
Moreover, \Refe{Wang2017loc1705.06728} also proposes a new systematic mechanism to obtain 
the \text{symmetry-extension} type of gapped boundaries, which includes
the  {{\bf\emph{{global-symmetry-extension}}}} for the bulk SPTs, or also includes
the  {{\bf\emph{{gauge-(symmetry)-extension}}}} for the bulk long-range entangled gauge theories,
with a schematic relation between them via gauging or {ungauging}:
\bea \label{eq:extension-1}
 \text{global-symmetry-extension} \stackrel{\text{un/gauging}}{\Longleftrightarrow} \text{gauge-(symmetry)-extension}.
\eea
%
Furthermore, 
we can generalize both the concepts of
the ordinary-symmetries (acting onsite, e.g., locally on the 0-simplex) to the generalized higher-$n$-symmetries
(acting on the extended $n$-dimensional objects, e.g., locally on the $n$-simplex)
\cite{Gaiotto2014kfa1412.5148}. Include the generalized higher-$n$-symmetries,
for the construction of gapped interfaces,
we can generalize \Eq{eq:breaking-1} to {{\bf\emph{{higher-symmetry breaking}}}}:
\bea   \label{eq:breaking-2}
 &&\hspace{-18mm}
 \text{higher-global-symmetry-breaking} 
 \stackrel{\text{un/gauging}}{\Longleftrightarrow} \text{higher-gauge-(symmetry)-breaking} \nn\\
 &&\Rightarrow
 \text{anyonic $n$-dimensional extended objects (e.g.~0-particles, 1-strings, 2-branes, etc.) condensation}. \quad
\eea
We can also generalize the symmetry extension \Eq{eq:extension-1} to {{\bf\emph{{higher-symmetry extension}}}}, 
which is firstly proposed in \cite{Wan2018djl1812.11955, Wan2018zql1812.11968} to construct gapped TQFTs or interfaces with higher 't Hooft anomalies:\footnote{By ``higher 't Hooft anomalies,'' we mean 
``the 't Hooft anomalies associated with higher global symmetries.''}
\bea  \label{eq:extension-2}
 \text{higher-global-symmetry-extension} \stackrel{\text{un/gauging}}{\Longleftrightarrow} \text{higher-gauge-(symmetry)-extension}.
\eea
While the (higher)-symmetry-breaking or (higher)-gauge-breaking gapped interfaces can be associated with
the phenomena of condensations of $n$-dimensional extended objects (0-particles, 1-strings, 2-branes, etc.) on the interfaces as in \Eq{eq:breaking-2},
the physical interpretations of (higher)-symmetry-extension are more subtle.
For example, 
\Refe{Wang2017loc1705.06728, Wang2018edf1801.05416} find that 1+1D gapped interfaces of 2+1D topological phases via the 
 symmetry-extension construction are in fact equivalent to (or dual to)
 another symmetry-breaking construction, thus which can be associated with the {anyon condensation} mechanism:
\bea \label{eq:extension-1-anyon}
 \text{2+1D bulk/1+1D interface: 
 global-symmetry-extension} \stackrel{\text{un/gauging}}{\Longleftrightarrow} \text{gauge-(symmetry)-extension} \nn\\
  \stackrel{\text{dual to}}{\Longleftrightarrow}  \text{gauge-(symmetry)-breaking}
   \Rightarrow
 \text{anyon condensation}.
\eea
Mysteriously \Refe{Wang2018edf1801.05416} also finds that \emph{certain specific}  higher-symmetry-extension (but may not be generic)
gapped interface constructions can be realized via the {{\bf\emph{{fuzzy-composite breaking}}}} mechanism.
By the {fuzzy-composite breaking}, we mean not the ordinary $n$-dimensional excitations alone are condensed on the interfaces,
but only the fuzzy-composite objects of extended operators can condense.
By fuzzy, we mean to emphasize the dimensionality differences between {\bf\emph{{the composite objects}}} and {{\bf\emph{{the quasi-excitation objects}}}}. More specifically, \Refe{Wang2017loc1705.06728} finds that a 2+1D higher-symmetry-extended gapped interface of 3+1D bulk
can be understood as a
new exotic topological interface on which neither 0-particle nor 1-string excitations alone condensed, but only
a 1-string-like composite object formed by a set of 0-particles can end on this special 2+1D boundary of 3+1D bulk.
In fact, such a composite 1-string formed by a set of 0-particles is analogous to the {\bf\emph{{particle-strings}}} or  {\bf\emph{{p-strings}}} 
that occurred in fracton 
orders.\footnote{The {\bf\emph{{particle-strings}}} or  {\bf\emph{{p-strings}}} are firstly proposed in \cite{Ma2017aogpstring1701.00747} for the coupled layer construction of fracton order.  
Via a different route,
\Refe{Wang2017loc1705.06728} finds a {\bf\emph{{composite string}}} condensation on the interface can give rise to
new exotic topological interfaces for bulk TQFTs. We will provide some examples in \Sec{sec:More-examples}.
} Thus for this example,
\emph{the composite object} is the {composite 1-string} (1+1D string worldsheet, by including time direction)
which is formed by \emph{the quasi-excitation objects} of 0-particles  (0+1D anyon worldline, by including time direction). We have:
\bea \label{eq:extension-p-string}
\hspace{-18mm}
 \text{3+1D bulk/2+1D interface: 
 higher-global-symmetry-extension} \stackrel{\text{un/gauging}}{\Longleftrightarrow} \text{higher-gauge-(symmetry)-extension} \nn\\
  \stackrel{\text{dual to}}{\Longleftrightarrow}  \text{fuzzy-composite-(symmetry)-breaking}
   \Rightarrow
 \text{composite string (p-string) condensation}.
\eea
More generally, 
\Refe{Wang2017loc1705.06728} 
proposes a generic construction of topological interfaces 
with a mixture of symmetry breaking, symmetry extension, and dynamical gauging.
We can generalize \Refe{Wang2017loc1705.06728} to the quantum systems with higher-$n$-symmetries
and the construction of topological interfaces with
a  {\bf\emph{{mixture of 
higher-symmetry breaking, higher-symmetry extension, and dynamical gauging}}} as:
\bea   \label{eq:mixed-breaking/extension}
 &&\hspace{-18mm}
 \text{higher-global-symmetry-mixed-breaking/extension} 
 \stackrel{\text{un/gauging}}{\Longleftrightarrow} \text{higher-gauge-(symmetry)-mixed-breaking/extension} \nn\\
 &&\Rightarrow
 \text{mixed condensations of anyonic $n$-dimensional extended objects (0-particles, 1-strings, 2-branes, etc.)}\nn\\
 && \text{and fuzzy composite extended objects + new sectors}.
\eea
Here the {\bf\emph{new sectors}} in \Eq{eq:mixed-breaking/extension} are associated with the emergent gauged sectors from the
symmetry extension construction \cite{Wang2017loc1705.06728}.\footnote{The symmetry extension construction \cite{Wang2017loc1705.06728} can be used to construct the 2+1D surface topological order of 3+1D SPTs.
The broad literature on the 2+1D surface topological order of 3+1D SPTs can be found in an excellent review  \cite{Senthil2014ooa1405.4015}.}

By combing the terminology of {non-liquid states} \cite{ZengWen20151406.5090, Swingle2014qpaMcGreevy1407.8203}
and {cellular topological states} \cite{Wen2020pri2002.02433}, we would like to focus on
the quantum phases of matter that we call
{\bf\emph{{Non-Liquid Cellular States}}},
which can be constructed via partitioning the space\footnote{It is also possible to generalize the space-partitioning to space-time partitioning.}
into several cells, while each cell (or the cell surface as in \cite{Wen2020pri2002.02433}) 
can host certain more familiar liquid topological phases (associated with TQFT data),
the cell and its neighbored cells can have overlapped interfaces.

\Refe{Wen2020pri2002.02433} provides a systematic construction of {\bf\emph{{gapped non-liquid cellular states}}} by gluing the anyon condensation induced gapped boundaries. In comparison, in our work, we would like to
glue the interfaces via the construction of the generic (global or gauge) higher-symmetry-breaking or symmetry-extension types
\cite{Wang2017loc1705.06728}, \cite{Wang2018edf1801.05416, Wan2018djl1812.11955, Wan2018zql1812.11968}. 
In fact, our construction of the interfaces via the symmetry-breaking or symmetry-extension construction may provide new insights
not only into {\bf\emph{{gapped non-liquid cellular states}}} but also 
into {\bf\emph{{gapless non-liquid cellular states}}}, because
some of our symmetry-breaking or symmetry-extended interfaces can also 
be \emph{gapless}.\footnote{However, the gapless non-liquid states are much more challenging than gapped non-liquid states.
For gapped non-liquid states, we can simply construct the gapped bulk cells and their gapped interfaces, with the topological consistency conditions on the gluing and the real-space renormalization.
For gapless non-liquid states, although we can obtain symmetry-breaking or symmetry-extended gapless interfaces, 
it is not entirely clear at this moment 
what are consistency conditions for the real-space renormalization for the entire bulk-interface gapless system, 
which we will leave some puzzles for gapless system for future work.}


\section{Topological, Geometrical, and Renormalization Formulas}

\subsection{Topological Consistency Criteria}

\subsubsection{Interface trivialization of cohomology/cobordism topological term of (higher)-classifying space}
\label{sec:trivialization-cohomology/cobordism-topological-term}

Here we aim to derive some consistency criteria
of the gapped interfaces via the construction of the generic (global or gauge) higher-symmetry-breaking or symmetry-extension types
\cite{Wang2017loc1705.06728}, \cite{Wang2018edf1801.05416, Wan2018djl1812.11955, Wan2018zql1812.11968}.
Suppose we consider $\tm$ multiple
of topological states (with its own label $j$, where $j=1,2,\dots, \tm$) in $d+1=D$ spacetime dimensions,
we assume that there is a topological term associated with each of the $j$-th topological states
 of some group $\mathbb{G}_j$.
Here $\mathbb{G}_j$ can be a higher group, thus $\B \mathbb{G}_j$ is its higher classifying space.\footnote{The
higher classifying space can include the Eilenberg-Maclane space $K(G,n)$.
For $n=1$, it is the ordinary classifying space $\B G=K(G,1)$.
For $n \geq 2$ and an abelian $G$, it is $\B^n G=K(G,n)$.}
We denote its generic topological term $\omega^{d+1}_j$,
which can be a generator of the cohomology group $\cH^{d+1}(\B\mathbb{G}_j,{U(1)})$
or cobordism group $\Hom(\Omega_{d+1}^X(\B\mathbb{G}_j), \U(1))$,\footnote{Here we need to choose the tangential structure $X$ of spacetime manifold, such as the $\SO/\Spin/\rO/\Pin^{\pm}$, their twisted cases and including higher symmetries, as in \cite{Wan2018bns1812.11967HAHSI}.
In our cobordism group formulation, we can consider not only the bosonic system without ($\SO$) or with ($\rO$) time-reversal symmetry,
 but also the fermionic system without ($\Spin$) or with ($\Pin^{\pm}$) time reversal symmetry.
See the systematic construction for fermionic systems in \cite{Guo2018vij1812.11959, Kobayashi2019lep1905.05391}.
Below we may leave the $X$ structure implicit, without written explicitly, unless we need to assume a specific structure of spacetime manifold.
}
\bea
\omega^{d+1}_{j}\in \cH^{d+1}(\B\mathbb{G}_j,{U(1)}), \quad\quad \text{or} \quad\quad \omega^{d+1}_j\in 
\Hom(\Omega^X_{d+1}(\B\mathbb{G}_j), \U(1)).
\eea
In other words, both the cohomology group classification and the cobordism group classification of 
topological phases can be directly implemented in our approach below.
Once we obtain the {\bf\emph{{topological term}}} (which corresponds to (higer-)iTQFT in math or (higer-)SPTs\footnote{By higher-iTQFT/SPTs, 
we mean the iTQFT/SPTs whose boundary has higher 't Hooft anomalies.} in condensed matter),
we can gauge the $\mathbb{G}_j$ in ${d+1}$D to obtain the generalized {\bf\emph{{higher-gauge theory}}} (possibly with higher $m$-gauge group).

In the more general $\mathbb{G}_j$, we denote $\mathbb{G}_j$ as a shorthand of
$$\mathbb{G}_j \equiv \mathbb{G}_{j,[0,1,\dots,n]},$$ 
where
the symmetry group $\mathbb{G}_j$ can include not just ordinary 0-symmetry acting on 0-points, but also 1-symmetry
${G}_{j,[1]}$ acting on 1-lines, 
2-symmetry
${G}_{j,[2]}$ acting on 2-faces, and so on to $n$-symmetry ${G}_{j,[n]}$ acting on $n$-simplex/simplices.
We can consider the general  $\mathbb{G}_j \equiv \mathbb{G}_{j,[0,1,\dots,n]}$,
where  $\B \mathbb{G}_j \equiv \B\mathbb{G}_{j,[0,1,\dots,n]}$ is a fibered higher classifying space as:\footnote{For instance, in the case of
having 0-symmetry and 1-symmetry together, we can consider:
$
\B^2{G}_{j,[1]} \hookrightarrow \B\mathbb{G}_{j,[0,1]}  \to \B{G}_{j,[0]}
$.}
\bea \label{eq:general-fiber}
\begin{tikzcd}
  \B^{n+1}{G}_{j,[n]} \arrow[hookrightarrow]{r}{}  \arrow[dash,dotted]{d} 
  &   \B\mathbb{G}_{j,[0,1,\dots,n]}  \arrow[dash,dotted]{d} & (\B\mathbb{G}_j  \equiv \B\mathbb{G}_{j,[0,1,\dots,n]})\\
  \B^3{G}_{j,[2]} \arrow[hookrightarrow]{r}{} &   \B\mathbb{G}_{j,[0,1,2]} \arrow[d] \\
  \B^2{G}_{j,[1]} \arrow[hookrightarrow]{r}{} &   \B\mathbb{G}_{j,[0,1]} \arrow[d] \\
    & \B{G}_{j,[0]}
\end{tikzcd}.
\eea
The gapped interface construction \cite{Wang2017loc1705.06728}, \cite{Wang2018edf1801.05416, Wan2018djl1812.11955, Wan2018zql1812.11968} requires the trivialization of $d+1$D 
topological term at the lower-dimensional interface (in $d$ spacetime dimension).
For example, we can consider the
{\bf\emph{{higher-symmetry breaking}}} construction by breaking all of $\B\mathbb{G}_j$  (again $j$ is the $j$-th layer index, $j=1,2,\dots, \tm$) 
to a common smaller space $\B\mathbb{G}_{\text{int}}$ where the lower subindex ``int'' meant for the ``interface'':
\bea
\B\mathbb{G}_{\text{int}} &\stackrel{\iota_1}{\longrightarrow}& \B\mathbb{G}_1, \nn \\
\B\mathbb{G}_{\text{int}}&\stackrel{\iota_2}{\longrightarrow}& \B\mathbb{G}_2,\nn\\
\vdots&\longrightarrow& \vdots \quad,\nn\\
\B\mathbb{G}_{\text{int}}&\stackrel{\iota_j}{\longrightarrow}& \B\mathbb{G}_j,\nn\\
\vdots&\longrightarrow& \vdots \quad,\nn\\
\B\mathbb{G}_{\text{int}}&\stackrel{\iota_m}{\longrightarrow}& \B\mathbb{G}_{\tm}. \label{eq:higher-symmetry-breaking}
\eea
The ${\iota_j}$ all are injective maps.
For another example, we can consider the
{\bf\emph{{higher-symmetry extension}}} 
construction by lifting all of $\B\mathbb{G}_j$  (again $j$ is the $j$-th layer index, $j=1,2,\dots, \tm$) 
to a common larger space $\B\mathbb{G}_{\text{int}}$ where the lower subindex ``int'' meant for the ``interface'':
\bea
\B \mathbf{N} \hookrightarrow&\B\mathbb{G}_{\text{int}} &\stackrel{r_1}{\longrightarrow} \B\mathbb{G}_1,\nn\\
\B \mathbf{N} \hookrightarrow&\B\mathbb{G}_{\text{int}}&\stackrel{r_2}{\longrightarrow} \B\mathbb{G}_2,\nn\\
\vdots \hookrightarrow& \vdots  &\stackrel{}{\longrightarrow} \vdots,\nn\\
\B \mathbf{N} \hookrightarrow&\B\mathbb{G}_{\text{int}} &\stackrel{r_j}{\longrightarrow} \B\mathbb{G}_{j},\nn\\
\vdots \hookrightarrow& \vdots  &\stackrel{}{\longrightarrow} \vdots,\nn\\
\B \mathbf{N} \hookrightarrow&\B\mathbb{G}_{\text{int}} &\stackrel{r_m}{\longrightarrow} \B\mathbb{G}_{\tm}.
\label{eq:higher-symmetry-extension}
\eea
The ${r_j}$ all are surjective maps.
Here $\B\mathbf{N}$ can be the (higher) classifying space of finite groups,
e.g. 
$$\B\mathbf{N} \equiv \B^{n+1}{N}_{[n]} (\rtimes   \dots \rtimes (\B^2{N}_{[1]} \rtimes \B{N}_{[0]}))$$
or a more general fibration similar to \Eq{eq:general-fiber}.\footnote{The semidirect product denotes a nontrivial
lower symmetry action on the higher symmetry. If the symmetry action is trivial between different $N_{[j]}$,
it will be a direct product structure.} For a more general example, we can consider the
{\bf\emph{{mixture of 
higher-symmetry breaking and higher-symmetry extension}}}  \cite{Wang2017loc1705.06728}, \cite{Wang2018edf1801.05416, Wan2018djl1812.11955}, mixing \Eq{eq:higher-symmetry-extension} and \Eq{eq:higher-symmetry-breaking}.
Now for either case of and other possible mixture of fibrations, including \Eq{eq:higher-symmetry-extension} or \Eq{eq:higher-symmetry-breaking},
we need to solve nontrivial 
{topological consistency criteria}:
\bea
&&\hspace{-18mm}\boxed{
\prod_{j=1}^\tm (r_j^* \omega^{d+1}_j (\{g_j\})) =
\prod_{j=1}^\tm (\omega^{d+1}_j (\{ r( {\bf g})\})) = 
\delta (\beta^{d}(\{ r({\bf g}) \})) 
 \simeq 1 \in \cH^{d+1}(\B\mathbb{G}_{\text{int}} ,{U}(1))
 \text{ or } 
 \Hom(\Omega_{d+1}(\B\mathbb{G}_{\text{int}}), \U(1))}, \quad\nn\\
&&g_j \in \mathbb{G}_j,   \quad\quad {\bf g} \in\mathbb{G}_{\text{int}},
\quad\quad \text{the map $r$ in \Eq{eq:higher-symmetry-extension} can be replaced to $\iota$ in \Eq{eq:higher-symmetry-breaking}}.
\eea
This formula says the joint gapped interface, associated with a higher classifying space $\B\mathbb{G}_{\text{int}}$,
needs to trivialize the product of all topological terms at once in the $\B\mathbb{G}_{\text{int}}$ by the pullback (the pullback is $r_j^*$
for the case of \Eq{eq:higher-symmetry-extension}, we should replace it to 
the pullback $\iota_j^*$ for the case of \Eq{eq:higher-symmetry-breaking}).

Once we obtain the gapped interface via
the higher-symmetry breaking/extension construction, by solving the suitable data 
$$\mathbb{G}_{\text{int}} \text{ and } \beta^{d},$$
we can perform the {\bf\emph{{dynamical gauging}}}
to get long-range entangled (LRE) topological states (see \cite{Wang2017loc1705.06728}, \cite{Wang2018edf1801.05416}). 

If there are $l$ lower-dimensional $d$D interfaces (labeled by $k$, where
$k=1,2,\dots, l$) also meet at say a further lower-dimensional $(d-1)$D junction,
then given $\mathbb{G}_{\text{int},k}$ {and} $\beta^{d}_k$ already solved above,
we need to again solve another set of nontrivial 
{topological consistency criteria} in one lower dimension:
\bea
\text{Analog of \Eq{eq:higher-symmetry-breaking} at one lower dimension} &:& 
\B \mathbf{N}' \hookrightarrow \B\mathbb{G}_{\text{junction}} \stackrel{r_k'}{\longrightarrow} \B\mathbb{G}_{{\text{int}}, k},\\
\text{Analog of \Eq{eq:higher-symmetry-extension} at one lower dimension} &:&  \B\mathbb{G}_{\text{junction}}\stackrel{\iota_k'}{\longrightarrow} \B\mathbb{G}_{{\text{int}}, k},
\eea
\bea
&&\hspace{-18mm}\boxed{
\prod_{k=1}^l ({r_k'}^* \beta^{d}_k (\{g'_k\})) =
\prod_{k=1}^l (\beta^{d}_k (\{ r( {\bf g}')\})) = 
\delta (\ga^{d-1}(\{ r({\bf g}') \})) 
 \simeq 1 \in \cH^{d}(\B\mathbb{G}_{\text{junction}} ,{U}(1))
 \text{ or } 
 \Hom(\Omega_{d}(\B\mathbb{G}_{\text{junction}}), \U(1))}, \quad\nn\\
&&g'_k \in \mathbb{G}_{\text{int}},   \quad\quad {\bf g}' \in\mathbb{G}_{\text{junction}},
\quad\quad \text{the map $r$ in \Eq{eq:higher-symmetry-extension} can be replaced to $\iota$ in \Eq{eq:higher-symmetry-breaking}}.
\eea
This formula says the joint gapped junction in $(d-1)$D, 
associated with a higher classifying space $\B\mathbb{G}_{\text{junction}}$,
needs to trivialize the product of all topological terms at once in the $\B\mathbb{G}_{\text{junction}}$ by the pullback.
We need to find appropriate solutions of
$$\mathbb{G}_{\text{junction}} \text{ and } \ga^{d-1}.$$
We can imagine more and more similar processes and so on, if there are more lower-dimensional junctions.

In fact, for many 1+1D gapped interfaces (when we have $d+1=2+1$),
the above set of data once are solved:
\bea \label{eq:trivialize-sol-data}
\mathbb{G}_j \text{ and } \omega^{d+1}_j, \quad \mathbb{G}_{\text{int}} \text{ and } \beta^{d},
\eea
we can convert them to a datum known as the tunneling matrix $\cW$  \cite{Lan2014uaa1408.6514} for
many 1+1D {\bf\emph{{dynamical gauged}}} gapped interfaces.
We will explain what this tunneling matrix $\cW$ is in the next subsection \Sec{sec:tunnelingW}.
We provide explicit examples of the data and the relations between \Eq{eq:trivialize-sol-data} and $\cW$ in \Sec{sec:Z2-Cellular}, 
 \Sec{sec:Z2t-Cellular} and others.
 The tunneling matrix $\cW$ plays the role of the set of anyon condensations on the interfaces,
 while the $\B\mathbb{G}_{\text{int}}\stackrel{\iota_j}{\longrightarrow} \B\mathbb{G}_j$ plays the role
 of gauge breaking, 
\bea \label{eq:breaking-intro-summary}
G_{\text{unbroken}} \stackrel{\iota}{\longrightarrow} G_{\text{original}}, \text{ or }  \quad G_{\text{to-be-condensed}}  \mapsto 0,
\eea
see the later  explanation in \Sec{sec:gapless-gap}.

\subsubsection{Generalized tunneling matrix $\cW$ as the interface tensor} 
\label{sec:tunnelingW}

Below we follow the formula derived in \cite{Lan2014uaa1408.6514} by Lan, the present author, and Wen.
\Refe{Lan2014uaa1408.6514} bootstraps topological 1+1D interfaces/boundaries of topological orders,
i.e., we can bootstrap topological 2-surface defects given by Abelian or non-Abelian TQFTs (with a finite number of types of line operators).
\Refe{Lan2014uaa1408.6514} labels the  2-surface defect (1+1D interface) as a \emph{tunneling matrix} $\cW$ that we recap below.

Given two unitary (non-)Abelian TQFTs (familiar topological orders described by the modular tensor category data) 
of $A$ and $B$, with the data of SL(2,$\Z$)-modular matrices\footnote{The mapping class group (MCG) of a 2-torus $T^2$ is 
$\rm{MCG}(T^2)=\rm{SL}(2,\Z)$, namely the rank-2 special linear group with $\Z$ coefficients, 
generated by
\bea
\hat{\cS}=
\bigl( {\begin{smallmatrix} 
 0&-1\\1&0
\end{smallmatrix}} \bigl),
\;\; \; \hat{\cT}=
\bigl( {\begin{smallmatrix} 
 1&1\\0&1
 \end{smallmatrix}}  \bigl).
\eea
For the $T^2$-coordinates $(x,y)$, the $\hat{\cS}$ maps $(x,y) \to (-y, x)$, 
while $\hat{\cT}$ maps $(x,y) \to (x+y, y)$. The modular data 
$\cS$ and $\cT$ is the quantum representation of $\hat{\cS}$ and $\hat{\cT}$
in terms of the Hilbert space basis of all of distinct anyons (associated with distinct line operators of TQFTs
that span the full ground-state Hilbert space of ${\bf Z}(M^{T^2} \times S^1)= \dim {\cal{H}}_{T^{2}}\equiv \GSD_{T^{2}}$).}
 and chiral central charges $c_-\equiv c_L - c_R$,\footnote{In fact, since the modular data (e.g. $\cS$ and $\cT$), and chiral central charge $c_-$, also play an important role in 2d CFT, the readers may digest that our way of constructing
 {\bf\emph{{liquid or non-liquid cellular states}}} may be understood as gluing CFTs on cells
 with overlapping patches via the CFT data (such as the CFT data in \cite{Moore1988qvSeiberg}).
}
$$\cS^A, \cT^A,c^A_- \text{ and } \cS^B,\cT^B,c^B_-.$$
Say we have $M$ and $N$ types of line operators (anyons) for TQFT$_A$ and TQFT$_B$,
then respectively the rank of SL(2,$\Z$)-modular matrices are $M$ and $N$.
In our treatment, we can first isolate the gapless sector (those chiral sectors cannot be gapped out)
away from the possible gappable sectors.
If $A$ and $B$ are connected by a gapped 2-surface defect, their 
net chiral central charges need to be  $c^A_--c^B_-=0$ required for the gappable sector.
Ref.\cite{Lan2014uaa1408.6514} introduces the interface defect labeled by 
a $N \times M$ \emph{tunneling matrix} $\cW$. Each entry $\cW_{Ia}$ represent \emph{fusion-space dimensions} with a
 non-negative integer $\mathbb{Z}_{\geq 0}$: 
\begin{align}
\cW_{Ia}\in\mathbb{Z}_{\geq 0},
 \label{Winteger}
\end{align}
satisfying
a \emph{commuting criterion} \Eq{commute}: 
\begin{align}
  \cW \cS^A = \cS^B \cW,\quad  \cW \cT^A = \cT^B \cW,
  \label{commute}
\end{align}
(\Eq{commute} imposes the consistency of anyon statistics to condense on a gapped interface,
analogous to Lagrangian subgroup later in \Sec{sec:gapless-gap}),
and a \emph{stable criterion} \Eq{stable}:
\begin{align}
  \cW_{ia}\cW_{jb}\leq\sum_{lc} (\cN^B)_{ij}^l \cW_{lc} (\cN^A)_{ab}^c\,.
  \label{stable}
\end{align}
Here 
$a,b,c,\dots$/$i,j,k,\dots$ are anyon (line operator) indices for TQFT$_A$/TQFT$_B$.
Given modular $\cS^A/\cS^B$, we can easily determine 
the fusion rules 
$(\cN^A)_{ab}^c$ and $(\cN^B)_{ij}^k$, for TQFT$_A$/TQFT$_B$ by Verlinde formula \cite{Verlinde:1988sn} 
$$
\cN_{ab}^c=\sum_{\al} \frac{\cS_{a \al} \cS_{b \al}{\cS_{c \al}^*}}{\cS_{1 \al}}\in \mathbb{Z}_{\geq 0}.
$$
The criteria whether there exists a topological 2-surface defect/interface, is equivalent to, 
whether there exists a non-zero (i.e., not all entries are zeros) solution $\cW$ under \Eq{Winteger},\Eq{commute} and \Eq{stable}. 
In principle, up to technical subtleties, we can bootstrap topological 2-surface defect/interface 
between two TQFTs by exhausting all solutions of $\cW$ analytically. 

A tunneling matrix entry $\cW_{ia}$ means that the anyon $a$ in
TQFT$_A$ has a number of
$\cW_{ia}$-splitting channels from $a$ to $i$ after going through an interface
 to TQFT$_B$. 
Moreover, it is well-known that we can use the {folding trick} to relate a gapped interface to a gapped boundary.
Thus we can bootstrap topological 2-surface defects both in the bulk interface and on the boundary.

Also by the \emph{folding trick}, we can rewrite a higher-rank interface tensor $\cW_{abcd\dots}$ to  
another rank-2 tunneling matrix $\cW_{a b'}$ between two TQFTs,
or another rank-1 tunneling matrix $\cW_{a' 1}$ between a TQFT to a trivial vacuum.
Therefore, below we will generally call the tunneling matrix as the interface tensor 
$\cW_{abcd\dots}$.\footnotemark 

%
%
\subsection{Geometrical and Renormalization Consistency Criteria: Crossing ``Symmetry''}
\label{sec:GeometricalRenormalizationConsistencyCriteria}

Below we consider the renormalization schemes and their formulas that are 
sensitive to the \emph{geometry} of sublattices. So these consistency formulas have their significant geometrical meanings behind. 

 \addtocounter{footnote}{-1} 
\footnotetext{It is possible to generalize the data to {higher-dimensional modular representation} 
\cite{Moradi2014kla1401.0518, Jiang2014ksa1404.1062, Wang2014oya1404.7854, Wang2019diz1901.11537}.
In the case of the unimodular group, 
there are the unimodular matrices of rank-$N$ matrix of GL$(N,\Z)$ form, which is the general linear group with $\Z$ coefficients.
\begin{itemize}
\item
GL$(N,\Z)$ is generated by 
$\hat{\cS}_\sfU$ and $\hat{\cT}_\sfU$ --- with determinants $\det(\hat{\cS}_\sfU)=-1$ and $\det(\hat{\cT}_\sfU)=1$ for
any general $N$:
\bea
&&\hat{\cS}_\sfU=\begin{pmatrix}
0 & 0 & 0 & \dots & (-1)^{N}\\
1 & 0 & 0 & \dots & 0 \\
0 & 1 & 0 & \dots & 0\\
\vdots & \vdots & \ddots & \dots & \vdots\\
0 & 0 & 0 & \dots & 0
\end{pmatrix}, 
\;\;\quad
\hat{\cT}_\sfU=\begin{pmatrix}
1 & 1 & 0 & \dots & 0\\
0 & 1 & 0 & \dots & 0\\
0 & 0 & 1 & \dots & 0\\
\vdots & \vdots & \vdots  & \ddots & \vdots\\
0 & 0 & 0  & \dots & 1
\end{pmatrix}.
\eea
Beware that $\det(\hat{\cS}_\sfU)=-1$ in order to generate both determinant 1 and $-1$ matrices.
\item  SL$(N,\Z)$ is generated by $\hat{\cS}$ and $\hat{\cT}$ ---
For the SL$(N,\Z)$ modular transformation, we denote their rank-$N$ generators as $\hat{\cS}$ and $\hat{\cT}$
with $\det(\hat{\cS})=\det(\hat{\cT})=1$:
\bea
&&\hat{\cS}=\begin{pmatrix}
0 & 0 & 0 & \dots & (-1)^{N-1}\\
1 & 0 & 0 & \dots & 0 \\
0 & 1 & 0 & \dots & 0\\
\vdots & \vdots & \ddots & \dots & \vdots\\
0 & 0 & 0 & \dots & 0
\end{pmatrix}, \;\;\quad 
\hat{\cT}=\hat{\cT}_\sfU.
\eea
\end{itemize}
\newpage
We then write down
the quantum amplitudes of the GL$(N,\Z)$ or SL$(N,\Z)$ generators projecting to degenerate ground states, say $|a\rangle$ and 
$|b\rangle$ as: 
\bea
(\cS_{\sfU})_{ab}=\langle a | \hat{\cS}_\sfU | b \rangle, \quad 
(\cT_{\sfU})_{ab}=\langle a | \hat{\cT}_\sfU | b \rangle;
\quad\quad\quad
\cS_{ab}=\langle a | \hat{\cS} | b \rangle, \quad 
\cT_{ab}=\langle a | \hat{\cT} | b \rangle.
\eea
}

\subsubsection{Hexagonal honeycomb column lattice}
\label{sec:Hexagonalhoneycombcolumnlattice}

\begin{figure}[!h]
(a)\includegraphics[width=0.16\columnwidth]{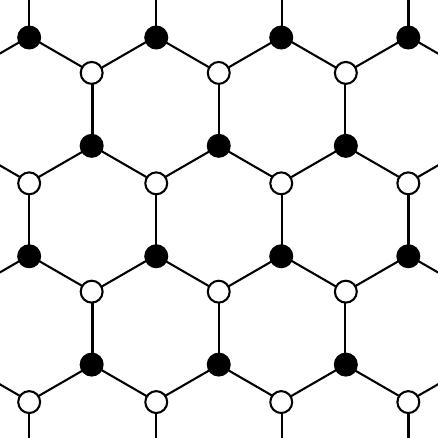}\;\;
(b)\includegraphics[height=0.15\columnwidth]{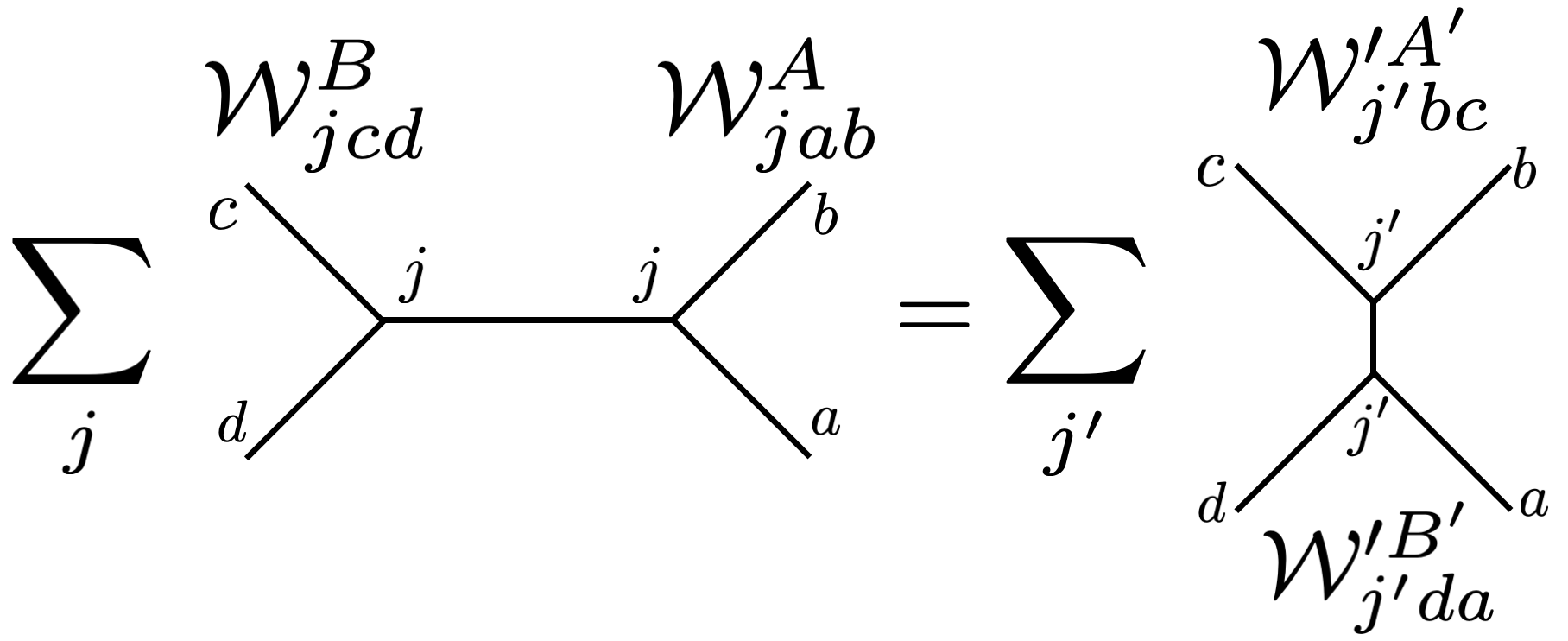} \\ 
(c)\includegraphics[height=0.15\columnwidth]{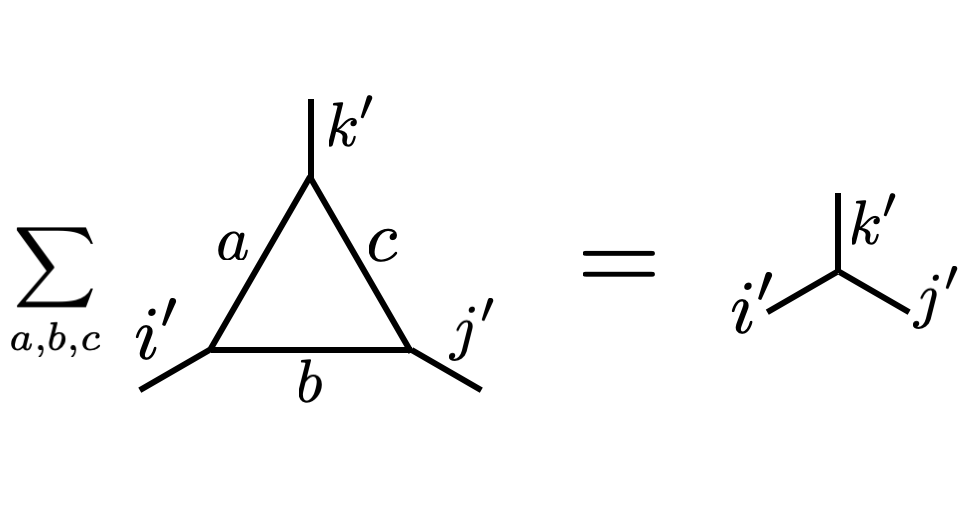}
\caption{(a) A hexagonal honeycomb lattice with two sublattices: black disk and white disk.
(b) The graphical interpretation of \eq{eq:hex-deformation}:
$\sum_j  \cW^A_{jab} \cW^B_{jcd} = \sum_{j'} (\cW'^{A'})_{j' bc} (\cW'^{B'})_{j'  da}$.
(c) The graphical interpretation of \eq{eq:hex-shrink-fuse}:
$ \sum_{a,b,c} (\cW'^{})_{ a i' b} (\cW'^{})_{b j' c} (\cW'^{})_{c k' a} = \t \cW^{}_{i' j' k'} $.}
\label{fig:Hexagonal}
\end{figure}

\begin{figure}[!h]
 \includegraphics[height=0.35\columnwidth]{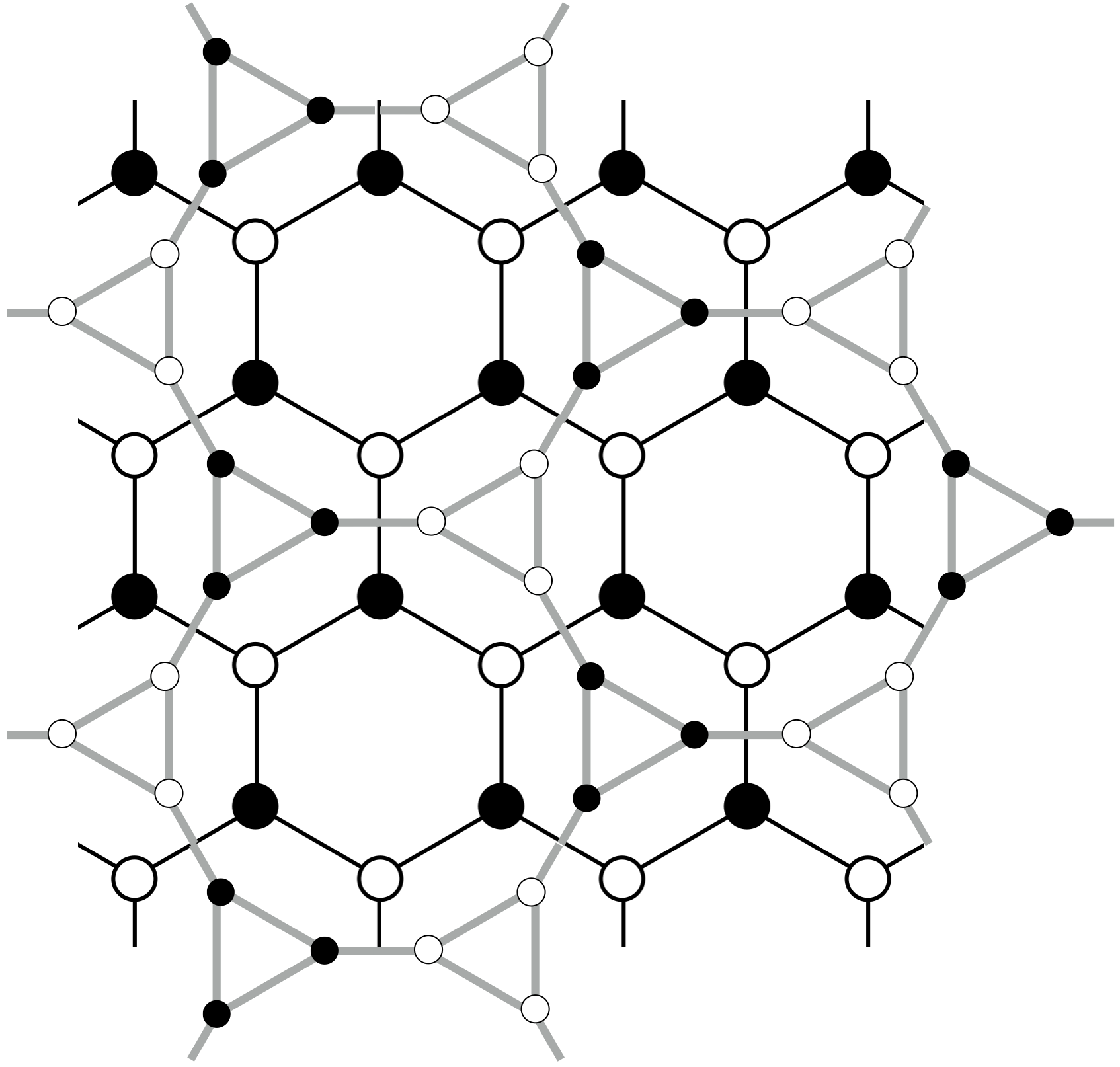}
\caption{The two-step RG on the real space on the honeycomb lattice.
The first step is via \Fig{fig:Hexagonal} (b), such that the black links on the honeycomb lattice are deformed to the gray links on the new lattice.
The second step is via \Fig{fig:Hexagonal} (c), any small triangle in the new lattice with gray links is shrunk to a point, so we obtain a new honeycomb lattice, 
with a larger new unit cell. There are still two kinds of sublattices (black and white) along the two-step RG deformation.}
\label{fig:Hexagonal-RG}
\end{figure}

For the hexagonal honeycomb column lattice (\Fig{fig:Hexagonal} (a)), there are two sublattices, let us label them $A$ and $B$ sublattices.
We perform the two-step RG on the real space.
Both steps have their significant geometry meanings behind because the renormalization scheme depends on
the \emph{geometry} of sublattices. (Of course there is also a slight topology meaning for it, because we need to obtain something stable.)
\begin{enumerate}[label=\textcolor{blue}{\arabic*.}, ref={},leftmargin=*]
\item
The first step is analogous to the crossing ``symmetry''\footnote{Here
the crossing ``symmetry'' is not the global symmetry in the sense of quantum matter and in the condensed matter terminology.
This crossing ``symmetry'' is merely a symmetry of the relabelings and rejoining external legs. 
} on the real space lattice (\Fig{fig:Hexagonal} (b)):
\bea
\label{eq:hex-deformation}
\sum_j  \cW^A_{jab} \cW^B_{jcd} = \sum_{j'} (\cW'^{A'})_{j' bc} (\cW'^{B'})_{j'  da},
\eea
while the $j$ and $j'$ are the two different ways of writing the internal channels connecting between external four legs $a,b,c, d$.
The sublattices change from $A$ and $B$ to $A'$ and $B'$.
The $\cW^A$ and $\cW^B$ are given and chosen by us,
while we need to solve $\cW'^{A'}$ and $\cW'^{B'}$.
 \Refe{Wen2020pri2002.02433} proposed to find 
 $\cW'^{A'}$ and $\cW'^{B'}$ that have the minimal quantum dimensions,
 that means that they have the lowest number of ground state subspace, thus GSD, say on the spatial torus.

\item The second step is the real space coarse graining ---  we enlarge the lattice cut-off scale by shrinking
several neighbor sites (or links/faces/n-simplices, etc.) and fuse them to a new single site (or a single link/face/n-simplex, etc.).
We call the three old neighbor sites $A'$,
and we call the three old neighbor sites $B'$;
they are fused to ${\t A}$ and ${\t B}$ with new interface tensors $\t \cW^{\t A}$ and $\t \cW^{\t B}$, see \Fig{fig:Hexagonal} (c):
\bea
\label{eq:hex-shrink-fuse}
 \sum_{a,b,c} (\cW'^{A'})_{ a i' b} (\cW'^{A'})_{b j' c} (\cW'^{A'})_{c k' a} = \t \cW^{\t A}_{i' j' k'}  ,
\nonumber\\
 \sum_{a,b,c} (\cW'^{B'})_{ a i' b} (\cW'^{B'})_{b j' c} (\cW'^{B'})_{c k' a} = \t \cW^{\t B}_{i' j' k'}.
\eea
\end{enumerate}
The new lattice cut off scale is $\sqrt{3}$ times larger than the 
old lattice cut off scale. The new unit cell is ${3}$ times larger than the old unit cell. 
See \Fig{fig:Hexagonal-RG}.

Then we can compare the old phases of matter with interface tensor data:
$\cW^A_{abc}$ and $\cW^B_{abc}$
to the new phases of matter with interface tensor data:
$\t \cW^{\t A}_{i' j' k'}$ and  $ \t \cW^{\t B}_{i' j' k'}$.
By reading and comparing these data, we can distinguish gapped phases whether they are liquids or non-liquids.
We leave the interpretations summarized in \Sec{sec:Interpretations}.

\subsubsection{Square column lattice}
\label{sec:squarecolumnlattice}

We can also consider the square column lattice. 
We perform the two-step RG on the real space.
\begin{figure}[!h]
(a)\;\includegraphics[width=0.35\columnwidth]{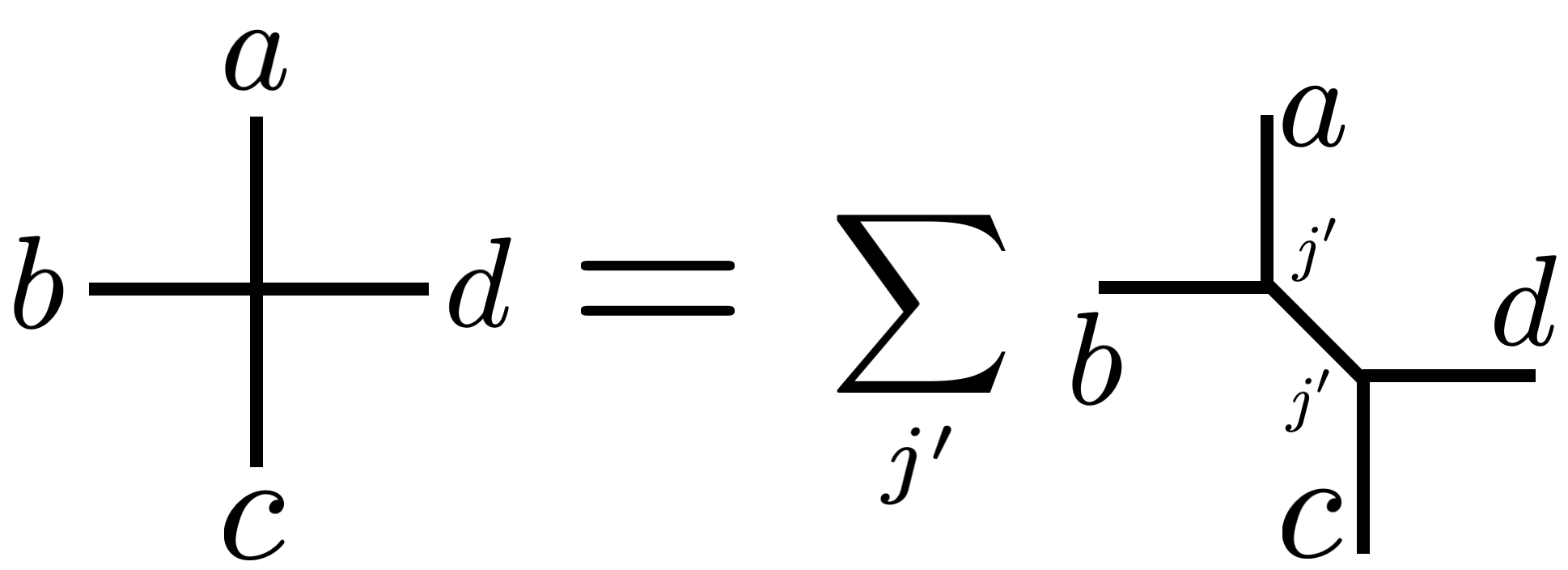}\;\;\;\;
(b)\;\includegraphics[width=0.35\columnwidth]{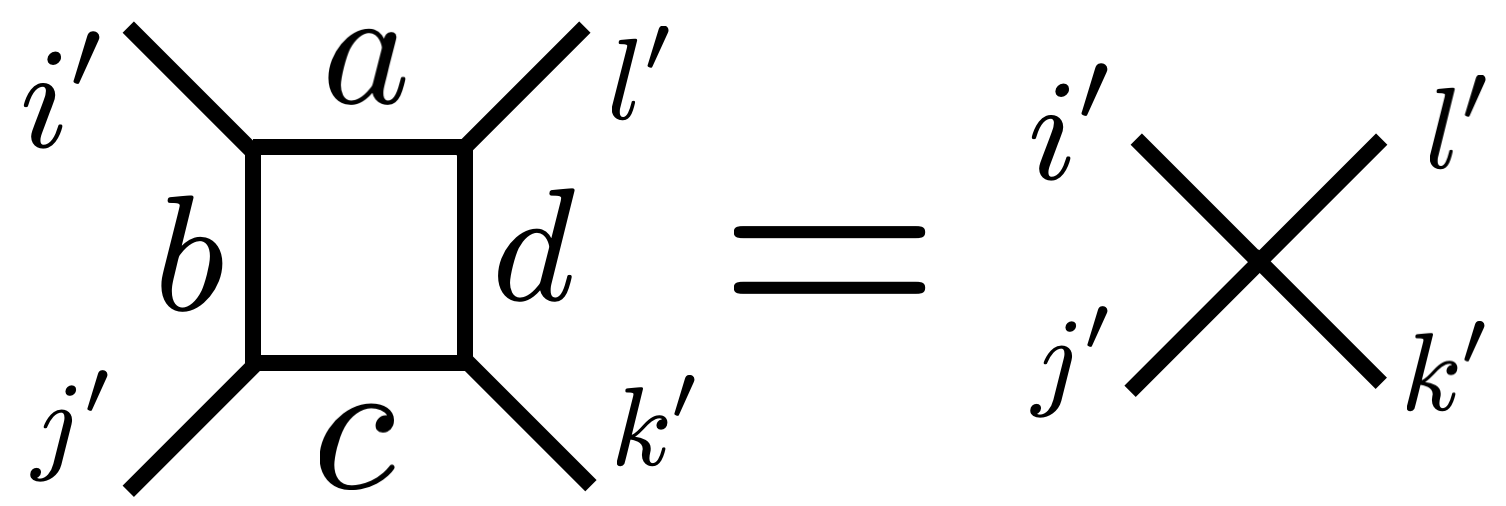}
\caption{
(a) The graphical interpretation of \eq{eq:sq-deformation}:
$\cW_{abcd}  = \sum_{j'} (\cW'^{})_{ab j'  } (\cW'^{})_{j'  cd}$.
(b) The graphical interpretation of \eq{eq:sq-shrink-fuse-single}:
$\sum_{a,b,c,d} (\cW'^{})_{ a i' b} (\cW'^{})_{b j' c} (\cW'^{})_{c k' d} (\cW'^{})_{d l' a} = \t \cW^{}_{i' j' k' l'}$.
}
\label{fig:square}
\end{figure}
\begin{figure}[!h]
(a)\;\includegraphics[height=0.35\columnwidth]{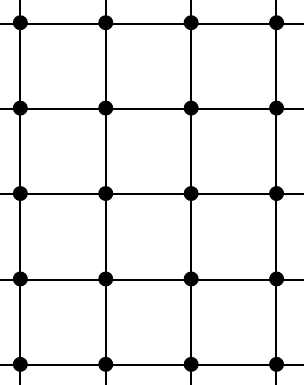}\;\;\;\;
(b)\;\includegraphics[height=0.35\columnwidth]{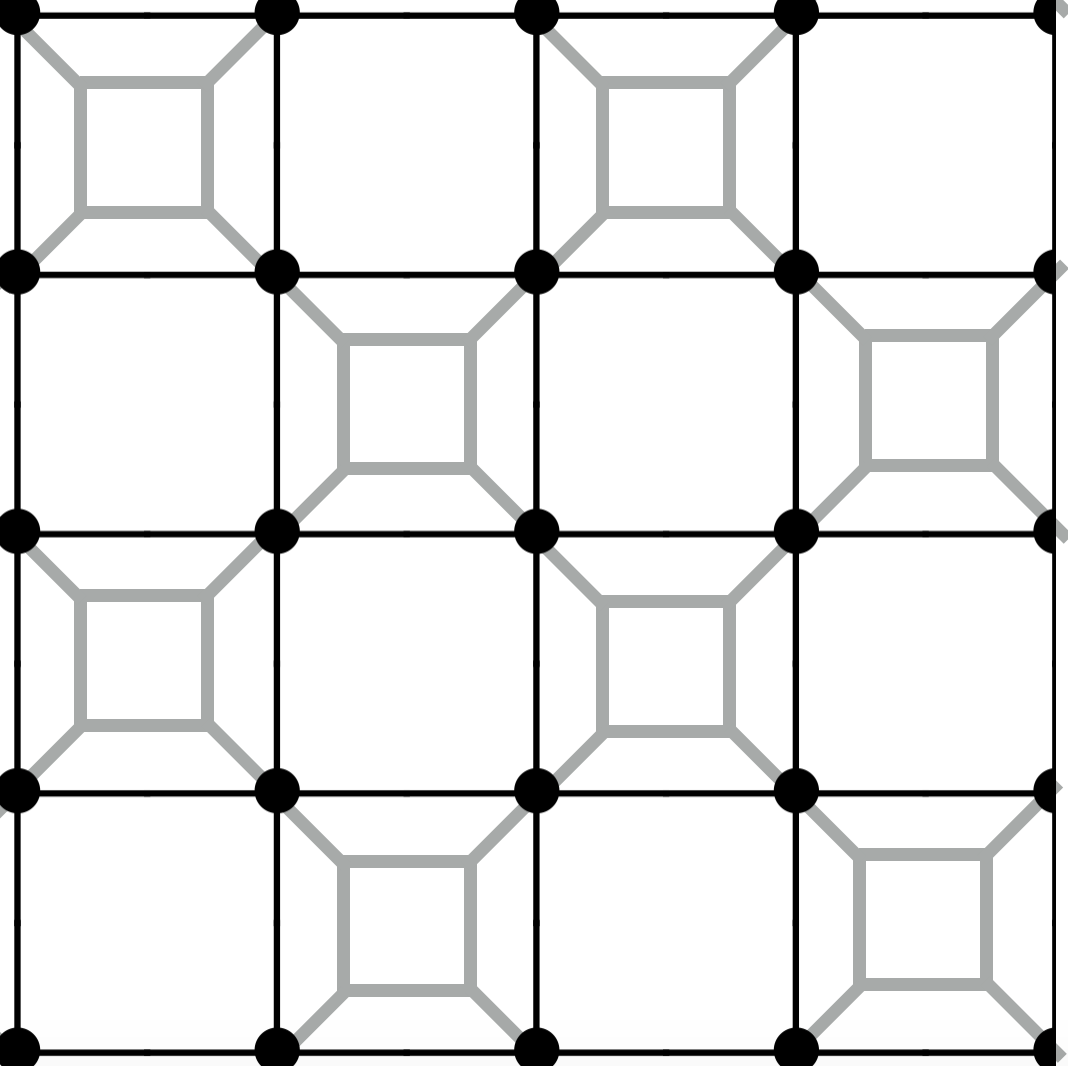}
\caption{(a) The square lattice with a single sublattice.
(b) {The two-step RG on the real space on the square lattice.
The first step is via \Fig{fig:square} (a), such that the black links on the square lattice are deformed to the gray links on the new lattice.
The second step is via \Fig{fig:square} (b), any small square in the new lattice with gray links is shrunk to a point, so we obtain a new square lattice, 
with a larger new unit cell.}
}
\label{fig:square-RG}
\end{figure}
\begin{enumerate}[label=\textcolor{blue}{\arabic*.}, ref={},leftmargin=*]
\item
The first step is analogous to the crossing ``symmetry'' on the real space lattice. We try to rewrite the 4-leg tensor
split to two 3-leg tensors via finding the appropriate interface tensors satisfy (\Fig{fig:square} (a)):
\bea
\label{eq:sq-deformation}
\cW_{abcd}  = \sum_{j'} (\cW'^{})_{ab j'  } (\cW'^{})_{j'  cd},
\eea
If there are two sublattices $A$ and $B$, then we have two sets:
\bea
\label{eq:deformation-AB}
 \cW^A_{abcd}  = \sum_{j'} (\cW'^{A'})_{ab j'  } (\cW'^{A'})_{j'  cd},\quad
  \cW^B_{abcd}  = \sum_{j'} (\cW'^{B'})_{ab j'  } (\cW'^{B'})_{j'  cd}.
\eea
\item The second step is the real space coarse graining --- we fuse four 
old neighbor sites to a new single site (or a single link/face/n-simplex, etc.).
If there is only one type of sublattice, we can fuse the deformed $\cW'^{}$ together to $\t \cW$, see \Fig{fig:square} (b): 
\bea
\label{eq:sq-shrink-fuse-single}
 \sum_{a,b,c,d} (\cW'^{})_{ a i' b} (\cW'^{})_{b j' c} (\cW'^{})_{c k' d} (\cW'^{})_{d l' a} = \t \cW^{}_{i' j' k' l'}.
\eea
If there are two types of sublattices $A$ and $B$ at the beginning, we may also have the 
two types of deformed sublattices $A'$ and $B'$.
In the tensor network language, the factorization can be obtained using the singular value decomposition (SVD).
We may {\bf\emph{{decorate}}}
 and {\bf\emph{{truncate}}} 
 \emph{some} of the $\cW'^{A'}$ and $\cW'^{B'}$ tensors by reducing the quantum dimensions (or the bond dimensions) following the improved tensor renormalization group approach (TRG) or the tensor network renormalization approach (TNR) on square lattices \cite{Levin2006jaiTensor0611687, Evenbly_2018.98.085155}.
We call those truncated tensors as $\cW''^{A''}$ and $\cW''^{B''}$.
Through the appropriate identification from the old sublattices to the new sublattices,
we can fuse their deformed $\cW'^{}$ or $\cW''$ together to $\t \cW$ as: 
\bea
\label{eq:sq-shrink-fuse-double}
 \sum_{a,b,c,d} (\cW'^{A'})_{ a i' b} (\cW'^{B'})_{b j' c} (\cW''^{A''})_{c k' d} (\cW''^{B''})_{d l' a} = \t \cW^{\t A}_{i' j' k' l'} ,
\nonumber\\
 \sum_{c,d,a,b} (\cW''^{A''})_{c k' d} (\cW''^{B''})_{d l' a}   (\cW'^{A'})_{ a i' b} (\cW'^{B'})_{b j' c}= \t \cW^{\t B}_{k' l' i' j' }.
\eea

\end{enumerate}

\subsubsection{Interpretations of renormalized interface tensors: $\cW'$ and $\t \cW$}
\label{sec:Interpretations}

Starting from the given data $\cW$,
since there are two steps above for the geometrical and renormalization consistency formulas,
we comment on the interpretations of 
$\cW'^{}$ and $\t \cW$.
\begin{enumerate}[label=\textcolor{blue}{\arabic*.}, ref={},leftmargin=*]
\item
The interpretation of the solution of the minimal total quantum dimensions of $\cW'$ by Wen \cite{Wen2020pri2002.02433} is the following.

 If only $\cW$ is renormalized, say in \Eq{eq:sq-deformation},
 $\cW 
 \Rightarrow \cW'
 $,
 then we compare $\cW'$ to $\cW$.
If there are multi-sublattices, say $\cW^A,\cW^B, \dots$ are renormalized, say 
 $\cW'^{A'},\cW'^{B'}, \dots$ in \Eq{eq:hex-deformation} or \Eq{eq:deformation-AB},
 then we compare the set of $\cW'$ (as the shorthand for the list $\cW'^{A'},\cW'^{B'}, \dots$) 
 to the set of $\cW$ (as the shorthand for the list $\cW^A,\cW^B, \dots$):\\ 
 (i) If $\cW'$ can be chosen to be either trivial or to remain the same as $\cW$,
 then the gapped quantum states are 
 {\bf\emph{{gapped liquid cellular states}}}.\\
 (ii) If $\cW'$ cannot be chosen to be either trivial nor to be $\cW$,
 then the gapped quantum states are 
 {\bf\emph{{gapped non-liquid cellular states}}}.
 
\item The interpretation of the second-step solution
$\t \cW$ is that,
if the set of $\t \cW$, say $\cW^{\t A}, \cW^{\t B}, \dots$ 
becomes the multiple n times of the original data $\cW^{ A}, \cW^{ B}, \dots$, namely
$$
(\cW^{\t A}, \cW^{\t B}, \dots) = \text{n}
(\cW^{ A}, \cW^{ B}, \dots),
$$
then we should test whether the increasing exponential degeneracy of the 
$$
 \text{n}^{\text{the number of sublattices in the volume}}
$$
is robust against the local perturbation. 
It may be the symmetry-breaking degeneracy. 
In most cases, this is just the accidental degeneracy 
not  robust against the local perturbation; thus the underlying data of coarse-grained renormalized
state is still the same  {\bf\emph{{gapped liquid cellular states}}} \cite{Wen2020pri2002.02433}.

In contrast, there are states that do not even allow such comparison, since $\cW'^{}$ and $\t \cW$ may be totally different from
the starting $\cW$, again those states can be  {\bf\emph{{gapped non-liquid cellular states}}}.

If the outcome state is a {\bf\emph{{gapped liquid cellular state}}}, 
we can read from the interface tensor $\cW$ to identify the potential corresponding TQFTs (say, 3+1D TQFT,
in the case of $\cW$ stands for the 1+1D interface between 2+1D cell TQFTs; the cellular TQFTs can be some dimensional higher than the original TQFTs on the cells).

If the outcome state is a {\bf\emph{{gapped non-liquid cellular state}}}, 
we can read from the interface tensor $\cW$ to identify the trajectory
of fractionalized (anyonic) excitations.
If certain fractionalized excitations cannot move freely crossing some of the interface tensor $\cW$,
then this confirms that the excitations have restricted mobility.
For the excitations having the restricted mobility:\\ 
$\bullet$ moving only along 2D spatial plane freely, the excitations are known as {\bf\emph{{planons}}}.\\
$\bullet$ moving only along 1D spatial line freely, the excitations are known as {\bf\emph{{lineons}}}.\\
$\bullet$ restricted entirely anywhere in the space, the excitations are known as {\bf\emph{{fractons}}}.\\

Moreover,
if there exist \emph{string operators} to move fractionalized excitations and change the ground state to different topological sectors, 
then we can determine this state is a {\bf\emph{{type-I fracton order}}}.

If there do not exist any \emph{string operators} to move fractionalized excitations and change the ground state to different topological sectors, 
then we can determine this state is a {\bf\emph{{type-II fracton order}}} \cite{Vijay2016phm1603.04442}.

A series of operators that generate any  {\bf\emph{{fractal}}} structure pattern
act in the configuration of the real-space wavefunction,
can still (indecisively) produce a {\bf\emph{{type-I}}} or {\bf\emph{{type-II fracton order}}} \cite{Yoshida1302.6248}.

\item At this moment, we do not know the transparent relation between the structure of 
the renormalization of the interface tensor performed in this section
and
the entanglement renormalization group (ERG, e.g., performed previously in \cite{Haah1310.4507, Shirley1712.05892, Dua1909.12304}).
It is an interesting direction for future works.

\end{enumerate}


\section{Gapless Conformal to Gapped Boundary Conditions}
\label{sec:gapless-gap}
For the gapped interfaces of 2+1D abelian topological orders, we can
obtain a field-theoretic understanding of their 1+1D boundaries/interfaces via 2+1D (3d) Chern-Simons gauge TQFTs (CS) \cite{Wang2012am1212.4863}. The 3d bulk Chern-Simons action 
\bea \label{eq:Sbulk}
S_{\text{3d bulk}} &=&
\frac{K_{IJ}}{4\pi}\int_{\mathcal{M}^3}  a_I \wedge \dd a_J
=
\frac{K_{IJ}}{4\pi}\int_{\mathcal{M}^3}\dd t\; \dd^2x \; \epsilon^{\mu\nu\rho} a_{I,\mu} \partial_\nu a_{J,\rho},
\eea
includes
the symmetric integer bilinear $N \times N$ matrix $K_{IJ}$,
where $a_I = a_{I,\mu} \dd x^\mu$ is the 1-form gauge field's $I$-th component in the multiplet.
In condensed matter, the $a$ gauge fields are emergent degrees of freedom after integrating out the bulk gapped matter fields.
If $\mathcal{M}^3$ has a boundary $(\partial \mathcal{M})^2$, 
what kinds of stable boundary conditions are allowed?
Consider arbitrary variations of gauge field $a$ on the boundary $(\partial \mathcal{M})^2$ as $a_{\partial} \to a_{\partial} + \delta a_{\partial}$,
where $a_{\partial}$ denotes the boundary 1-form gauge field.
Then we require the variation of $S_{\text{3d bulk}}$ on $(\partial \mathcal{M})^2$ to be vanished \cite{1008.0654KapustinSaulina}:
\bea
\delta_{\text{2d, bdry}} (S_{\text{3d bulk}})
= 
\frac{K_{IJ}}{4\pi}\int_{(\partial \mathcal{M})^2} 
a_{\partial, I} \wedge \delta a_{\partial, J}.
\eea
The differential $\delta$ of this variation is in fact a symplectic form
\bea
\omega_{\text{Sp}} = \frac{K_{IJ}}{4\pi}\int_{(\partial \mathcal{M})^2} 
\delta a_{\partial, I} \wedge \delta a_{\partial, J}
\eea
on the space of boundary gauge fields. Consistent stable boundary conditions on $(\partial \mathcal{M})^2$
defines a Lagrangian submanifold with respect to the symplectic form
$\omega_{\text{Sp}}$.

\begin{enumerate}[label=\textcolor{blue}{\arabic*.}, ref={},leftmargin=*]
\item One way to define a consistent boundary condition is setting one component of $a_{\partial}$ vanished on $(\partial \mathcal{M})^2$,
such as
\bea
\big( K_{IJ} a_{J, t}-V_{IJ} a_{J, x}\big) \bigg\rvert_{(\partial M)^2}=0, 
\eea
where $V_{IJ}$ is also a symmetric integer bilinear $N \times N$ matrix, known as the velocity matrix.
The $V_{IJ}$ is positive definite for the \emph{potential energy} in the Hamiltonian term to be bounded from below.
Then we find the boundary action can be:
\bea  \label{eq:Sedge}
S_{\partial}&=& \frac{1}{4\pi} \int_{\partial \mathcal{M}} \dd t \; \dd x \; ( K_{IJ} \partial_t \phi_{I} \partial_x \phi_{J} -V_{IJ}\partial_x \phi_{I}   \partial_x \phi_{J} ),
\eea
where the boundary variations of
$a_{\partial, I}$ is meant to be cancelled by the variations of $\dd \phi_{I}$.
The theory given by the action \Eq{eq:Sedge}
is known as the 1+1D $K$-matrix (non-)chiral bosons or generalized Luttinger liquids in condensed matter.
With a proper choice of velocity matrix with a rescaled ``speed of light,''
the 1+1D gapless theory is a conformal field theory (CFT).
 This theory describes the 2d gapless CFT.
 Each gapless mode associated with  $\phi_{I}$ has an individual chiral central charge 1.
The number of left moving modes subtracts the number of right moving modes giving the total chiral central charge
 $$c_- \equiv c_L-c_R= \text{signature}(K),$$
is the signature of $K$ matrix given by the numbers of positive eigenvalues subtracting negative eigenvalues.

 \item We can also approach from a gapless boundary to a topological gapped boundary.
From \cite{Wang2018edf1801.05416}, we can set certain boundary gauge degrees of freedom to vanish,
\bea \label{eq:bdry-cond-a}
\ell_{\ra,I}  a_I \bigg\rvert_{\partial \mathcal{M}} =0.
\eea
The boson modes $\phi_{I}$, originally related by the gauge transformation $a_I \to a_I + \dd \lambda_I$ and $\phi_{I} \to \phi_{I}- \lambda_I$,
now {may} condense on the boundary, which means the nonzero vacuum expectation value
\bea
\langle \exp[\ii  (\ell_{I}^{} \cdot\phi_{I})] \rangle \bigg\rvert_{\partial \mathcal{M}} \neq 0, \;\;\;\; \text{more precisely,  indeed }
\langle \exp[\ii  (\frac{\ell_{I}}{|\gcd({\ell_{I}}) |} \cdot\phi_{I})]  \rangle \bigg\rvert_{\partial \mathcal{M}} \neq 0,
\eea
where $\gcd({\ell_{I}}) \equiv \gcd(\ell_1, \ell_2, \dots, \ell_N)$ is the greatest common divisor (gcd) of the all components of $\ell$.
This condensation can be triggered by the sine-Gordon cosine term at strong coupling $g$\footnote{We will
interchangeably write the $\ell$ vector in terms of its transpose vector form $\ell^T$.
It should not cause any ambiguity, as long as we fit the $\ell$ vector to its proper form.
For example,
$\cos(\ell_{I}^{} \cdot\phi_{I})$ is meant to be
$\cos(\ell_{}^{T} \cdot\phi_{})$, but
we can interchangeably write the transpose row vector $\ell^T$ as the column vector $\ell$ sometimes.
\label{footnote:L}
}
\bea
g \int_{\partial \mathcal{M}} \dd t \; \dd x   \cos(\ell_{I}^{} \cdot\phi_{I}).
\eea
Suppose we add a set of interactions
\bea  \label{eq:Sedge-int}
S_{\partial,\text{interaction}}&=&
 \int_{\partial \mathcal{M}} \dd t \; \dd x\;  \sum_{\ra} g_{\ra}  \cos(\ell_{\ra,I}^{} \cdot\phi_{I}),
\eea
where ``$\ra$'' labels different 
components, such that they satisfy the 
topological gapping conditions (or known as Lagrangian subgroup conditions):
There exists a $N \times N/2$-component matrix $\mathbf{L}$, such that
$\mathbf{L}$ is formed by ${N/2}$ column vectors of these  linear-independent vectors $\ell_\ra$ ($\ra=1,2,\dots, {N/2}$),
\bea
\mathbf{L} \equiv \Big(\ell_{1}, \ell_{2}, \dots,  \ell_{N/2}  \Big).
\eea
By saying so, this only makes sense that
$N$ is an even integer,
also it turns out that it must be $c_- \equiv c_L-c_R= \text{signature}(K)=0$ thus free from perturbative gravitational anomalies.
By \emph{anyon condensations} on the boundary, 
we mean the set of \emph{condensed anyons} should be generated by a subset of
\bea \label{eq:anyon-line}
\ell_{I}' = n \frac{\ell_{I}}{|\gcd({\ell_{I}}) |}, \quad n \in \Z_{|\gcd({\ell_{I}}) |}.
\eea
\emph{Anyons} are labeled by $\ell_{I}'$ corresponding to the line operator $\exp(\ii \ell_{\ra,I}' \int a_I)$ in the bulk.
The anyons who live on the open ends of this line operator can end thus condense on the boundary as \eqn{eq:bdry-cond-a}. 
We define 
the abelian mutual/self statistics phase of anyons, given the abelian Chern-Simons theory \eqn{eq:Sbulk}, 
associated with the line operator $\exp(\ii {\ell'}_{\ra,I} \int a_I)$, as
\begin{equation}
\begin{array}{rcl}
\exp[\ii \theta_{\text{mutual}}]  &=&\exp[\ii \theta_{\ra\rb}]= \exp[ \ii \, 2\pi \, {\ell'}_{\ra,I}^{} K^{-1}_{IJ} {\ell'}_{\rb,J}^{}],\;\;\\
\exp[\ii \theta_{\text{self}}]   &=&\exp[\ii \frac{\theta_{\ra\ra}}{2}]= \exp[ \ii \pi \, {\ell'}_{\ra,I}^{} K^{-1}_{IJ} {\ell'}_{\ra,J}^{}].\;\;
\end{array}
\end{equation}
This can be derived as the path integral of Hopf link of two line operators labeled by $ \ell_{\ra}$/$\ell_{\rb}$  with a proper normalization.

{\bf Topological gapping conditions} are equivalent to that the existence of such set of $\mathbf{L}$ satisfies:
\begin{itemize}
\item
Trivial self statistics for each anyon type ``$\ra$'':
$${\ell'}_{\ra,I}^{} K^{-1}_{IJ} {\ell'}_{\ra,J}^{} \in 2 \mathbb{Z}$$ 
even integers for bosonic systems (non-spin TQFTs), or in 
$${\ell'}_{\ra,I}^{} K^{-1}_{IJ} {\ell'}_{\ra,J}^{} \in \mathbb{Z}$$ 
odd integers for fermionic systems (spin TQFTs).
This means that the self-statistics of ${\ell'}_{a}$ line operator is bosonic/fermionic, with $\theta_{\text{self}}$ a multiple $2\pi$ or $\pi$ phase.
\item Trivial mutual statistics between two anyons ${\ell'}_{\ra}$ and ${\ell'}_{\rb}$:
$${\ell'}_{\ra,I}^{} K^{-1}_{IJ} {\ell'}_{\rb,J}^{} \in \mathbb{Z}$$
thus yield a trivial mutual bosonic statistical phase.
\end{itemize}
\end{enumerate}

The above theories are in fact 
RG fixed point 3d TQFT \Eq{eq:Sbulk} 
in the bulk, the 2d gapless theory \Eq{eq:Sedge} or  2d gapped boundaries \Eq{eq:Sedge-int} on the edge.
Importantly, as noticed in \Refe{Wang2017loc1705.06728} (Appendix E and F), many such
anyon-condensation induces gapped interfaces can be achieved by the dynamically gauging of trivialization of the bulk topological term
on the boundary via the (gauge-)symmetry-breaking mechanism: 
\bea \label{eq:breaking}
G_{\text{unbroken}} \stackrel{\iota}{\longrightarrow} G_{\text{original}},
\eea
where  $G_{\text{original}}$ is the original (gauge or global symmetry) group,
while $G_{\text{unbroken}}$ is the unbroken subgroup, the map $\iota$ is an injective map.
Alternatively, we have a surjective map where all-to-be-condensed excitations in $G_{\text{to-be-condensed}}$
will be broken completely to nothing (the trivial group):
\bea  \label{eq:breaking-0}
 G_{\text{to-be-condensed}}  \mapsto 0,
\eea
many elements map to the trivial element.
We will demonstrate many examples, including higher-symmetry cases in the following sections.


\section{Cellular States from $\Z_2$-Gauge Theory ($\Z_2$-Toric Code)}

\label{sec:Z2-Cellular}

\subsection{$\Z_2$-gauge theory: 2 types of gapped boundaries} 
\label{sec:Z2-1}
We start from 2+1D $\Z_2$-toric code (TC) also known as 3d $\Z_2$-gauge theory at low energy.
The TQFT data of $\Z_2$-toric code are given by the representation of the modular SL(2,$\Z$), 
generated by $\cS,\cT$ matrices:\footnote{The mapping class group (MCG) of $d$-torus $\mathbb{T}^d=(S^1)^d$
is \bea
\text{MCG}(\mathbb{T}^d)= \text{SL}(d,\Z). 
\eea For topological order, we study the Hilbert space of generate ground states on $(\mathbb{T}^d)$, 
and the associated representations of the mapping class group (MCG) of $\mathbb{T}^d$. 
\label{footnote:MCG}
}
\begin{align}
  \cS&=\frac{1}{2}\begin{pmatrix}
    1&1&1&1\\
    1&1&-1&-1\\
    1&-1&1&-1\\
    1&-1&-1&1
  \end{pmatrix},\\
    \cT&=\diag(1,1,1,-1),
\end{align}
where each row and column index entry runs from the anyon sectors $1, e, m,$ and $em$.
Since $em$ is fermionic (with its $T$ matrix eigenvalue $-1$), we also write $f=em$.
They are the trivial vacuum sector, the electric particle $e$ (of $\Z_2$-Wilson line),
 the magnetic particle $m$ (of $\Z_2$-'t Hooft line, as the $\Z_2$-dual object),
and the dyon particle $em$.
There are two types of gapped boundaries solved from \cite{Lan2014uaa1408.6514}:\footnote{For the convenience of readers, we
may sometimes denote the tunneling matrix say 
$\cW=( 1  \dots  \dots  \dots )$ with explicit column and row labels as
$\left(
\begin{array}{cccc c c}
1 &  a_1 & a_2 & \dots   & & \\
\hline
 1 & \dots & \dots & \dots & \;\vline & 1 
\end{array}
\right)$,
where 1 means the trivial vacuum,
while the $a_j$ is the $j$-th anyonic sector (also called the topological super-selection sector).
}
\begin{align}
  \cW^\text{TC}_e&=\begin{pmatrix} 1&1&0&0 \end{pmatrix}=
\left(
\begin{array}{cccc c c}
1 &  e & m & em   & & \\
\hline
 1 & 1 & 0 & 0 & \;\vline & 1 
\end{array}
\right),
  \\
  \cW^\text{TC}_m&=\begin{pmatrix} 1&0&1&0 \end{pmatrix}
  =
\left(
\begin{array}{cccc c c}
1 &  e & m & em   & & \\
\hline
 1 & 0 & 1 & 0 & \;\vline & 1 
\end{array}
\right).
\end{align}
We can also organize the tunneling matrix $\cW$ data via the condensed anyons (i.e., anyons that are allowed to be condensed) 
\bea 
  \cW^\text{TC}_e&:& 1, e.\\
  \cW^\text{TC}_m&:&1, m.
\eea
We can also interpret it as the gauge-breaking data, where the right hand side is the
$\Z_2^{e} \times \Z_2^{m}$ including the electric gauge group $\Z_2^{e}$ and the dual magnetic group $\Z_2^{m}$.
This $\Z_2^{e} \times \Z_2^{m}$ group also is the group of fusion algebra of the TQFT.
In terms of the breaking notation\footnote{In the following discussions, we may provide
$G_{\text{unbroken}}$ and $G_{\text{original}}$ as the electric gauge group $G_e$, or the magnetic gauge group $G_m$, or both the electric and 
magnetic gauge groups. For example, we have the original 
$G_e=  \Z_2^{e}$,
$G_m=  \Z_2^{m}$,
and
$G_e \times G_m=  \Z_2^{e} \times \Z_2^{m}$.}
in \Eq{eq:breaking}'s
$G_{\text{unbroken}} \stackrel{\iota}{\longrightarrow}  G_{\text{original}},$
and \Eq{eq:breaking-0}'s $G_{\text{to-be-condensed}}  \mapsto 0$,
we find $\cW^\text{TC}_e$ is equivalent to the gauge-breaking pattern: 
\bea
\Z_2^{m} & \stackrel{\iota}{\longrightarrow}& \Z_2^{e} \times \Z_2^{m},\\
 \Z_2^{e} &\mapsto& 0. 
\eea
We find $\cW^\text{TC}_m$ is equivalent to the gauge-breaking pattern:  
\bea
\Z_2^{e} & \stackrel{\iota}{\longrightarrow}& \Z_2^{e} \times \Z_2^{m},\\
 \Z_2^{m} &\mapsto&  0 . 
\eea
This $\Z_2$-gauge theory also has the 3d CS \Eq{eq:Sbulk} description with
\bea
K= \bigl( {\begin{smallmatrix} 
0 &2 \\
2 & 0  
\end{smallmatrix}} \bigl)
.\eea
We can also write the same data of $\cW^\text{TC}_e$ and $\cW^\text{TC}_m$ in terms of the 1+1D sine-Gordon cosine terms 
\Eq{eq:Sedge-int} that
can gap the gapless CFT to obtain the $e$- and $m$-type gapped boundaries \cite{Wang2012am1212.4863}.
There are two types:\footnote{See footnote \ref{footnote:L}, we 
interchangeably write the transpose row vector $\ell^T$ as the column vector $\ell$ sometimes.
}
\bea
\ell^{  \cW^\text{TC}_e}=2(1,0), \\
\ell^{  \cW^\text{TC}_m}=2(0,1),
\eea
which satisfy the gapping rules in \Sec{sec:gapless-gap}.
This also corresponds to the rough and smooth boundary of $\Z_2$-toric code on the lattice, see 
 \cite{9811052BravyiKitaev, 1104.5047KitaevKong}, as an electric and magnetic type of gapped boundaries \cite{Wang2012am1212.4863}.

\subsection{Two $\Z_2$ gauge theories: 6 types of gapped interfaces}
\label{subsection:Z2two}

There are 6 types of gapped interfaces between two $\Z_2$ gauge theories (toric codes)  \cite{Lan2014uaa1408.6514}. 
The first two are labeled by:\footnote{We define $\mathbb{I}_n$ as a rank-$n$ identity matrix. \label{footnote:I_n}}
\begin{align}
  \cW^\mathrm{TC|TC}=\begin{pmatrix}
    1&0&0&0\\
    0&1&0&0\\
    0&0&1&0\\
    0&0&0&1
  \end{pmatrix} \equiv \mathbb{I}_4. \quad\quad\quad 
  \cW^\mathrm{TC|TC}_{e\leftrightarrow m}=\begin{pmatrix}
    1&0&0&0\\
    0&0&1&0\\
    0&1&0&0\\
    0&0&0&1
  \end{pmatrix}.
  \end{align}
The rest 4 are the compositions of the previous gapped boundaries from \Sec{sec:Z2-1},
i.e.~
\bea
(\cW^\text{TC}_{e_1})^\dag\cW^\text{TC}_{e_2}, \quad 
(\cW^\text{TC}_{e_1})^\dag\cW^\text{TC}_{m_2}, \quad
(\cW^\text{TC}_{m_1})^\dag\cW^\text{TC}_{e_2},\quad
(\cW^\text{TC}_{m_1})^\dag\cW^\text{TC}_{m_2}.
\eea
The 6 gapped interfaces have the following correspondence to the set of anyon condensations:\footnote{Here
we label the set of anyons that are the generators 
without using the bracket notations, while those anyons that can be generated by other earlier generators 
have the bracket around as $[...]$. 
For example, in the first line, the
$1, e_1, e_2, [e_1 e_2]$ means that 
the generators are the $e_1$ and $e_2$, while $[e_1 e_2]$ can be generated by fusing the two generators together.
In the later sections, we may also denote $[\dots]$ as other
condensed anyons that can be generated from the previously given generators.
\label{footnote:bracket}}
\bea
(\cW^\text{TC}_{e_1})^\dag\cW^\text{TC}_{e_2} &:& 1, e_1, e_2, [e_1 e_2].\\
(\cW^\text{TC}_{m_1})^\dag\cW^\text{TC}_{m_2} &:& 1, m_1, m_2, [m_1 m_2]\\
(\cW^\text{TC}_{e_1})^\dag\cW^\text{TC}_{m_2} &:& 1, e_1, m_2, [e_1 m_2].\\
(\cW^\text{TC}_{m_1})^\dag\cW^\text{TC}_{e_2} &:& 1, m_1, e_2, , [m_1 e_2].
\eea
\bea
  \cW^\mathrm{TC|TC} &:& 1, e_1 e_2, m_1 m_2, [e_1 e_2 m_1 m_2].\\
  \cW^\mathrm{TC|TC}_{e\leftrightarrow m} &:& 1, e_1 m_2, m_1 e_2, [e_1 e_2 m_1 m_2].
\eea
The above data can also be understood in terms of $G_{\text{unbroken}} \stackrel{\iota}{\longrightarrow}  G_{\text{original}},$ we have:
\bea
(\cW^\text{TC}_{e_1})^\dag\cW^\text{TC}_{e_2} &:&
 \Z_2^{m_1} \times \Z_2^{m_2}  \stackrel{\iota}{\longrightarrow} \Z_2^{e_1} \times \Z_2^{m_1} \times  \Z_2^{e_2} \times \Z_2^{m_2},\\
(\cW^\text{TC}_{m_1})^\dag\cW^\text{TC}_{m_2}  &:& 
\Z_2^{e_1} \times \Z_2^{e_2}  \stackrel{\iota}{\longrightarrow} \Z_2^{e_1} \times \Z_2^{m_1} \times  \Z_2^{e_2} \times \Z_2^{m_2},\\
(\cW^\text{TC}_{e_1})^\dag\cW^\text{TC}_{m_2} &:& 
   \Z_2^{m_1} \times \Z_2^{e_2}  \stackrel{\iota}{\longrightarrow} \Z_2^{e_1} \times \Z_2^{m_1} \times  \Z_2^{e_2} \times \Z_2^{m_2},\\
(\cW^\text{TC}_{m_1})^\dag\cW^\text{TC}_{e_2} &:& 
  \Z_2^{e_1} \times \Z_2^{m_2}  \stackrel{\iota}{\longrightarrow} \Z_2^{e_1} \times \Z_2^{m_1} \times  \Z_2^{e_2} \times \Z_2^{m_2},\\
  \cW^\mathrm{TC|TC} &:&    
  \overline{\Z_2^{{e_1 e_2}} \times \Z_2^{{m_1 m_2}}}  \stackrel{\iota}{\longrightarrow}
   \Z_2^{e_1} \times \Z_2^{m_1} \times  \Z_2^{e_2} \times \Z_2^{m_2},\\
  \cW^\mathrm{TC|TC}_{e\leftrightarrow m} &:& 
         \overline{ \Z_2^{{e_1 m_2}} \times \Z_2^{{m_1 e_2}} }
         \stackrel{\iota}{\longrightarrow}
        {\Z_2^{e_1} \times \Z_2^{m_1} \times  \Z_2^{e_2} \times \Z_2^{m_2}}.
\eea
Here we define $\Z_2^{{a_1 a_2}} \equiv \diag(\Z_2^{a_1}, \Z_2^{a_2}  )$, for example,
$\Z_2^{{e_1 e_2}} \equiv \diag(\Z_2^{e_1}, \Z_2^{e_2}  )$,
$\Z_2^{{m_1 m_2}} \equiv \diag(\Z_2^{m_1}, \Z_2^{m_2}  )$,
$\Z_2^{{e_1 m_2}} \equiv \diag(\Z_2^{e_1}, \Z_2^{m_2}  )$,
and $\Z_2^{{m_1 e_2}} \equiv \diag(\Z_2^{m_1}, \Z_2^{e_2}  )$, etc.
Here the overline notation, such as
$\overline{\Z_2^{{e_1 e_2}} \times \Z_2^{{m_1 m_2}}}$,
means the complement subgroup in ${\Z_2^{e_1} \times \Z_2^{m_1} \times  \Z_2^{e_2} \times \Z_2^{m_2}}$
but excluding those overlap with ${\Z_2^{{e_1 e_2}} \times \Z_2^{{m_1 m_2}}}$.
Namely, the last two $\cW^\mathrm{TC|TC}$ and
$\cW^\mathrm{TC|TC}_{e\leftrightarrow m}$ can also be understood as the following breaking for  $G_{\text{to-be-condensed}}  \mapsto 0$:
\bea
  \cW^\mathrm{TC|TC} &:&         \Z_2^{{e_1 e_2}} \times \Z_2^{{m_1 m_2}} \mapsto 0,  \\
  \cW^\mathrm{TC|TC}_{e\leftrightarrow m} &:&      \Z_2^{{e_1 m_2}} \times \Z_2^{{m_1 e_2}} \mapsto 0,   
\eea
such that the right hand side groups are broken completely on the left hand side (as 0).
This $(\Z_2)^2$-gauge theory also has the 3d CS \Eq{eq:Sbulk} description with
\bea
K= {\begin{pmatrix} 
0 &2 \\
2 & 0  
\end{pmatrix}} 
\oplus
 {\begin{pmatrix} 
0 &2 \\
2 & 0  
\end{pmatrix}}. 
\eea
We can also write the same data in terms of the 1+1D sine-Gordon cosine terms 
\Eq{eq:Sedge-int} with
$
\int \dd t \dd x  \big( g_{{\mathbf{1}}}  \cos( \ell_{\mathbf{1},I} \cdot \Phi^{}_{ I} )+  g_{\mathbf{2}}  \cos( \ell_{\mathbf{2},I} \cdot \Phi^{}_{I} ) \big)
$
that
can gap the gapless CFT to obtain gapped interfaces:
\bea
(\cW^\text{TC}_{e_1})^\dag\cW^\text{TC}_{e_2} &:& \ell_{\mathbf{1}}=2(1,0,0,0),\quad \ell_{\mathbf{2}}=2(0,0,1,0),\\
(\cW^\text{TC}_{m_1})^\dag\cW^\text{TC}_{m_2} &:& \ell_{\mathbf{1}}=2(0,1,0,0),\quad\ell_{\mathbf{2}}=2(0,0,0,1),\\
(\cW^\text{TC}_{e_1})^\dag\cW^\text{TC}_{m_2} &:&  \ell_{\mathbf{1}}=2(1,0,0,0),\quad\ell_{\mathbf{2}}=2(0,0,0,1),\\
(\cW^\text{TC}_{m_1})^\dag\cW^\text{TC}_{e_2}  &:&  \ell_{\mathbf{1}}=2(0,1,0,0),\quad\ell_{\mathbf{2}}=2(0,0,1,0),\\
  \cW^\mathrm{TC|TC} &:& \ell_{\mathbf{1}}=2(1,0,1,0),\quad\ell_{\mathbf{2}}=2(0,1,0,1),\\
  \cW^\mathrm{TC|TC}_{e\leftrightarrow m} &:& \ell_{\mathbf{1}}=2(1,0,0,1),\quad\ell_{\mathbf{2}}=2(0,1,1,0).
\eea
They satisfy the gapping rules in \Sec{sec:gapless-gap}.


\subsection{Three $\Z_2$ gauge theories: Liquid and Non-Liquid Cellular states}
\label{subsection:Z2three}

Now let us consider gapped interfaces between three $\Z_2$ gauge theories: 
$(\Z_2)^3$ gauge theories as a whole.
The fusion algebra has $((\Z_2)^3)^2=\Z_2^6$ structure. 
There are finitely many types of gapped interfaces that can be bootstrapped by the method of \cite{Lan2014uaa1408.6514}. 
We expect that some subset of anyons (with a number $|(\Z_2)^3|=8$) can condense on gapped interfaces.
For example, we can consider the following 8 types of gapped interfaces via 8 types of anyon condensations,
where $a_j$ is chosen to be either the electric anyon $e_j$ or the magnetic anyon $m_j$ of the $j$-th sector:
\bea
\text{8 types}: 
&&1, a_1, a_2, a_3, \label{eq:Z2^3-1}  \\
(\text{more precisely}: 
&&1, a_1, a_2, a_3, [a_1 a_2],    [a_1 a_3],  [a_2 a_3],  [a_1 a_2 a_3].)        
\nn
\eea
where $[...]$ means that can be composed of the generators from the left side of generating anyons.
By using
the $(\Z_2)^3$-gauge theory with 3d CS \Eq{eq:Sbulk} description with
\bea
 {\begin{pmatrix} 
0 &2 \\
2 & 0  
\end{pmatrix}} 
\oplus
 {\begin{pmatrix} 
0 &2 \\
2 & 0  
\end{pmatrix}} 
\oplus
 {\begin{pmatrix} 
0 &2 \\
2 & 0  
\end{pmatrix}}, 
\eea
gapping by the 1+1D sine-Gordon cosine terms 
\Eq{eq:Sedge-int}, 
we require that
\bea
&&\ell_{\mathbf{1}}=2(e_1,m_1,0,0,0,0), \quad \ell_{\mathbf{2}}=2(0,0,{e_2},{m_2},0,0), 
\quad \ell_{\mathbf{3}}=2(0,0,0,0,{e_3},{m_3}).
\eea
Here we abuse the notation $e_j,m_j \in \{0,1\}$  meaning the coefficient of $\ell$ vectors.
When $a_j=e_j$, we set the coefficient as $(e_j,m_j) = (1,0)$.
When $a_j=m_j$, we set the coefficient as $(e_j,m_j) = (0,1)$.
The above {8 types} of data can also be understood in terms of $G_{\text{unbroken}} \stackrel{\iota}{\longrightarrow}  G_{\text{original}}$ as previous cases, 
we have to choose the unbroken sector on the left hand side:
\bea
 \Z_2^{{e_1/m_1}} \times \Z_2^{{e_2/m_2}} \times \Z_2^{{e_3/m_3}} & \stackrel{\iota}{\longrightarrow}& \Z_2^{e_1} \times \Z_2^{m_1} \times  \Z_2^{e_2} \times \Z_2^{m_2}  \times  \Z_2^{e_3} \times \Z_2^{m_3},
\eea
We can also write the {8 types} data as those groups being completely broken (those anyons can condense on the interface) as
$G_{\text{to-be-condensed}}  \mapsto 0$:
\bea
  \Z_2^{a_1} \times  \Z_2^{a_2} \times \Z_2^{a_3} \mapsto 0.
\eea
However, the above gapped interfaces are rather not useful, because they are formed by tensor products of
decoupled gapped boundaries of each $\Z_2$ gauge theory. 

We find there are at least 10 types of gapped interfaces that cannot be decoupled to the tensor product of some of individual $\Z_2$ gauge theories. 
We can write down their tunneling matrix $\cW$ solved from \cite{Lan2014uaa1408.6514}.
More easily, we can also write down the set of condensed anyons on gapped interfaces via anyon condensations  (which thus shows the nonzero element of $\cW$ matrix),
\bea
\cW^{\I} &: 1, e_1 e_2, e_2 e_3, m_1 m_2 m_3, [e_1 e_3], [f_1 f_2 m_3], [m_1 f_2 f_3], [f_1 m_2 f_3],
\nonumber\\
\cW^{\II} &: 
1, f_1 f_2, f_2 f_3, m_1 m_2 m_3, [f_1 f_3], [e_1 e_2 m_3], [m_1 e_2 e_3], [e_1 m_2 e_3]. 
\nonumber\\
\cW^{\III} &:  
1, m_1 m_2, m_2 m_3, e_1 e_2 e_3, [m_1 m_3], [f_1 f_2 e_3], [e_1 f_2 f_3], [f_1 e_2 f_3]. 
\nonumber\\
\cW^{\IV} &: 1, f_1 f_2, f_2 f_3, e_1 e_2 e_3,  [f_1 f_3], [m_1 m_2 e_3], [e_1 m_2 m_3], [m_1 e_2 m_3]. 
\nonumber\\
\cW^{\V} &: 1, e_1 m_2, e_1 e_3,  m_1 e_2 m_3, [m_2 e_3], [m_1 f_2 f_3], [f_1 e_2 f_3], [f_1 f_2 m_3]. 
\nonumber\\
\cW^{\VI} &: 1, m_2 e_3, 
m_1 m_2, 
e_1 e_2 m_3,  [e_1 f_2 f_3], [m_1 e_3], 
[f_1 e_2 f_3], [f_1 f_2 m_3]. 
\nonumber\\
\cW^{\VII} &:
1,   e_2 m_3, e_1 e_2, m_1 m_2 e_3, [e_1 m_3], [m_1 f_2 f_3], [f_1 m_2 f_3], [f_1 f_2 e_3].
\nonumber\\
\cW^{\VIII} &: 
1,   m_1 e_2, m_1 m_3,  e_1 m_2 e_3, [e_2 m_3],  [e_1 f_2 f_3], [f_1 m_2 f_3], [f_1 f_2 e_3].
\nonumber\\
\cW^{\IX} &: 1,  m_2 m_3,  e_1 m_2, m_1 e_2 e_3, [e_1 m_3], [m_1 f_2 f_3], [ f_1 e_2 f_3], [ f_1  f_2 e_3].
\nonumber\\
\cW^{\X} &: 
1,  e_2 e_3,  m_1 e_2, e_1 m_2 m_3, [m_1 e_3], [e_1 f_2 f_3], [ f_1 m_2 f_3], [ f_1  f_2 m_3].
\eea
Note that the pair $\cW^{\V}$ and $\cW^{\VIII}$,
 the pair $\cW^{\VI}$ and $\cW^{\VII}$, 
and also  the pair $\cW^{\IX}$ and $\cW^{\X}$,
are related by exchanging $e$ and $m$ labels.
Note that 
\begin{itemize}[leftmargin=-2mm] 
\item 
$\cW^{\I}, \cW^{\II},\cW^{\III}$, and $\cW^{\IV}$
are \emph{fully symmetric} under exchanging any layer indices $(1,2,3)$.
\item
$\cW^{\IV}, \cW^{\V}, \cW^{\VI}, \cW^{\VII}, \cW^{\VIII},
\cW^{\IX}$, and $\cW^{\X}$
are \emph{not fully symmetric},
\emph{nor cyclic symmetric}
under exchanging any layer indices $(1,2,3)$.
But $\cW^{\V}$ and $\cW^{\VIII}$
 are
\emph{symmetric} under exchanging the layer indices 1 and 3.
$\cW^{\VI}$ and $\cW^{\VII}$ are
\emph{symmetric} under exchanging the layer indices 1 and 2.
$\cW^{\IX}$ and $\cW^{\X}$
are
\emph{symmetric} under exchanging the layer indices 2 and 3.
\end{itemize}
 The $\ell$ vector can be mapped to the coefficient of each of the above generators of the condensed anyons,
say $\ell =2({e_1},{m_1},{e_2},{m_2},{e_3},{m_3})$,
also recall that $f_j=e_j m_j$,
we obtain 1+1D sine-Gordon cosine gapping terms 
\Eq{eq:Sedge-int} as
\bea
\cW^{\I} &: 
\ell_{\mathbf{1}}
=2({1},0,{1},0,0,0),\quad
\ell_{\mathbf{2}}
=2(0,0,{1},0,{1},0), \quad
 \ell_{\mathbf{3}}
 =2(0,{1},0,{1},{0},{1}). \nn\\
\cW^{\II} &:  \ell_{\mathbf{1}}
=2({1},1,{1},1,0,0),
\quad
\ell_{\mathbf{2}}
=2(0,0,{1},1,{1},1), \quad
 \ell_{\mathbf{3}}
 =2(0,{1},0,{1},{0},{1}).\nn\\
\cW^{\III} &: \ell_{\mathbf{1}} 
=2(0,{1},0,{1},0,0),\quad
\ell_{\mathbf{2}} 
=2(0,0,0,{1},0,{1}), \quad
 \ell_{\mathbf{3}} 
 =2({1},0,{1},{0},{1},0).\nn\\
\cW^{\IV} &:  \ell_{\mathbf{1}} 
=2({1},1,{1},1,0,0),\quad
\ell_{\mathbf{2}} 
=2(0,0,{1},1,{1},1), \quad
 \ell_{\mathbf{3}} 
 =2({1},0,{1},{0},{1},0).\nn\\
\cW^{\V} &:\ell_{\mathbf{1}} 
=2(1,0,0,1,0,0),\quad
 \ell_{\mathbf{2}} 
 =2(1,0,0,0,1,0), \quad
\ell_{\mathbf{3}} 
=2(0,1,1,0,0,1).\nn\\
\cW^{\VI} &:\ell_{\mathbf{1}}
=2(0,0,0,1,1,0),\quad
 \ell_{\mathbf{2}} 
  =2(0,1,0,1,0,0), \quad
\ell_{\mathbf{3}} 
=2(1,0,1,0,0,1).
\nn\\
\cW^{\VII} &:\ell_{\mathbf{1}}
=2(0,0,1,0,0,1),\quad
 \ell_{\mathbf{2}} 
  =2(1,0,1,0,0,0), \quad
\ell_{\mathbf{3}} 
=2(0,1,0,1,1,0).
\nn\\
\cW^{\VIII} &:\ell_{\mathbf{1}}
=2(0,1,1,0,0,0),\quad
 \ell_{\mathbf{2}} 
  =2(0,1,0,0,0,1), \quad
\ell_{\mathbf{3}} 
=2(1,0,0,1,1,0).\nn\\
\cW^{\IX} &:\ell_{\mathbf{2}}
=2(0,0,0,{1},0,{1}),\quad
 \ell_{\mathbf{2}} 
 =2(1,0,0,1,0,0),\quad
 \ell_{\mathbf{3}} 
   =2(0,1,1,0,1,0).
\nn\\
\cW^{\X} &:
\ell_{\mathbf{1}}
=2(0,0,{1},0,{1},0),\quad
 \ell_{\mathbf{2}} 
 =2(0,1,1,0,0,0),\quad
 \ell_{\mathbf{3}} 
=2(1,0,0,1,0,1).
\eea

To make comparison,  
we can use $1,2,3,4$ to label the $1,e,m,em$ anyon sectors respectively,
this matches the notations of \cite{Wen2020pri2002.02433}. We write the nonzero components of $\cW$ tensors: 
\bea
 && \text{
 We write the tensor components with
 the lower index relabelings $1,e,m,f \Leftrightarrow \mi{1,2,3,4}$ with $f=em$}, \nn\\
\cW^{\I} &:& 
\cW^{\I} _{\mi{111}}=\cW^{\I}_{\mi{122}}=\cW^{\I}_{\mi{212}}=\cW^{\I}_{\mi{221}}=\cW^{\I}_{\mi{333}}=\cW^{\I}_{\mi{344}}=\cW^{\I}_{\mi{434}}=\cW^{\I}_{\mi{443}}=1. 
\nonumber\\
\cW^{\II} &:&  \cW^{\II}_{\mi{111}}= \cW^{\II}_{\mi{144}}= \cW^{\II}_{\mi{223}}= \cW^{\II}_{\mi{232}}=\cW^{\II}_{\mi{322}}= \cW^{\II}_{\mi{333}}= \cW^{\II}_{\mi{414}}= \cW^{\II}_{\mi{441}}=1. 
\nonumber\\
\cW^{\III} &: & \cW^{\III}_{\mi{111}} = \cW^{\III}_{\mi{133}}= \cW^{\III}_{\mi{222}}= \cW^{\III}_{\mi{244}}= \cW^{\III}_{\mi{313}}= \cW^{\III}_{\mi{331}}= \cW^{\III}_{\mi{424}}= \cW^{\III}_{\mi{442}}=1. 
\nonumber\\
\cW^{\IV} &: &\cW^{\IV}_{\mi{111}}= \cW^{\IV}_{\mi{144}}=\cW^{\IV}_{\mi{222}}=\cW^{\IV}_{\mi{233}}=\cW^{\IV}_{\mi{323}}=\cW^{\IV}_{\mi{332}}
=\cW^{\IV}_{\mi{414}}=\cW^{\IV}_{\mi{441}}=1. 
\nonumber\\
\cW^{\V} &:& \cW^{\V}_{\mi{111}}=\cW^{\V}_{\mi{132}}=\cW^{\V}_{\mi{212}}=\cW^{\V}_{\mi{231}}=\cW^{\V}_{\mi{323}}=\cW^{\V}_{\mi{344}}=\cW^{\V}_{\mi{424}}
=\cW^{\V}_{\mi{443}}=1. 
\nonumber\\
\cW^{\VI} &:& 
\cW^{\VI}_{\mi{111}}=\cW^{\VI}_{\mi{132}}=\cW^{\VI}_{\mi{223}}=\cW^{\VI}_{\mi{244}}=\cW^{\VI}_{\mi{312}}=\cW^{\VI}_{\mi{331}}=\cW^{\VI}_{\mi{424}}=\cW^{\VI}_{\mi{443}}=1. 
\nonumber\\
\cW^{\VII} &:& \cW^{\VII}_{\mi{111}}=\cW^{\VII}_{\mi{123}}=\cW^{\VII}_{\mi{213}}=\cW^{\VII}_{\mi{221}}=\cW^{\VII}_{\mi{332}}=\cW^{\VII}_{\mi{344}}=\cW^{\VII}_{\mi{434}}=\cW^{\VII}_{\mi{442}}=1. 
\nonumber\\
\cW^{\VIII}&:&
\cW^{\VIII}_{\mi{111}}=
\cW^{\VIII}_{\mi{123}}=\cW^{\VIII}_{\mi{232}}=\cW^{\VIII}_{\mi{244}}=\cW^{\VIII}_{\mi{313}}=\cW^{\VIII}_{\mi{321}}=\cW^{\VIII}_{\mi{434}}=\cW^{\VIII}_{\mi{442}}=1. 
\nonumber\\
\cW^{\IX} &:& \cW^{\IX}_{\mi{111}}=\cW^{\IX}_{\mi{133}}=\cW^{\IX}_{\mi{213}}=\cW^{\IX}_{\mi{231}}
=\cW^{\IX}_{\mi{322}}=\cW^{\IX}_{\mi{344}}=\cW^{\IX}_{\mi{424}}=\cW^{\IX}_{\mi{442}}=1. 
\nonumber\\
\cW^{\X} &:& \cW^{\X}_{\mi{111}}=\cW^{\X}_{\mi{122}}=\cW^{\X}_{\mi{233}}=\cW^{\X}_{\mi{244}}=\cW^{\X}_{\mi{312}}=\cW^{\X}_{\mi{321}}=\cW^{\X}_{\mi{434}}=\cW^{\X}_{\mi{443}}=1. \label{eq:WZ2-tensor}
\eea
In fact, the earlier result matches the result of \cite{Wen2020pri2002.02433}.
 Follow the interpretations in \cite{Wen2020pri2002.02433} and \Sec{sec:Interpretations},
 we see that
 these are {\bf\emph{{gapped liquid cellular states}}} \cite{Wen2020pri2002.02433}: 
\bea
(\cW^{\I}, \cW^{\I}), (\cW^{\II}, \cW^{\II}), (\cW^{\III}, \cW^{\III}), (\cW^{\IV}, \cW^{\IV}), (\cW^{\II}, \cW^{\IV}),
\eea
because they can be coarse-grained and renormalized to themselves (up to accidental copies and accidental degeneracy 
that can be broken by local perturbations).

We see that
 these are
{\bf\emph{{gapped non-liquid cellular states}}} \cite{Wen2020pri2002.02433}: 
\bea
(\cW^{\I}, \cW^{\II}), (\cW^{\I}, \cW^{\III}), (\cW^{\I}, \cW^{\IV}), (\cW^{\II}, \cW^{\III}), (\cW^{\III}, \cW^{\IV}),
\eea
because they cannot be coarse-grained and renormalized to themselves. 
To change ground state sectors, the require spatial pattern of generated operators acting on the wavefunction needs to be fractal
(with small triangle patterns on the honeycomb lattice). 
In order to decide whether they are type I or type II fracton orders,
we need to know if there exist string operators or not to move fractionalized excitations and change the ground states to different topological sectors.
The generic construction here may contain
lineons, planons, or fractons, depending on how we extend the honeycomb lattice to the third spatial dimension 
(the $z$ axis).

We do not discuss the phases of matter constructed out of 
gapped interfaces $\cW^{\IV}, \cW^{\V}, \cW^{\VI}, \cW^{\VII}, \cW^{\VIII}, \cW^{\IX}$, and $\cW^{\X}$
since they are \emph{not fully symmetric},
\emph{nor cyclic symmetric} under exchanging any layer indices $(1,2,3)$,\footnote{The readers should not be confused with the layer indices $1,2,3$, and the other anyon labelings
$\mi{1,2,3,4} \Leftrightarrow 1,e,m,f$ with $f=em$.}
so these either construct
{\bf\emph{{anisotropic}}}  phases of matter,
or they can be coarse-grained and renormalized to 
\emph{isotropic} only at the larger length scale.

\subsection{Four $\Z_2$ gauge theories: Liquid and Non-Liquid Cellular states}

 \label{subsection:Z2four}

Consider gapped interfaces between four $\Z_2$ gauge theories: 
$(\Z_2)^4$ gauge theories as a whole.
The fusion algebra has $((\Z_2)^4)^2=\Z_2^8$ structure. 
There are finitely many types of gapped interfaces that can be bootstrapped by the method of \cite{Lan2014uaa1408.6514}. 
We expect that some subset of anyons (with a number $|(\Z_2)^4|=8$) can condense on gapped interfaces.
Three particular interesting gapped interfaces (those are not the tensor product of individual gauge theories) are pointed out by \cite{Wen2020pri2002.02433} (R, B, G for red, blue, and green, labels for interfaces):
\bea
\cW^{R} :&& 
1, e_1 e_2, e_1 e_3, e_1 e_4, m_1 m_2 m_3 m_4, 
[f_1 f_2 f_3 f_4], 
[e_2 e_3],[e_2 e_4],[e_3 e_4], 
[e_1 e_2 e_3 e_4],\nn\\
&&[f_1 f_2 m_3 m_4],
[f_1  m_2 f_3 m_4], [f_1 m_2 m_3  f_4],
[m_1 f_2 f_3 m_4],[m_1 f_2  m_3 f_4],
[m_1   m_2 f_3 f_4]. \nn\\
\cW^B: && 
1, m_1 m_2, m_1 m_3, m_1 m_4, e_1 e_2 e_3 e_4, 
[f_1 f_2 f_3 f_4], 
[m_2 m_3], [m_2 m_4], [m_3 m_4], 
[m_1 m_2 m_3 m_4],\nn\\
&&[f_1 f_2 e_3 e_4],
[f_1  e_2 f_3 e_4], [f_1 e_2 e_3  f_4],
[e_1 f_2 f_3 e_4], [e_1 f_2  e_3 f_4],
[e_1   e_2 f_3 f_4]. \nn\\
\cW^G:  &&
1, f_1 f_2, f_1 f_3, f_1 f_4, m_1 m_2 m_3 m_4, 
[e_1 e_2 e_3 e_4],
[f_1 f_2 f_3 f_4], 
[f_2 f_3],[f_2 f_4],[f_3 f_4],\nn\\
&&[e_1 e_2 m_3 m_4],
[e_1  m_2 e_3 m_4], [e_1 m_2 m_3  e_4],
[m_1 e_2 e_3 m_4], [m_1 e_2  m_3 e_4],
[m_1   m_2 e_3 e_4].
\eea
again while those anyons that can be generated by other earlier generators 
have the bracket around as $[...]$, see footnote \ref{footnote:bracket}. 
Note that we can do the following anyon relabels to obtain another interface tensor,
we can switch {$\cW^B \sim \cW^R (e \to m, m \to e,  f \to f)$,
$\cW^G \sim \cW^R (e \to f, f \to m, m \to e)$.
}
These sets of condensed anyons also mean the nonzero ($=1$) component
for each of the 4-leg tensors:
$\cW^{R}_{abcd}, \cW^{B}_{abcd}$ and $\cW^{G}_{abcd}$.
We can obtain the 1+1D sine-Gordon cosine gapping terms for the above
\Eq{eq:Sedge-int} as
\bea
&\hspace{-18mm}
\cW^{R} : 
\ell_{\mathbf{1}}=
2({1},0,{1},0,0,0,0,0),\quad
\ell_{\mathbf{2}}
=2({1},0,0,0,{1},0,0,0), \quad
 \ell_{\mathbf{3}}
 =2({1},0,0,0,0,0,{1},0), \quad 
 \ell_{\mathbf{4}}=2(0,{1},0,{1},{0},{1},{0},{1}). \nn\\
&\hspace{-18mm}
\cW^{B} :  \ell_{\mathbf{1}}
=2(0,{1},0,{1},0,0,0,0),
\quad
\ell_{\mathbf{2}}
=2(0,{1},0,0,0,{1},0,0), \quad
 \ell_{\mathbf{3}}
 =2(0,{1},0,0,0,0,0,{1}), \quad 
 \ell_{\mathbf{4}}=2({1},0,{1},{0},{1},{0},{1},0). \nn\\
 &\hspace{-18mm}
\cW^{G} :  \ell_{\mathbf{1}}
=2({1},1,{1},1,0,0,0,0),
\quad
\ell_{\mathbf{2}}
=2(1,1,0,0,1,1,0,0), \quad
 \ell_{\mathbf{3}}
 =2(1,1,0,0,0,0,1,1), \quad 
 \ell_{\mathbf{4}}=2(0,{1},0,{1},{0},{1},{0},{1}). \nn\\
\eea
Using the geometrical/renormalization consistency criteria in \Sec{sec:GeometricalRenormalizationConsistencyCriteria},
in the first step, we can show that
$\cW^{}_{abcd}$ can be renormalized to two 3-leg $\cW'^{}_{abc}$ in \Sec{subsection:Z2three}
\bea
\label{eq:deformation-WRBG}
&&  \cW^{R}_{abcd}  = \sum_{j'} (\cW'^{\I})_{ab j'  } (\cW'^{\I})_{j'  cd},\quad
  \cW^B_{abcd}  = \sum_{j'} (\cW'^{{\III}})_{ab j'  } (\cW'^{{\III}})_{j'  cd}. \nn\\
&&  \cW^G_{abcd}  = \sum_{j'} (\cW'^{{\II}})_{ab j'  } (\cW'^{{\II}})_{j'  cd}= \sum_{j'} (\cW'^{{\IV}})_{ab j'  } (\cW'^{{\IV}})_{j'  cd}.
\eea
For the second step criteria in \Sec{sec:GeometricalRenormalizationConsistencyCriteria},
we have
\bea
\label{eq:sq-shrink-fuse-Z2-four}
&& \sum_{a,b,c,d} (\cW'^{\I})_{ a i' b} (\cW'^{\I})_{b j' c} (\cW'^{\I})_{c k' d} (\cW'^{\I})_{d l' a} =2 \t \cW^{R}_{i' j' k' l'}.\nonumber\\
&&  \sum_{a,b,c,d} (\cW'^{\III})_{ a i' b} (\cW'^{\III})_{b j' c} (\cW'^{\III})_{c k' d} (\cW'^{\III})_{d l' a} =2 \t \cW^{B}_{i' j' k' l'}.\\
  &&  \sum_{a,b,c,d} (\cW'^{\II})_{ a i' b} (\cW'^{\II})_{b j' c} (\cW'^{\II})_{c k' d} (\cW'^{\II})_{d l' a} =
        \sum_{a,b,c,d} (\cW'^{\IV})_{ a i' b} (\cW'^{\IV})_{b j' c} (\cW'^{\IV})_{c k' d} (\cW'^{\IV})_{d l' a} =
2    \t \cW^{G}_{i' j' k' l'}.\nonumber
\eea
So this means that
if there is only {\bf\emph{{one sub-lattice}}} for the
{square column lattice},
we obtain 3+1D 
{\bf\emph{{gapped liquid cellular states}}}. 

However, if there are {\bf\emph{{two sub-lattices}}} for the
{square column lattice},
and we choose {two sub-lattices} with different interface tensor data 
(say out of $\cW^{R}$, $\cW^{B}$ and $\cW^{G}$),
we find they cannot be renormalized back to themselves.
So the physics interpretations in \Sec{sec:Interpretations}
show that they are 3+1D 
{\bf\emph{{gapped non-liquid cellular states}}} with type-I fracton order.

We can consider {\bf\emph{{three sub-lattices}}} for the
 {\bf\emph{{cubic lattice}}},
and we choose {three sub-lattices} with different interface tensor data 
($\cW^{R}$, $\cW^{B}$ and $\cW^{G}$),
 \Refe{Wen2020pri2002.02433} finds
 that they are 3+1D 
{\bf\emph{{gapped non-liquid cellular states}}} as well.

\section{Cellular States from twisted $\Z_2$-Gauge Theory ($\Z_2$-Double-Semion)}
\label{sec:Z2t-Cellular}

\subsection{Twisted $\Z_2$-gauge theory: 1 type of gapped boundary} 
\label{sec:Z2t-1}

For 2+1D $\Z_2$-double-semion model (DS, $s$ and $\bar{s}$), also known as 3d twisted $\Z_2$-gauge theory,
we have $\cS,\cT$ matrices (see footnote \ref{footnote:MCG}):
\begin{align*}
  \cS&=\frac{1}{2}
  \begin{pmatrix}
    1&1&1&1\\
    1&-1&1&-1\\
    1&1&-1&-1\\
    1&-1&-1&1
  \end{pmatrix}.\\
    \cT&=\diag(1,\ii,-\ii,1).
\end{align*}
where each row and column index entry runs from the anyon sectors $1, s, \bar{s}$, and $s\bar{s}$.
They are the trivial vacuum sector, the semion $s$ (of $\Z_2$-semion line),
 the anti-semion $\bar{s}$ (of $\Z_2$-anti-semion line, as the $\Z_2$-dual object),
and the double-semion $s\bar{s}$.
There is only one type of gapped boundary solved from \cite{Lan2014uaa1408.6514}:\footnote{We
can also interpret the gapped boundary data between the $\Z_2$-double-semion model
to the trivial vacuum as the
gapped interface between 
the
$\Z_2$-semion (S) to the
$\Z_2$-semion (S)
as 
$$
\cW^\text{S$\mid$S}=
\mathbb{I}_2=
\bigl( {\begin{smallmatrix} 
1 & 0 \\
0 & 1  
\end{smallmatrix}} \bigl)
  =
\left(
\begin{array}{ccc c}
1 &  s & & \\
\hline
 1 & 0 & \;\vline & 1  \\ 
 0 & 1 & \;\vline & s
\end{array}
\right).
$$
This is a trivial gapped interface sending $s$ to $s$.
See footnote \ref{footnote:I_n}.
}
\begin{align} \label{eq:WDS}
\cW^\text{DS}&=\begin{pmatrix} 1&0&0&1 \end{pmatrix}
  =
\left(
\begin{array}{cccc c c}
1 &  s & \bar{s} & s \bar{s}   & & \\
\hline
 1 & 0 & 0 & 1 & \;\vline & 1 
\end{array}
\right).
\end{align}
We can also organize the tunneling matrix $\cW$ data via the condensed anyons (i.e., anyons that allowed to be condensed) 
\bea 
\cW^\text{DS}&:& 1,  s \bar{s}.
\eea
We can also interpret it as the gauge-breaking data, where the right hand side is the
$\Z_2^{s} \times \Z_2^{ \bar{s}}$ including the gauge group $\Z_2^{s}$ and the dual group $\Z_2^{ \bar{s}}$.
This $\Z_2^{s} \times \Z_2^{{ \bar{s}}}$ group also is the group of fusion algebra of the TQFT.
Follow the notations in \Sec{sec:Z2-Cellular}, in terms of the breaking notation
$G_{\text{unbroken}} \stackrel{\iota}{\longrightarrow}  G_{\text{original}},$
or $G_{\text{to-be-condensed}}  \mapsto 0$,
we find that $\cW^\text{DS}$ is equivalent to the gauge-breaking pattern: 
\bea \label{eq:ssbar-breaking}
  \overline{\Z_2^{ s\bar{s}}} & \stackrel{\iota}{\longrightarrow}& \Z_2^{s} \times \Z_2^{ \bar{s}},\\
 \Z_2^{ s\bar{s}}  &\mapsto& 0. 
\eea
This twisted $\Z_2$-gauge theory also has the 3d CS \Eq{eq:Sbulk} description with\footnote{Under the GL(2,$\Z$), we have
$\bigl( {\begin{smallmatrix} 
0 &2 \\
2 & 2  
\end{smallmatrix}} \bigl) \simeq
 \bigl( {\begin{smallmatrix} 
2 & 0 \\
0 & -2  
\end{smallmatrix}} \bigl)
$,
this is the 3d $\U(1)_2 \times \U(1)_{-2}$ CS theory, 
or a 2+1D $\U(1)_2 \times \U(1)_{-2}$ non-chiral fractional quantum Hall state.
\label{footnote:3dCSDS}
}
\bea
K= \bigl( {\begin{smallmatrix} 
2 & 0 \\
0 & -2  
\end{smallmatrix}} \bigl)
.\eea
We can also write the same data of $\cW^\text{DS}$ in terms of the 1+1D sine-Gordon cosine terms 
\Eq{eq:Sedge-int} that
can gap the gapless CFT to obtain the $s \bar{s}$-type gapped boundaries \cite{Wang2012am1212.4863} (See footnote \ref{footnote:L}):
\bea
\ell^{  \cW^\text{DS}}=2(1,1) \simeq 2(1,-1),
\eea
which satisfies the gapping rules in \Sec{sec:gapless-gap}.

The gapped interface can also be understood as the dynamically gauging of the
 trivialization of the nontrivial 3-cocycle $\omega^3$ in $\cH^{3}(\B\mathbb{Z}_2 ,{U}(1))$ \cite{Wang2017loc1705.06728}:
\footnote{The
$\Sq^j$ denotes the $j$-th Steenrod square, and the
$\cup$ is the cup product, see of an introduction \cite{Wan2018bns1812.11967HAHSI}.
We may omit the $\cup$ product to make it implicit later.
We see that 
$(-1)^{\int_{M^3} A\cup \Sq^1 A} =(-1)^{\int_{M^3} A\cup \frac{1}{2} \delta A} =(-1)^{\int_{M^3} A\cup A \cup A}
=(-1)^{\int_{M^3} A^3}$.
The $\delta$ is a coboundary operator, which sends $A \in \cH^1(M^3 ,\Z_2)$ to 
$\Sq^1 A =  \frac{1}{2} \delta A  \in \cH^2(M^3,\Z_2)$.
In the last expression, we convert to the group cocycle in \cite{1405.7689}, we write
the 3-cocycle $\omega^3 : G^3 \mapsto \U(1)$, mapping $(g_a,g_b,g_c) \in G^3 = \Z_2^3$ to U(1).
The 3-cocycle as group-cocycle is solved as $\omega^3(g_a,g_b,g_c) = (-1)^{\int_{M^3} g_a \cup g_b \cup g_c}$  \cite{1405.7689}.
\label{footnote:Steenrod}
 }
\bea
\omega^3=(-1)^{\int_{M^3} A\cup A \cup A} =(-1)^{\int_{M^3} A\cup \Sq^1 A}   \to (-1)^{\int_{M^3} g_a \cup g_b \cup g_c},
\eea
either by the breaking in \Eq{eq:ssbar-breaking}, 
\begin{equation} 	\label{eq:0Z2}
	 0_{\text{boundary}}^{G'} \to{\mathbb{Z}_2^G}_{\text{bulk}},
\end{equation}
or extensions, such as \cite{Wang2017loc1705.06728} \cite{Wang2018edf1801.05416}
\begin{equation}
	0\to {\mathbb{Z}_2^\mathbf{N}}_{\text{boundary}}\to {\mathbb{Z}_4^\mathbb{G} }_{\text{boundary}} \to{\mathbb{Z}_2^G}_{\text{bulk}} \to 0.
	\label{eq:Z2Z4Z2}
\end{equation}
\Refe{Wang2017loc1705.06728, Wang2018edf1801.05416} finds that in 2+1D bulk and 1+1D gapped interface, in fact both breaking and extension constructions are equivalent when ${\mathbb{Z}_2^G}_{\text{bulk}}$ is dynamically gauged. (See especially Sec.~7.1 of  \cite{Wang2018edf1801.05416}.)

\subsection{Two twisted $\Z_2$-gauge theories: 2 types of gapped interfaces}
\label{subsection:Z2two}

We obtain 2 types of gapped interfaces between two twisted $\Z_2$ gauge theories, solved from \cite{Lan2014uaa1408.6514}'s formula 
(or a gapped boundary  between two double-semion models to a trivial vacuum\footnote{We can rewrite the gapped interface
$\cW_{s \leftrightarrow {s}}^\mathrm{DS|DS}$
in terms of the gapped boundary 
$\cW_{s_i \bar{s}_j \leftrightarrow 1}^{\mathrm{DS}^2}$$=$
$\left(
\begin{array}{cccc cccc  cccc cccc c  c}
1 &  s_1 & \bar{s}_1 & s_1 \bar{s}_1   &
s_2 &  s_1 s_2 & \bar{s}_1 s_2 & s_1 \bar{s}_1 s_2
&
\bar{s}_2 &  s_1 \bar{s}_2  & \bar{s}_1 \bar{s}_2  & s_1 \bar{s}_1 \bar{s}_2  
&
s_2 \bar{s}_2 &  s_1 s_2 \bar{s}_2  & \bar{s}_1  s_2\bar{s}_2  & s_1 \bar{s}_1 s_2\bar{s}_2 & &   \\
\hline
 1 & 0 & 0 & 0 &  0 & 0 & 1 & 0 & 0  &  1 & 0 & 0 & 0  & 0 & 0 & 1  &  \;\vline  & 1  \\
 \end{array}
\right)
 $, which is the boundary 
 between two double-semion models to a trivial vacuum.
Notice that the folding trick flips the semion $s_j$ on one side to its
complex-conjugated semion $\bar{s}_j$, and vice versa.
(The $\bar{s}_j$ is the time-reversal partner of $s_j$.)
The reader should beware the flipping of 
 $s_j$ and  $\bar{s}_j$, between
$\cW_{s \leftrightarrow {s}}^\mathrm{DS|DS}$ and $\cW_{s_i \bar{s}_j \leftrightarrow 1}^{\mathrm{DS}^2}$.
\label{footnote:W}
}):
\bea
\cW^\mathrm{DS|DS}&=&
\left(
\begin{array}{cccc c c}
1 &  s_1 & \bar{s}_1 & s_1 \bar{s}_1   & & \\
\hline
 1 & 0 & 0 & 1 & \;\vline & 1  \\
  0 & 0 & 0 & 0 & \;\vline & s_2  \\
 0 & 0 & 0 & 0 & \;\vline &  \bar{s}_2    \\
 1 & 0 & 0 & 1 & \;\vline & s_2 \bar{s}_2  
\end{array}
\right)=(\cW^\text{DS}_{})^\dag\cW^\text{DS}_{},\nn\\
\cW_{s \leftrightarrow {s}}^\mathrm{DS|DS}&=&
\mathbb{I}_4=
\left(
\begin{array}{cccc c c}
1 &  s_1 & \bar{s}_1 & s_1 \bar{s}_1   & & \\
\hline
 1 & 0 & 0 & 0 & \;\vline & 1  \\
  0 & 1 & 0 & 0 & \;\vline & s_2  \\
 0 & 0 & 1 & 0 & \;\vline &  \bar{s}_2    \\
 0 & 0 & 0 & 1 & \;\vline & s_2 \bar{s}_2  
\end{array}
\right).
\eea
In terms of the set of condensed anyons for the above two interfaces we have,
\bea
\cW^\mathrm{DS|DS} &: 1, s_1 \bar{s}_1, s_2 \bar{s}_2, [s_1 \bar{s}_1 s_2 \bar{s}_2].\\
\cW^\mathrm{DS|DS}_{s \leftrightarrow {s}} &:  1, s_1 \bar{s}_2, s_2 \bar{s}_1, [s_1 \bar{s}_1 s_2 \bar{s}_2].
\eea
This $(\Z_2)^2$-gauge theory also has the 3d CS \Eq{eq:Sbulk} description with
\bea
K=
\bigl( {\begin{smallmatrix} 
2 &0 \\
0 & -2  
\end{smallmatrix}} \bigl)
\oplus
\bigl( {\begin{smallmatrix} 
2 &0 \\
0 & -2  
\end{smallmatrix}} \bigl). 
\eea
We can also write the same data in terms of the 1+1D sine-Gordon cosine terms 
\Eq{eq:Sedge-int} with
$
\int \dd t \dd x  \big( g_{{\mathbf{1}}}  \cos( \ell_{\mathbf{1},I} \cdot \Phi^{}_{ I} )+  g_{\mathbf{2}}  \cos( \ell_{\mathbf{2},I} \cdot \Phi^{}_{I} ) \big)
$
that
can gap the gapless CFT to obtain gapped interfaces:
\bea
\cW^\mathrm{DS|DS} &:& \ell_{\mathbf{1}}=2(1,1,0,0),\quad \ell_{\mathbf{2}}=2(0,0,1,1).\\
\cW_{s \leftrightarrow {s}}^\mathrm{DS|DS}&:& \ell_{\mathbf{1}}=2(1,0,0,1),\quad\ell_{\mathbf{2}}=2(0,1,1,0).
\eea
They satisfy the gapping rules in \Sec{sec:gapless-gap}.
They can also be understood as the breaking construction via the dynamically gauging of the trivialization of the 3-cocycle \cite{Wang2017loc1705.06728, Wang2018edf1801.05416}
at the interface $G_{\text{to-be-condensed}}  \mapsto 0$:
\bea
\cW^\mathrm{DS|DS} &:&   \Z_2^{{s_1 {\bar{s}_1}}} \times \Z_2^{{s_2 {\bar{s}_2}}}  \mapsto 0, \\
\cW_{s \leftrightarrow {s}}^\mathrm{DS|DS} \text{ or } {\cW_{s_i \bar{s}_j \leftrightarrow 1}^{\mathrm{DS}^2}}&:&
 \Z_2^{{s_1 {\bar{s}_2}}} \times \Z_2^{{s_2 {\bar{s}_1}}}  \mapsto 0.
\eea
Here $\cW_{s \leftrightarrow {s}}^\mathrm{DS|DS}$ and ${\cW_{s_i \bar{s}_j \leftrightarrow 1}^{\mathrm{DS}^2}}$
are related by the folding trick \ref{footnote:W}.


\subsection{Three twisted $\Z_2$ gauge theories: Liquid and Non-Liquid Cellular states}
\label{subsection:Z2t-three}

We obtain $3!=6$ types of gapped interfaces between three twisted $\Z_2$ gauge theories, solved from \cite{Lan2014uaa1408.6514}'s formula.
Each of them have $2^3$ types of anyons condensed on its gapped interface:
\begin{align}
\cW^{\text{DS}^3_{0}} 
\equiv \cW^{0} 
&: 1, s_1 \bar{s}_1, s_2 \bar{s}_2, s_3 \bar{s}_3, [ s_1 \bar{s}_1 s_2 \bar{s}_2], 
[s_1 \bar{s}_1  s_3 \bar{s}_3], [s_2 \bar{s}_2 s_3 \bar{s}_3],
[s_1 \bar{s}_1 s_2 \bar{s}_2 s_3 \bar{s}_3].\nn\\
\cW^{\text{DS}^3_{\ri}} 
\equiv
\cW^{\ri} 
&: 1, s_1 \bar{s}_2, s_2 \bar{s}_1, s_3 \bar{s}_3, [ s_1 \bar{s}_1 s_2 \bar{s}_2], 
[s_1 \bar{s}_2  s_3 \bar{s}_3], [s_2 \bar{s}_1 s_3 \bar{s}_3],
[s_1 \bar{s}_1 s_2 \bar{s}_2 s_3 \bar{s}_3].\nn\\
\cW^{\text{DS}^3_{\rii}} 
\equiv\cW^{\rii} 
&: 1,  s_2 \bar{s}_3, s_3 \bar{s}_2, s_1 \bar{s}_1, [ s_1 \bar{s}_1 s_2 \bar{s}_3], 
[s_1 \bar{s}_1  s_3 \bar{s}_2], [s_2 \bar{s}_2 s_3 \bar{s}_3],
[s_1 \bar{s}_1 s_2 \bar{s}_2 s_3 \bar{s}_3].\nn\\
\cW^{\text{DS}^3_{\riii}} 
\equiv
\cW^{\riii} 
&: 1, s_3 \bar{s}_1, s_1 \bar{s}_3,  s_2 \bar{s}_2,
[s_1 \bar{s}_1 s_3 \bar{s}_3], [s_2 \bar{s}_2 s_1 \bar{s}_3],
[s_2 \bar{s}_2 s_3 \bar{s}_1],[s_1 \bar{s}_1 s_2 \bar{s}_2 s_3 \bar{s}_3].\nn\\
\cW^{\text{DS}^3_{\I_{s \bar{s}}} } 
\equiv
\cW^{\I_{s \bar{s}}} 
&:1, s_1 \bar{s}_2, s_2 \bar{s}_3, s_3 \bar{s}_1, 
 [s_2 \bar{s}_2 s_1 \bar{s}_3],
[ s_1 \bar{s}_1 s_3 \bar{s}_2], [ s_3 \bar{s}_3 s_2 \bar{s}_1], 
[s_1 \bar{s}_1 s_2 \bar{s}_2 s_3 \bar{s}_3]. \nn\\
\cW^{\text{DS}^3_{\II_{s \bar{s}}} }
\equiv
\cW^{\II_{s \bar{s}}}
&: 1, s_1 \bar{s}_3, s_3 \bar{s}_2, s_2 \bar{s}_1, 
[ s_3 \bar{s}_3 s_1 \bar{s}_2],
[ s_1 \bar{s}_1 s_2 \bar{s}_3],
 [s_2 \bar{s}_2 s_3 \bar{s}_1],
 [s_1 \bar{s}_1 s_2 \bar{s}_2 s_3 \bar{s}_3]. \label{eq:3-couple-semion}
\end{align}
%
Here we use the \emph{subscripts} 1,2,3 to denote 
the \emph{layer indices} in \Eq{eq:3-couple-semion}: the 1st layer,
the 2nd layer, and the 3rd layer of 2+1D TQFTs.
The gapped interface can also be understood as the dynamically gauging of the boundary trivialization of the bulk topological term \cite{Wang2017loc1705.06728},
\bea
(-1)^{\int_{M^3} (A_1)^3+(A_2)^3+(A_3)^3} =(-1)^{\int_{M^3} \sum_{j=1}^3 A_j\cup \Sq^1 A_j},
\eea
with the lower 
subscripts of $A_j$ here meaning the $j$-th layer index.

Of course, we can also write down their tensor expressions. For the rank-3 tensor with 3-legs, we write one lower subindex on the right corner of the rank-2 matrix, e.g., 
\begin{equation}\hspace{-18mm}
\cW^0
=
\left(
\begin{array}{cccc c c}
1 &  s_1 & \bar{s}_1 & s_1 \bar{s}_1   & & \\
\hline
 1 & 0 & 0 & 1 & \;\vline & 1  \\
  0 & 0 & 0 & 0 & \;\vline & s_2  \\
 0 & 0 & 0 & 0 & \;\vline &  \bar{s}_2    \\
 1 & 0 & 0 & 1 & \;\vline & s_2 \bar{s}_2  
\end{array}
\right)_1,
\left(
\begin{array}{cccc c c}
1 &  s_1 & \bar{s}_1 & s_1 \bar{s}_1   & & \\
\hline
 0 & 0 & 0 & 0 & \;\vline & 1  \\
  0 & 0 & 0 & 0 & \;\vline & s_2  \\
 0 & 0 & 0 & 0 & \;\vline &  \bar{s}_2    \\
 0 & 0 & 0 & 0 & \;\vline & s_2 \bar{s}_2  
\end{array}
\right)_{s_3},
\left(
\begin{array}{cccc c c}
1 &  s_1 & \bar{s}_1 & s_1 \bar{s}_1   & & \\
\hline
 0 & 0 & 0 & 0 & \;\vline & 1  \\
  0 & 0 & 0 & 0 & \;\vline & s_2  \\
 0 & 0 & 0 & 0 & \;\vline &  \bar{s}_2    \\
 0 & 0 & 0 & 0 & \;\vline & s_2 \bar{s}_2  
\end{array}
\right)_{\bar{s}_3},
\left(
\begin{array}{cccc c c}
1 &  s_1 & \bar{s}_1 & s_1 \bar{s}_1   & & \\
\hline
 1 & 0 & 0 & 1 & \;\vline & 1  \\
  0 & 0 & 0 & 0 & \;\vline & s_2  \\
 0 & 0 & 0 & 0 & \;\vline &  \bar{s}_2    \\
 1 & 0 & 0 & 1 & \;\vline & s_2 \bar{s}_2  
\end{array}
\right)_{s_3\bar{s}_3} 
\end{equation}
and others.
The $\cW^{0}$ is fully symmetric, but the three 2+1D TQFTs are totally decoupled --- 
$\cW^{0}$ is the tensor product of three copies of \Eq{eq:WDS}.
 Follow the renormalization process of hexagonal honeycomb column lattice in \Sec{sec:Hexagonalhoneycombcolumnlattice}, we find 
 \bea
\label{eq:hex-deformation-DS3-0}
&&\sum_j  \cW^0_{jab} \cW^0_{jcd} = \sum_{j'} (\cW'^{0})_{j' bc} (\cW'^{0})_{j'  da}.\\
&&\sum_{a,b,c} (\cW'^{0})_{ a i' b} (\cW'^{0})_{b j' c} (\cW'^{0})_{c k' a} = 8\t \cW^{0}_{i' j' k'}.
\eea
Here $\cW^{\text{DS}^3_{0}} 
\equiv  \cW^0 =\cW'^{0}= \t \cW^{0}$,\footnote{In the following, we also set $\cW^{\#}=\cW'^{\#}=\tilde{\cW}^{\#}$ for the generic type of interface indices ${\#} \in 
\{0, \ri, \rii, \riii, {\I_{s \bar{s}}}, {\II_{s \bar{s}}}\}$ in \Eq{eq:3-couple-semion}. 
Namely $W'$ and $\tilde{W}$ indices are only meant to indicate those $\cW$ are in the procedure of renormalization.} 
following the interpretations in \Sec{sec:Interpretations}, the factor 8 is due to accidental degeneracy  \cite{Wen2020pri2002.02433}. This implies that $\cW^{0}$ can give rise to 
 3+1D {\bf\emph{{gapped liquid cellular states}}}.

The
$\cW^{\ri}, \cW^{\rii}, \cW^{\riii}$ are non-symmetric and non-cyclic symmetric;
one out of three 2+1D TQFTs are decoupled from the other two. A simple interpretation of this type of gapped interface, say $\cW^{\ri}$ 
is that the bound $s_1\bar{s}_1$ from the 1st layer can move to the interface, while
the semion $s_2$ from the 2nd layer can move into the 3rd layer as the complex-conjugation of $\bar s_3$ (i.e., $s_3$).
Namely the semion $s_2$ can cross the interface and become $s_3$, and the
$\bar{s}_2$ can cross the interface and become $\bar{s}_3$: and vice versa, see footnote \ref{footnote:W}.

The $\cW^{\I_{s \bar{s}}}$ and $\cW^{\II_{s \bar{s}}}$ are cyclic symmetric ($1\Rightarrow 2 \Rightarrow 3 \Rightarrow 1$)
and anti-cyclic symmetric ($1\Leftarrow 2 \Leftarrow 3 \Leftarrow 1$) under the \emph{layer subscripts} 1,2,3 in \Eq{eq:3-couple-semion}; 
nicely, three 2+1D TQFTs are totally coupled together. We can also write down
$\cW^{\I_{s \bar{s}}}$ and $\cW^{\II_{s \bar{s}}}$  tensors in terms of their nonzero components similar to \Eq{eq:WZ2-tensor}:
\bea
 && \text{We can also write the tensor components with the lower index labelings $1,s,\bar{s}, s\bar{s} \Leftrightarrow \mi{1,2,3,4}$}, \nn\\
\cW^{\I_{s \bar{s}}} &:& 
\cW^{\I_{s \bar{s}}}_{\mi{111}}=\cW^{\I_{s \bar{s}}}_{\mi{231}}=\cW^{\I_{s \bar{s}}}_{\mi{123}}=\cW^{\I_{s \bar{s}}}_{\mi{312}}=\cW^{\I_{s \bar{s}}}_{\mi{432}}=\cW^{\I_{s \bar{s}}}_{\mi{243}}=\cW^{\I_{s \bar{s}}}_{\mi{324}}=\cW^{\I_{s \bar{s}}}_{\mi{444}}=1. 
\nonumber\\
\cW^{\II_{s \bar{s}}} &:&  \cW^{\II_{s \bar{s}}}_{\mi{111}}= \cW^{\II_{s \bar{s}}}_{\mi{321}}= \cW^{\II_{s \bar{s}}}_{\mi{213}}= \cW^{\II_{s \bar{s}}}_{\mi{132}}=\cW^{\II_{s \bar{s}}}_{\mi{234}}= \cW^{\II_{s \bar{s}}}_{\mi{342}}= \cW^{\II_{s \bar{s}}}_{\mi{423}}= \cW^{\II_{s \bar{s}}}_{\mi{444}}=1. 
\eea


We are interested in constructing nontrivial gapped quantum states out of fully coupled gapped interfaces
from any choice of $\cW^{\I_{s \bar{s}}}$ and $\cW^{\II_{s \bar{s}}}$. Interestingly,
for the hexagonal lattice, checking the first step (``crossing symmetry'') renormalization \Eq{eq:hex-deformation},
we find
\bea
\sum_j  \cW^{A_1}_{jab} \cW^{B_1}_{jcd} \neq \sum_{j'} (\cW'^{A'_2})_{j' bc} (\cW'^{B'_2})_{j'  da},
\eea
for any ${A_1},{B_1} \in \{{\I_{s \bar{s}}}, {\II_{s \bar{s}}}\}$
and
for any ${A'_2},{B'_2} \in \{0,{\I_{s \bar{s}}}, {\II_{s \bar{s}}}\}$.
Furthermore, we check that
for any ${A_1},{B_1} \in \{{\I_{s \bar{s}}}, {\II_{s \bar{s}}}\}$,
they cannot satisfy any of the analogous formulas\footnote{The symbol, ``$\not\propto$,'' in the renormalization formula,
particularly mean that the left hand side
is not equivalent ($\neq$) nor proportional up to an integer constant factor to the right hand side formula. For example,
\Eq{eq:DS3-WAB-1}
means that
$\sum_j  \cW^{A_1}_{jab} \cW^{B_1}_{jcd} \neq  \text{n} \sum_{j'} (\cW'^{A'_2})_{j' bc} (\cW'^{B'_2})_{j'  da}$
and
$ \text{n} \sum_j  \cW^{A_1}_{jab} \cW^{B_1}_{jcd} \neq  \sum_{j'} (\cW'^{A'_2})_{j' bc} (\cW'^{B'_2})_{j'  da}$
for any integer n.
\label{footnote:notpropto}
} 
\bea
\sum_j  \cW^{A_1}_{jab} \cW^{B_1}_{jcd} \not \propto \sum_{j'} (\cW'^{A'_2})_{j' bc} (\cW'^{B'_2})_{j'  da}, \label{eq:DS3-WAB-1}\\
\sum_j  \cW^{A_1}_{jab} \cW^{B_1}_{jcd} \not \propto \sum_{j'} (\cW'^{A'_2})_{j' bc} (\cW'^{B'_2})_{j'  ad}, \label{eq:DS3-WAB-2}\\
\sum_j  \cW^{A_1}_{jab} \cW^{B_1}_{jcd} \not \propto \sum_{j'} (\cW'^{A'_2})_{j' cb} (\cW'^{B'_2})_{j'  da}, \label{eq:DS3-WAB-3}\\
\sum_j  \cW^{A_1}_{jab} \cW^{B_1}_{jcd} \not \propto \sum_{j'} (\cW'^{A'_2})_{j' cb} (\cW'^{B'_2})_{j'  ad}, \label{eq:DS3-WAB-4}
\eea
for any ${A'_2},{B'_2} \in \{0,\ri, \rii, \riii, {\I_{s \bar{s}}}, {\II_{s \bar{s}}}\}$ including those \emph{non-cyclic symmetric} interface tensors, i.e.,
$\cW^{\ri}$, $\cW^{\rii}$,  and $\cW^{\riii}$. 
%
From \Eq{eq:DS3-WAB-1} to \Eq{eq:DS3-WAB-4}, on the right hand side,
we consider all possible permutations of interface subscripts ($bcda$,  $cbda$, $bcad$, and $cbad$),
because we also need to check the non-cyclic symmetric interfaces on the right hand side whose subscript-ordering is crucial.
Thus, by  \Eq{eq:DS3-WAB-1}-\Eq{eq:DS3-WAB-4}, 
we exclude the possibility of having any pair of interface tensor $\cW$, from ${A_1},{B_1} \in \{{\I_{s \bar{s}}}, {\II_{s \bar{s}}}\}$,
proportional up to an integer constant factor to the right hand side formula from any 
proposed renormalized 3-leg tensors $\cW'^{\text{DS}^3}$.
 Follow the interpretations in \Sec{sec:Interpretations},
 we see that
 all these interfaces on two sublattices $A$ and $B$, denoted $(\cW^A, \cW^B)$, on a {hexagonal honeycomb column lattice} 
 may be used to construct 3+1D {\bf\emph{{gapped non-liquid cellular states}}}: 
\bea
(\cW^{\I_{s \bar{s}}}, \cW^{\I_{s \bar{s}}}), \quad (\cW^{\I_{s \bar{s}}}, \cW^{\II_{s \bar{s}}}), \quad (\cW^{\II_{s \bar{s}}}, \cW^{\I_{s \bar{s}}}), \quad
(\cW^{\II_{s \bar{s}}}, \cW^{\II_{s \bar{s}}}),
\eea
because they cannot be coarse-grained and renormalized to themselves.
Moreover, from \Eq{eq:3-couple-semion}, we see that in general,
the semion $s_a$ (anyon) crossing the gapped interface has to go either clockwise or counter-clockwise direction;
the $\bar{s}_a$  crossing the gapped interface has to go the opposite direction.
Therefore, both semions and anti-semions have restricted mobility to cross the interfaces.
While the double-semion anyon $s_c \bar{s}_c$ can cross the gapped interface, the 
$s_c \bar{s}_c$ needs to be split into 
the semion $s_a$ 
and anti-semion $\bar{s}_b$ ---  the $s_a$ goes to the $a$ side and 
the $\bar{s}_b$ goes to the $b$ side.
Therefore, the anyon-splitting and restricted mobility
behavior indicates possible 
lineons, planons, or fractons --- depending on how do we extend the lattice to the third spatial dimension (the $z$ axis).


\subsection{Four twisted $\Z_2$ gauge theories: Liquid and Non-Liquid Cellular states}
\label{subsection:Z2t-four}

Similar to \Eq{eq:3-couple-semion},
we can at least obtain $4!=24$ types of gapped interfaces between four twisted $\Z_2$ gauge theories, 
each of them have $2^4$ types of anyons condensed on its gapped interface, let us call it $\cW^{ijkl_{s \bar{s}}}$ 
(see footnote \ref{footnote:bracket}):
\bea \label{eq:WDS4ijkl} 
\cW^{\text{DS}^4_{ijkl_{s \bar{s}}} } 
\equiv \cW^{ijkl_{s \bar{s}}} &: 1, s_1 \bar{s}_i, s_2 \bar{s}_j, s_3 \bar{s}_k, s_4 \bar{s}_l, [\dots].
\eea
Here $i,j,k,l$ are all permutations of 1,2,3,4.
For example,
$$
\cW^{2341_{s \bar{s}}} : 1, s_1 \bar{s}_2, s_2 \bar{s}_3, s_3 \bar{s}_4, s_4 \bar{s}_1, [\dots].
$$
We can also find other exotic gapped interface (with explicit 16 types of anyons that can condense on this specific interface, see footnote 
\ref{footnote:bracket}):\footnote{To obtain this gapped interface \Eq{eq:WDS4al},
we can add four linear independent cosine terms to gapless 1+1D CFTs on the boundary of
3d CS \Eq{eq:Sbulk} description with
$
K=
\bigl( {\begin{smallmatrix} 
2 &0 \\
0 & -2  
\end{smallmatrix}} \bigl)
\oplus
\bigl( {\begin{smallmatrix} 
2 &0 \\
0 & -2  
\end{smallmatrix}} \bigl)
\oplus
\bigl( {\begin{smallmatrix} 
2 &0 \\
0 & -2  
\end{smallmatrix}} \bigl)
\oplus
\bigl( {\begin{smallmatrix} 
2 &0 \\
0 & -2  
\end{smallmatrix}} \bigl)
$
following \Sec{sec:gapless-gap}:
$$
\ell_{\mathbf{1}}=2(1, 0, 1, 0, 1, 0, 1, 0),\quad \ell_{\mathbf{2}}=2(0, 1, 0, 1, 0, 1, 0, 1)
,\quad \ell_{\mathbf{3}}=2(1, 0, 1, 0, 0, 1, 0, 1),
\quad \ell_{\mathbf{4}}=2(0, 1, 1, 0, 1, 0, 0, 1).
$$
} 
\bea
&&\hspace{-20mm}
\cW^{\text{DS}^4_\al} : 1, s_1 s_2 s_3 s_4, \bar{s}_1 \bar s_2 \bar s_3  \bar s_4, {s}_1 s_2  \bar s_3  \bar s_4, {s}_2 s_3  \bar s_4  \bar s_1, 
[ s_1 s_2 \bar s_1  \bar s_2], 
[ s_1 s_3 \bar s_1  \bar s_3],  [ s_1 s_4 \bar s_1  \bar s_4],
 [ s_2 s_3 \bar s_2  \bar s_3],    [ s_2 s_4 \bar s_2  \bar s_4],   [ s_3 s_4 \bar s_3  \bar s_4], \nn\\
 &&
[ s_3   s_4 \bar{s}_1 \bar  s_2 ],  [ s_1   s_4 \bar{s}_2 \bar  s_3 ], [ s_2   s_4 \bar{s}_1 \bar  s_3 ], 
[ s_1   s_3 \bar{s}_2 \bar  s_4 ], [s_1 s_2 s_3 s_4 \bar{s}_1 \bar s_2 \bar s_3  \bar s_4]. \label{eq:WDS4al} 
 \eea
\bea
 && \hspace{-20mm}
 \text{We can also write the tensor components of $\cW^{\text{DS}^4_\al}$ with the lower index relabelings $1,s,\bar{s}, s\bar{s} \Leftrightarrow \mi{1}, \mi{2}, \mi{3},\mi{4}$, then} \nn\\
\cW^{\text{DS}^4_\al} &:& 
\cW^{\text{DS}^4_\al}_{\mi{1111}}=\cW^{\text{DS}^4_\al}_{\mi{2222}}=\cW^{\text{DS}^4_\al}_{\mi{3333}}=\cW^{\text{DS}^4_\al}_{\mi{2233}}=\cW^{\text{DS}^4_\al}_{\mi{3223}}=\cW^{\text{DS}^4_\al}_{\mi{4411}}=\cW^{\text{DS}^4_\al}_{\mi{4141}}=\cW^{\text{DS}^4_\al}_{\mi{4114}} 
\nonumber\\
&&  
=\cW^{\text{DS}^4_\al}_{\mi{1441}}= \cW^{\text{DS}^4_\al}_{\mi{1414}}= \cW^{\text{DS}^4_\al}_{\mi{1144}}= \cW^{\text{DS}^4_\al}_{\mi{3322}}=\cW^{\text{DS}^4_\al}_{\mi{2332}}= \cW^{\text{DS}^4_\al}_{\mi{3232}}= \cW^{\text{DS}^4_\al}_{\mi{2323}}= \cW^{\text{DS}^4_\al}_{\mi{4444}}=1. 
\eea
Here $\cW^{\text{DS}^4_\al}$ is fully symmetric and cyclic symmetric,
so this gapped interface can be helpful to
construct an isotropic phase of matter.
 Follow the renormalization process on a square column lattice in \Sec{sec:squarecolumnlattice}, we find 
 \bea
\label{eq:deformation-Walpha-0}
&&  \cW^{\text{DS}^4_\al}_{abcd}  \not\propto 
\sum_{j'} (\cW'^{\text{DS}^3_{0}})_{ab j'  } (\cW'^{\text{DS}^3_{0}})_{j'  cd}
=
\sum_{j'} (\cW'^{0})_{ab j'  } (\cW'^{0})_{j'  cd}.
\eea
Here $\cW'^{\text{DS}^3_{0}} = \cW^{\text{DS}^3_{0}} \equiv \cW^{0}$ is given by \Eq{eq:3-couple-semion}.
We further check for arbitrary $\cW^{A'_2}$ and $\cW^{B'_2}$ with ${A'_2},{B'_2} \in \{0,\ri, \rii, \riii, {\I_{s \bar{s}}}, {\II_{s \bar{s}}}\}$
given by all possible 
3-leg tensors in
\Eq{eq:3-couple-semion}, we find
\bea \label{eq:deformation-Walpha-all}
&&  \cW^{\text{DS}^4_\al}_{abcd}  \not\propto
\sum_{j'} (\cW^{A'_2})_{ab j'  } (\cW^{B'_2})_{j'  cd}.
\eea
By the symbol, ``$\not\propto$,''
we mean that $\cW^{\text{DS}^4_\al}_{}$
is not equivalent ($\neq$) nor proportional up to an integer constant factor to the right hand side formula for some 
proposed renormalized 3-leg tensors $\cW'^{\text{DS}^3}$. (See footnote \ref{footnote:notpropto}.) 

Furthermore, by comparing \Eq{eq:WDS4ijkl}'s and \Eq{eq:WDS4al}'s interface tensors,
$\cW^{\text{DS}^4_{ijkl_{s \bar{s}}} }$ and $\cW^{\text{DS}^4_\al}$, 
we notice that the semion \emph{can} 
cross the interface $\cW^{\text{DS}^4_{ijkl_{s \bar{s}}} }$ to any neighbor layers (any layer of 2+1D double-semion TQFT labeled by $i,j,k,l \in \{1,2,3,4\}$) 
without the fracton behavior as long as the semion moves to each layer via the ordering
$i \to j \to k \to l$.
In contrast, the semions \emph{cannot} freely cross the interface $\cW^{\text{DS}^4_\al}$
unless they form a bound state $s_i s_j$ or $s_i \bar{s}_j$ from the neighbored layers ($i,j$),
or unless a semion $s_i$ splits to three semions and/or anti-semions after crossing the interface $\cW^{\text{DS}^4_\al}$.

In summary, based on \Eq{eq:deformation-Walpha-0} and \Eq{eq:deformation-Walpha-all}, and the similar reasonings in \Sec{sec:Interpretations}, we expect to construct
3+1D {\bf\emph{{gapped non-liquid cellular states}}} on a \emph{square column lattice}
via the $\cW^{\text{DS}^4_\al}_{}$ tensor.
Similarly, we expect to construct
3+1D {\bf\emph{{gapped non-liquid cellular states}}} on a \emph{cubic lattice}
via the $\cW^{\text{DS}^4_\al}_{}$ tensor.

Lastly, we remark on an alternative approach on gapped interfaces 
based on \Sec{sec:trivialization-cohomology/cobordism-topological-term} method:
the trivialization of topological term from cohomology/cobordism data on the interface.
The gapped interface can also be understood as the dynamically gauging of the trivialization on the topological term \cite{Wang2017loc1705.06728},
with the lower subindices of $A_j$ as the $j$-th layer index,
\bea
(-1)^{\int_{M^3} (A_1)^3+(A_2)^3+(A_3)^3+(A_3)^4} =(-1)^{\int_{M^3} \sum_{j=1}^4 A_j\cup \Sq^1 A_j}  .
\eea
In general, for the $n$-layers coupled interface, say $j=1,2,\dots, n$, we have the topological term
\bea
(-1)^{\int_{M^3} \sum_{j=1}^n A_j\cup \Sq^1 A_j} ,
\eea 
The product of groups for all layers is $\prod_{j=1}^{n} \mathbb{G}_j=\prod_{j=1}^{n} (\mathbb{Z}_2)_j$.
We can trivialize by this topological term either lifting (pullback as the extension) as
the following fibration extension
\cite{Wang2017loc1705.06728}
\cite{Wang2018edf1801.05416} analogous to \Eq{eq:general-fiber}:
\bea \label{eq:Z2Z4Z2j}
(\B\Z_2)_j \hookrightarrow (\B\Z_4)_j \to  (\B\Z_2)_j;
\eea
or we can trivialize the topological term by breaking to set
\bea \label{eq:sum-A=0}
\sum_{j=1}^n A_j = A_1+A_2+\dots +  A_n=0 \mod 2.
\eea
To show \Eq{eq:sum-A=0} can trivialize the topological term ${\int_{M^3} \sum_{j=1}^n A_j\cup \Sq^1 A_j}$,
we plug (\ref{eq:sum-A=0}) into
\bea
\sum_{j=1}^n  A_j \Sq^1 A_j  &=& A_1 \Sq^1 A_1+\sum_{j=2}^n  A_j \Sq^1 A_j 
= (\sum_{j=2}^n  A_j) \Sq^1 (\sum_{j=2}^n  A_j)+\sum_{j=2}^n  A_j \Sq^1 A_j 
\nn\\
&=& \sum_{i < j \in \{ 2, \dots, n\}}  A_j ( \Sq^1 A_i) +A_i (\Sq^1 A_j)=\sum_{i < j \in \{ 2, \dots, n\}}   \Sq^1 (A_i A_j)=0  \mod 2.
\eea
The above equalities all are mod 2 relations, see also footnote \ref{footnote:Steenrod} and \Refe{Wan2018bns1812.11967HAHSI} for more math backgrounds.
This concludes that the 
breaking construction via \Eq{eq:sum-A=0} on the 1+1D interface of ${\int_{M^3} \sum_{j=1}^n A_j\cup \Sq^1 A_j}$ can
obtain a gapped interface.
In summary, these two approaches in \Eq{eq:Z2Z4Z2j} and \Eq{eq:sum-A=0} can be used for constructing gapped interfaces.
%


\section{Cellular States from double-Fibonacci anyon models}
\label{sec:Fin-Cellular}

We comment on some construction of non-abelian cellular states using double-Fibonacci anyons (abbreviated as double-Fib, DFib, or DF).
The $\cS,\cT$ matrices  of a double-Fibonacci model are (set the golden ratio constant $\gamma \equiv \dfrac{1+\sqrt{5}}{2}$):
\begin{align*}
  \cS&=\frac{1}{1+\gamma^2}
  \begin{pmatrix}
    1&\gamma&\gamma&\gamma^2\\
    \gamma &-1&\gamma^2&-\gamma\\
    \gamma&\gamma^2&-1&-\gamma\\
    \gamma^2&-\gamma&-\gamma&1
  \end{pmatrix}.\\
    \cT&=\diag(1,\re^{\frac{\ii 4\pi }{5}},\re^{-\frac{\ii 4\pi }{5}},1).
\end{align*}
There is only one type of gapped boundary to the trivial vacuum,
\begin{align*}
  \cW^\text{DFib}=\begin{pmatrix}
    1&0&0&1
  \end{pmatrix}
    =
\left(
\begin{array}{cccc c c}
1 &  \tau & \bar{\tau} & \tau \bar{\tau}   & & \\
\hline
 1 & 0 & 0 & 1 & \;\vline & 1 
\end{array}
\right).
\end{align*}

\subsection{
Two double-Fibonacci anyon models:
2 types of gapped interfaces}
\label{sec:DFib-2}

We find $2!=2$ types of gapped interfaces between two double-Fibonacci models, solved from the \cite{Lan2014uaa1408.6514}'s formula:
\bea
\cW^{\text{DFib}^2_{\ri}}_{} &:&  1, \tau_1 \bar{\tau}_1, \tau_2 \bar{\tau}_2, [\tau_1 \bar{\tau}_1 \tau_2 \bar{\tau}_2].\\
\cW^{\text{DFib}^2_{\rii}}_{} &:&  1, \tau_1 \bar{\tau}_2, \tau_2 \bar{\tau}_1, [\tau_1 \bar{\tau}_1 \tau_2 \bar{\tau}_2].
\eea
Note that $\cW^{\text{DFib}^2_{\ri}}_{} =
 ( \cW^\text{DFib})^\dagger  \cW^\text{DFib}$ has two decoupled sectors, while
$\cW^{\text{DFib}^2_{\rii}}_{}$ has two layers coupled together.
 
\subsection{
Three double-Fibonacci anyon models: 6 types  of gapped interfaces}

We find $3!=6$ types of gapped interfaces between three double-Fibonacci models:
\begin{align}
\cW^{\text{DFib}^3_{0}}_{} &:  1, \tau_1 \bar{\tau}_1, \tau_2 \bar{\tau}_2, \tau_3 \bar{\tau}_3, [\dots]. \label{eq:DFib0}\\
\cW^{\text{DFib}^3_{\ri}}_{} &:  1, \tau_1 \bar{\tau}_2, \tau_2 \bar{\tau}_1, \tau_3 \bar{\tau}_3, [\dots]. \label{eq:DFibi}\\
\cW^{\text{DFib}^3_{\rii}}_{} &:  1,  \tau_1 \bar{\tau}_1, \tau_2 \bar{\tau}_3, \tau_3 \bar{\tau}_2, [\dots]. \label{eq:DFibii}\\
\cW^{\text{DFib}^3_{\riii}}_{} &:  1, \tau_1 \bar{\tau}_3,  \tau_2 \bar{\tau}_2, \tau_3 \bar{\tau}_1 , [\dots]. \label{eq:DFibiii}\\
\cW^{\text{DFib}^3_{\I}}_{} &:  1, \tau_1 \bar{\tau}_2, \tau_2 \bar{\tau}_3, \tau_3 \bar{\tau}_1, [\dots]. \label{eq:DFibI}\\
\cW^{\text{DFib}^3_{\II}}_{} &:  1, \tau_1 \bar{\tau}_3, \tau_3 \bar{\tau}_2, \tau_2 \bar{\tau}_1,[\dots]. \label{eq:DFibII}
\end{align}

 
Since the gapped interface structures of double-Fibonacci anyon models
are very similar to the previous double-semion models
in \Sec{sec:Z2t-Cellular} (except the Fibonacci anyon model is a non-abelian TQFT), 
we expect the construction of 3+1D {\bf\emph{{gapped cellular states}}}
can be worked out based on the similar reasonings in  \Sec{sec:Interpretations}. 

{To construct {\bf\emph{{gapped cellular states}}} from Fibonacci anyon models,
we check that on a two-sublattice {hexagonal honeycomb column lattice}, 
for any two sublattices ${A_1},{B_1} \in \{{{\text{DFib}^3_{\I}}}, {{\text{DFib}^3_{\II}}}\}$
of (DFib)$^3$ interfaces given in \Eq{eq:DFibI} and \Eq{eq:DFibII},
they cannot satisfy any of the analogous formulas (see footnote \ref{footnote:notpropto}): 
\bea
&&\sum_j  \cW^{A_1}_{jab} \cW^{B_1}_{jcd} \not \propto \sum_{j'} (\cW'^{A'_2})_{j' bc} (\cW'^{B'_2})_{j'  da},  
\quad \sum_j  \cW^{A_1}_{jab} \cW^{B_1}_{jcd} \not \propto \sum_{j'} (\cW'^{A'_2})_{j' bc} (\cW'^{B'_2})_{j'  ad}, 
\nn
\\
&&\sum_j  \cW^{A_1}_{jab} \cW^{B_1}_{jcd} \not \propto \sum_{j'} (\cW'^{A'_2})_{j' cb} (\cW'^{B'_2})_{j'  da}, 
\quad \sum_j  \cW^{A_1}_{jab} \cW^{B_1}_{jcd} \not \propto \sum_{j'} (\cW'^{A'_2})_{j' cb} (\cW'^{B'_2})_{j'  ad}, 
\eea
for any ${A'_2},{B'_2} \in \{\text{DFib}^3_0,\text{DFib}^3_\ri, \text{DFib}^3_\rii, \text{DFib}^3_\riii, \text{DFib}^3_{\I_{s \bar{s}}}, \text{DFib}^3_{\II_{s \bar{s}}}\}$, 
given in \Eq{eq:DFib0}-\Eq{eq:DFibII},
including those \emph{non-cyclic symmetric} interface tensors. However, 
it is more challenging to identify the consistent 3+1D non-abelian {\emph{{gapped cellular states}}} from
the given 2+1D non-abelian topological orders and their interface tensors, compared to the construction of abelian counterparts.
It will be illuminating to obtain a Hamiltonian construction of such {gapped cellular states} in the future.}

\section{Cellular States from Ising anyon models}

Let us consider some possible constructions 
of non-abelian cellular states using Ising anyon models (abbreviated as Ising).
The $\cS,\cT$ matrices of the Ising anyon model are
\begin{align*}
  \cS&=\frac{1}{2}
  \left(
  \begin{array}{ccccccccc}
    1 & \sqrt{2} & 1 \\
    \sqrt{2} & 0 & -\sqrt{2}  \\
    1 & -\sqrt{2} & 1 
  \end{array}
  \right).  \\
    \cT&=\diag(1,\re^{\frac{ \ii \pi}{8}},-1).
\end{align*}
The $\cS,\cT$ matrices of doubled-Ising anyon model (abbreviated as DIsing or Ising $\times$ $\overline{\text{Ising}}$) are
\begin{align*}
  \cS&=\frac{1}{4}
  \left(
  \begin{array}{ccccccccc}
    1 & \sqrt{2} & 1 & \sqrt{2} & 2 & \sqrt{2} & 1 & \sqrt{2} & 1 \\
    \sqrt{2} & 0 & -\sqrt{2} & 2 & 0 & -2 & \sqrt{2} & 0 & -\sqrt{2} \\
    1 & -\sqrt{2} & 1 & \sqrt{2} & -2 & \sqrt{2} & 1 & -\sqrt{2} & 1 \\
    \sqrt{2} & 2 & \sqrt{2} & 0 & 0 & 0 & -\sqrt{2} & -2 & -\sqrt{2} \\
    2 & 0 & -2 & 0 & 0 & 0 & -2 & 0 & 2 \\
    \sqrt{2} & -2 & \sqrt{2} & 0 & 0 & 0 & -\sqrt{2} & 2 & -\sqrt{2} \\
    1 & \sqrt{2} & 1 & -\sqrt{2} & -2 & -\sqrt{2} & 1 & \sqrt{2} & 1 \\
    \sqrt{2} & 0 & -\sqrt{2} & -2 & 0 & 2 & \sqrt{2} & 0 & -\sqrt{2} \\
    1 & -\sqrt{2} & 1 & -\sqrt{2} & 2 & -\sqrt{2} & 1 & -\sqrt{2} & 1 \\
  \end{array}
  \right).  \\
    \cT&=\diag(1,\re^{\frac{\ii \pi}{8}},-1,\re^{-\frac{\ii \pi
  }{8}},1,-\re^{-\frac{\ii \pi}{8}},-1,-\re^{\frac{\ii \pi}{8}},1).
\end{align*}

\subsection{
Two Ising anyon models: 1 type of gapped interface}
\label{sec:Ising-2}
We find one type of gapped boundary between a double-Ising model and a trivial vacuum, 
solved from \cite{Lan2014uaa1408.6514}'s formula:
\begin{align} \label{eq:WDIsing}
  \cW^\text{DIsing}=\begin{pmatrix}
    1&0&0&0&1&0&0&0&1
  \end{pmatrix}
    =
\left(
\begin{array}{ccccccccc c c}
1 &  {\sigma} & {\psi} &  \bar{\sigma} &   \bar{\sigma} \sigma &  \bar{\sigma}  {\psi} &  \bar{\psi} &   \bar{\psi} \sigma &  \bar{\psi}  {\psi}  & \\
\hline
   1&0&0&0&1&0&0&0&1 & \;\vline & 1 
\end{array}
\right).
\end{align}
Via the folding trick, we can also convert this $\cW^\text{DIsing}$ to the only one type of gapped interface between two Ising anyon models as $ \cW^{\text{Ising}\mid\text{Ising}}$:
\bea
 \cW^{\text{Ising}\mid\text{Ising}}=\mathbb{I}_3=\left(
\begin{array}{ccccc}
1 &  \sigma & \psi   & \\
\hline
 1 & 0 & 0  \;\vline & 1 \\
 0 & 1 &  0  \;\vline & \sigma\\
  0 & 0 & 1  \;\vline & \psi
\end{array}
\right).
\eea
\subsection{
Four Ising anyon models: ``non-abelian'' gapped interfaces}

Now we classify gapped interfaces between four copies of Ising anyon models.
We can regard such a gapped interface either as the interface between
(Ising $\times$ Ising) and (Ising $\times$ Ising),
or as an interface between
(Ising $\times$ $\overline{\text{Ising}}$) and (Ising $\times$ $\overline{\text{Ising}}$).
We find several types of gapped interfaces between (Ising $\times$ Ising) and (Ising $\times$ Ising):\footnote{In fact, the similar type of gapped interfaces 
are studied previously in the percolation system of non-abelian Pfaffian and anti-Pfaffian states in \cite{Lian2018xepWang1801.10149}. 
}
\label{sec:Ising-2}
\bea
&&
\cW^{\text{Ising}^4}_{\I}=\mathbb{I}_9.\\
&& 
\cW^{\text{Ising}^4}_{\II}=\left(
\begin{array}{cccccccccc}
1 &  \sigma_1 & \psi_1 & \sigma_2 & \sigma_2  \sigma_1 & \sigma_2 \psi_1 & \psi_2 & \psi_2  \sigma_1 & \psi_2  \psi_1   & \\
\hline
 1 & 0 & 0 & 0 & 0 & 0 & 0 & 0 & 1 \;\vline & 1 \\
 0 & 0 & 0 & 0 & 0 & 0 & 0 & 0 & 0  \;\vline & \sigma_3\\
 0 & 0 & 1 & 0 & 0 & 0 & 1 & 0 & 0  \;\vline &\psi_3\\
 0 & 0 & 0 & 0 & 0 & 0 & 0 & 0 & 0  \;\vline &\sigma_4\\
 0 & 0 & 0 & 0 & 2 & 0 & 0 & 0 & 0  \;\vline &\sigma_4  \sigma_3\\
 0 & 0 & 0 & 0 & 0 & 0 & 0 & 0 & 0  \;\vline  &\sigma_4 \psi_3\\
 0 & 0 & 1 & 0 & 0 & 0 & 1 & 0 & 0  \;\vline &\psi_4\\
 0 & 0 & 0 & 0 & 0 & 0 & 0 & 0 & 0  \;\vline &\psi_4  \sigma_3\\
 1 & 0 & 0 & 0 & 0 & 0 & 0 & 0 & 1  \;\vline &\psi_4  \psi_3
\end{array}
\right).
\eea
There is also an 
interface tensor 
$\cW^{\text{Ising}^4}_{\I'}=\left(
\begin{array}{ccccccccc c}
1 &  \sigma_1 & \psi_1 & \sigma_2 & \sigma_2  \sigma_1 & \sigma_2 \psi_1 & \psi_2 & \psi_2  \sigma_1 & \psi_2  \psi_1   & \\
\hline
 1 & 0 & 0 & 0 & 0 & 0 & 0 & 0 & 0  \;\vline & 1 \\
 0 & 0 & 0 & 1 & 0 & 0 & 0 & 0 & 0 \;\vline & \sigma_3\\
 0 & 0 & 0 & 0 & 0 & 0 & 1 & 0 & 0 \;\vline  &\psi_3\\
 0 & 1 & 0 & 0 & 0 & 0 & 0 & 0 & 0 \;\vline  &\sigma_4\\
 0 & 0 & 0 & 0 & 1 & 0 & 0 & 0 & 0 \;\vline  &\sigma_4  \sigma_3\\
 0 & 0 & 0 & 0 & 0 & 0 & 0 & 1 & 0  \;\vline  &\sigma_4 \psi_3\\
 0 & 0 & 1 & 0 & 0 & 0 & 0 & 0 & 0 \;\vline  &\psi_4\\
 0 & 0 & 0 & 0 & 0 & 1 & 0 & 0 & 0 \;\vline  &\psi_4  \sigma_3\\
 0 & 0 & 0 & 0 & 0 & 0 & 0 & 0 & 1 \;\vline  &\psi_4  \psi_3
\end{array}
\right)
$
based on the relabeling
via $1 \leftrightarrow 2$ (or $3 \leftrightarrow 4$); thus we shall identify 
$\cW^{\text{Ising}^4}_{\I'}$ as the same as the rank-9 identity matrix $\cW^{\text{Ising}^4}_{\I}=\mathbb{I}_9$ (up to the relabeling).

We can also convert the
$\cW^{\text{Ising}^4}$ to $\cW^{\text{DIsing}^2}$,
by regarding them as two types of gapped interfaces between (Ising $\times$ $\overline{\text{Ising}}$) and (Ising $\times$ $\overline{\text{Ising}}$), 
using \Eq{eq:WDIsing}:\footnote{There is also one more data $\cW^{\text{DIsing}^2}_{\I'}$ with the same matrix form as $\cW^{\text{Ising}^4}_{\I'}$ that we omit.}
\bea
&&
\cW^{\text{DIsing}^2}_{\I}\equiv ( \cW^\text{DIsing})^\dagger  \cW^\text{DIsing}
=\left(
\begin{array}{cccccccccc}
1 &  {\sigma} & {\psi} &  \bar{\sigma} &   \bar{\sigma} \sigma &  \bar{\sigma}  {\psi} &  \bar{\psi} &   \bar{\psi} \sigma &  \bar{\psi}  {\psi}  & \\
\hline
 1 & 0 & 0 & 0 & 1 & 0 & 0 & 0 & 1  \;\vline & 1 \\
 0 & 0 & 0 & 0 & 0 & 0 & 0 & 0 & 0  \;\vline & {\sigma}' \\
 0 & 0 & 0 & 0 & 0 & 0 & 0 & 0 & 0  \;\vline & {\psi}' \\
 0 & 0 & 0 & 0 & 0 & 0 & 0 & 0 & 0  \;\vline &\bar{\sigma}' \\
 1 & 0 & 0 & 0 & 1 & 0 & 0 & 0 & 1   \;\vline &   \bar{\sigma}' \sigma' \\
 0 & 0 & 0 & 0 & 0 & 0 & 0 & 0 & 0  \;\vline  & \bar{\sigma}'  {\psi}' \\
 0 & 0 & 0 & 0 & 0 & 0 & 0 & 0 & 0  \;\vline &  \bar{\psi}' \\
 0 & 0 & 0 & 0 & 0 & 0 & 0 & 0 & 0  \;\vline &   \bar{\psi}' \sigma'  \\
 1 & 0 & 0 & 0 & 1 & 0 & 0 & 0 & 1   \;\vline & \bar{\psi}'  {\psi}' \\
\end{array}
\right).
\eea
\bea
&&\cW^{\text{DIsing}^2}_{\II}
=\left(
\begin{array}{cccccccccc}
1 &  {\sigma} & {\psi} &  \bar{\sigma} &   \bar{\sigma} \sigma &  \bar{\sigma}  {\psi} &  \bar{\psi} &   \bar{\psi} \sigma &  \bar{\psi}  {\psi}  & \\
\hline
 1 & 0 & 0 & 0 & 0 & 0 & 0 & 0 & 1  \;\vline & 1 \\
 0 & 0 & 0 & 0 & 0 & 0 & 0 & 0 & 0  \;\vline & {\sigma}' \\
 0 & 0 & 1 & 0 & 0 & 0 & 1 & 0 & 0  \;\vline & {\psi}' \\
 0 & 0 & 0 & 0 & 0 & 0 & 0 & 0 & 0  \;\vline &\bar{\sigma}' \\
 0 & 0 & 0 & 0 & 2 & 0 & 0 & 0 & 0   \;\vline &   \bar{\sigma}' \sigma' \\
 0 & 0 & 0 & 0 & 0 & 0 & 0 & 0 & 0  \;\vline  & \bar{\sigma}'  {\psi}' \\
 0 & 0 & 1 & 0 & 0 & 0 & 1 & 0 & 0  \;\vline &  \bar{\psi}' \\
 0 & 0 & 0 & 0 & 0 & 0 & 0 & 0 & 0  \;\vline &   \bar{\psi}' \sigma'  \\
 1 & 0 & 0 & 0 & 0 & 0 & 0 & 0 & 1   \;\vline & \bar{\psi}'  {\psi}' \\
\end{array}
\right).
\eea

Now let us discuss the construction of {\bf\emph{{cellular states}}} out of the above data.
We can consider the gapped interfaces on the 
{\bf\emph{{square column lattice}}} (as in \Sec{sec:squarecolumnlattice})
or on the {\bf\emph{{cubic lattice}}}.
On each 1D interface, we can place four 2+1D Ising models on the four 2D faces neighbored to this 1D interface.
There are two types of ways to assign the gapped interfaces:
\begin{enumerate}[label=\textcolor{blue}{\arabic*.}, ref={},leftmargin=*]
\item The first way uses $\cW^{\text{Ising}^4}_{\I}$ (i.e., $\cW^{\text{DIsing}^2}_{\I}$) above. In this case, 
the four neighbored 2+1D Ising models on four 2D faces actually \emph{decouple} to two
sets of 
 (Ising $\times$ $\overline{\text{Ising}}$) and (Ising $\times$ $\overline{\text{Ising}}$).
 Each set of (Ising $\times$ $\overline{\text{Ising}}$) has a trivial gapped interface as  $\cW^\text{DIsing}$
in \Eq{eq:WDIsing}. The cellular states constructed from only this type of interface can only be {\bf\emph{liquid cellular states}}.

\item The second way uses $\cW^{\text{Ising}^4}_{\II}$ (i.e., $\cW^{\text{DIsing}^2}_{\II}$) above.
In this case, 
the four neighbored 2+1D Ising models on four 2D faces \emph{fully couple} together. The 
$\cW^{\text{Ising}^4}_{\II}$ cannot be decoupled to tensor product ($\otimes$) structures.
We can express the $\cW^{\text{Ising}^4}_{\II}$ tensor as the
tunneling data between two (Ising $\times$ Ising) TQFTs.
We can express the $\cW^{\text{DIsing}^2}_{\II}$ tensor as the
tunneling data between two (Ising $\times$ $\overline{\text{Ising}}$) TQFTs.
Instead, we can express both tensors, $\cW^{\text{Ising}^4}_{\II}$ or $\cW^{\text{DIsing}^2}_{\II}$,
in terms of the set of anyons that can condense on this gapped interface: 
\bea \label{eq:Ising4-condensed}
\begin{array}{l}
1 \oplus \psi_1 \psi_2\oplus \psi_1 \bar{\psi}_3 
\oplus \psi_2 \bar{\psi}_3 
\oplus   (2 \sigma_1 \sigma_2  \bar{\sigma}_3 \bar{\sigma}_4)
\oplus \psi_1 \bar{\psi}_4
\oplus \psi_2 \bar{\psi}_4
\oplus \bar{\psi}_3  \bar{\psi}_4 
\oplus \psi_1 \psi_2 \bar{\psi}_3  \bar{\psi}_4. 
\end{array}
\eea
This set of condensed anyons may be regarded as the {\bf\emph{non-abelian version of anyon condensations}}.
The $\oplus$ sum means that the vacuum associated to this 1+1D gapped interface (i.e., the lowest energy state),
can create or annihilate this set of anyons, without causing energy.
The $\oplus$ sum also implies the  trivial sector 1 in the trivial non-topological gapped vacuum once crosses this
gap interface can split to any of anyon among this set of anyons.
Remarkably, the factor 2 in 
$$(2 \sigma_1 \sigma_2  \bar{\sigma}_3 \bar{\sigma}_4)=
(\sigma_1 \sigma_2  \bar{\sigma}_3 \bar{\sigma}_4) \oplus (\sigma_1 \sigma_2  \bar{\sigma}_3 \bar{\sigma}_4)
$$ is quite distinct from the other condensed anyons.
This implies the energy degeneracy splitting to a two-fold degeneracy 
for this channel $(2 \sigma_1 \sigma_2  \bar{\sigma}_3 \bar{\sigma}_4)$.
The two-fold degeneracy also implies the quantum dimension associated to this
condensed anyon $\sigma_1 \sigma_2  \bar{\sigma}_3 \bar{\sigma}_4$ is 2 instead of 1 for the familiar abelian case, 
which implies the \emph{non-abelian-ness} of this gapped interface  $\cW^{\text{Ising}^4}_{\II}$or $\cW^{\text{DIsing}^2}_{\II}$. 
In fact, such a 2-surface defect may be an example of \emph{beyond-symmetry topological 2-surface defect} in 3d TQFT, analogous to that of 
\emph{beyond-symmetry topological topological 1-line defect} in 2d CFT studied in \cite{Chang2018iay1802.04445}.

The condensed anyon $\sigma_1 \sigma_2  \bar{\sigma}_3 \bar{\sigma}_4$ on this interface also indicates that any single one of
the sigma anyon $\sigma_j$ cannot cross the interface \Eq{eq:Ising4-condensed}, unless
the $\sigma_j$ splits to three anyons on the neighbored layers.
The upshot is that we may use this \emph{non-abelian} gapped interface to construct 
3+1D {\bf\emph{non-liquid cellular states}} which can be intrinsically non-abelian.
The $\sigma$ anyon has restricted mobility (that can be planon, lineon, or fracton, 
depending on how we construct the full lattice from the interface. 

\end{enumerate}
However, importantly, a 2+1D Ising model by itself with a boundary to the vacuum has the boundary chiral central charge $c_-=c_L-c_R=1/2.$
Therefore, there are boundary \emph{gapless chiral edge modes}, such as the 1+1D chiral Majorana-Weyl fermion which is a real-valued chiral fermion $\chi_L$
with an action
$\int \dd t \dd x  \; \chi_L  (\ii \partial_t -\ii v_{L} \partial_x) \chi_L$ for some velocity constant $v_{L}$.
Thus, the caveat is that if we design such 3+1D {\emph{non-liquid cellular states}} on a 3D spatial lattice with a 2D spatial boundary,
the 2+1D boundary can have \emph{gapless} modes, although the 3+1D bulk is still \emph{fully gapped}.
On the other hand, if
we design such 3+1D {\emph{non-liquid cellular states}} on a 3D spatial periodic lattice without any spatial boundary, 
then we may have a \emph{fully gapped} 3+1D bulk only.
It will be illuminating to obtain a Hamiltonian construction of such {cellular states}, for the system with or without spatial boundaries, in the future.

\section{Cellular States of higher-symmetries and higher-dimensions}

Below we consider higher-symmetry and
higher-dimensional generalization of
{\bf{\emph{cellular states}}}
via gluing gapped interfaces, see also the useful background information in \cite{Delcamp2018wlb1802.10104, Benini2018reh1803.09336, 
Delcamp2019fdp1901.02249, Wan2019oyr1904.00994, Bullivant2019tbp, Tsui2019ykk1908.02613, Cordova2019bsd1910.04962}.\footnote{
The author thanks  Zheyan Wan and Yunqin Zheng 
for previous collaborations and inspiring discussions on the related issues.}
We will focus on constructing gapped interfaces first, then later make comments on applying the interpretations in \Sec{sec:Interpretations} for constructing {cellular states}.

\subsection{2-form gauge field, 1-form symmetry, and semionic string worldsheets}

Consider the 4+1D (5d) bulk on the $j$-th layer with a 5-cocycle of cohomology group or cobordism group, 
\bea
\omega^{5}_{j}\in \cH^{5}(\B^2(\Z_2)_j,{\U(1)}), \quad\quad \text{or} \quad\quad 
\omega^{5}_j\in  \Hom(\Omega_{5}^{\SO}(\B^2(\Z_2)_j), \U(1)).  
\eea
This describes a higher-SPTs protected by 1-form $\Z_2$ global symmetry \cite{Gaiotto2014kfa1412.5148}, commonly denoted as
$\Z_{2,[1]}$.
The classification of this $\Z_{2,[1]}$-symmetry is given by a cobordism group, e.g. in \cite{Wan2018bns1812.11967HAHSI},
as 
$$\Hom(\Omega_{5}^{\SO}(\B^2(\Z_2)), \U(1))=(\Z_2)^2.$$ 
The two generators for each of $(\Z_2)^2$ are given in \cite{Wan2018bns1812.11967HAHSI} as:
\bea
&&\left\{\begin{array}{l} 
\exp( \ii \pi \int_{M^5} B\cup\Sq^1B ) = 
\exp( \ii \pi \int_{M^5} \Sq^2\Sq^1 B).\\ 
\exp( \ii \pi \int_{M^5}  w_2(TM^5)w_3(TM^5)).
\end{array}\right.
\label{eq:bordism5SOB2Z2}
\eea
The
$\Sq^j$ denotes the $j$-th Steenrod square, the $\cup$ is the cup product which
we may omit $\cup$ making it implicit. The $w_j(TM)$ is the $j$-th Stiefel-Whitney class
of the spacetime tangent bundle over $M$.  
See an introduction in \cite{Wan2018bns1812.11967HAHSI}.
We can also later gauge the $\Z_{2,[1]}$-symmetry by coupling to 2-form or 2-cocycle gauge field $B$ and make the $B$ field dynamical, 
i.e., summing over all gauge configurations of $B$ in the path integral on $M=M^5$, see \cite{Wan2019oyr1904.00994}:
\bea
\bZ_{}^{5\text{d}}[M]
&\equiv&\frac{|\cH^0(M,\Z_2)|}{|\cH^1(M,\Z_2)|} \sum_{B\in\cH^2(M^5,\Z_2)} {\rm e}^{\ii\pi\int_{M^5}B\Sq^1B  }
\label{eq:5dSET-all}\\
&=&\frac{|\cH^0(M,\Z_2)|}{|\cH^1(M,\Z_2)|} \sum_{{{}\atop{B, b, h\in \rm{C}^2( M^5, \Z_2)}}\atop{{c \in \rm{C}^3(  M^5, \Z_2)}}}
{\exp}\Big(\ii 
\pi \int_{M^5}   
b \delta B  {+B\Sq^1B }\Big)
\nn\\
&&\cong
\int [\cD B][\cD b]{\exp}\Big(\ii 
\pi \int_{M^5}  b \mathrm{d} B  {+B\frac{1}{2} \mathrm{d} B}\Big).
\nn
\eea
In the last step (under the symbol $\cong$), we  {convert} the 5d higher-cochain TQFT to 5d higher-form gauge field continuum TQFT.\footnote{Notice that
in the last expression 
$\bZ=\int [\cD B][\cD b]{\exp}\Big(\ii 
\pi \int_{M^5}  b \mathrm{d} B  {+B\frac{1}{2} \mathrm{d} B}\Big)$, we set $B$ and $b$ with mod 2 valued periodicity.
In the usual TQFT in terms of differential forms, we should rewrite it as $\cB \sim \pi B$ and $\mathscr{C} \sim \pi b$ with mod 2$\pi$ valued periodicity,
then the 5d partition function becomes
\bea  \label{eq:5dTQFT-DS}
\bZ=\int [\cD \cB][\cD \mathscr{C}]{\exp}\Big(\ii 
\frac{2}{2\pi} \int_{M^5}  \mathscr{C} \mathrm{d} \cB  {+\cB\frac{1}{2} \mathrm{d} \cB}\Big).
\eea
This 5d TQFT is analogous to the 3d TQFT of \Sec{sec:Z2t-1},   
\bea \label{eq:3dTQFT-DS}
\bZ=\int [\cD \cA][\cD a]{\exp}\Big(\ii 
\frac{2}{2\pi} \int_{M^3}  a \mathrm{d} \cA  {+\cA\frac{1}{2} \mathrm{d} \cA}\Big)
=\int [\cD a_1][\cD a_2]
{\exp}\Big(\ii 
\frac{1}{2\pi} \int_{M^3}  (a_1, a_2)^T
 \bigl( {\begin{smallmatrix} 
2 & 0 \\
0 & -2  
\end{smallmatrix}} \bigl)
 \bigl( {\begin{smallmatrix} 
 a_1\\
a_2  
\end{smallmatrix}} \bigl) \Big)
.
\eea
The second expression of 3d TQFT is the rewriting, via the semion ($s$) gauge field $a + \cA \to a_1$, the $\overline{\text{semion}}$ ($\bar{s}$) gauge field $a  \to a_2$,
and the semion-$\overline{\text{semion}}$ (s$\bar{s}$) gauge field $\cA  \to a_1 + a_2$,
to the abelian $\Z_2$ double-semion theory as a 3d $\U(1)_2 \times \U(1)_{-2}$ CS theory in footnote \ref{footnote:3dCSDS}).
To characterize the braiding statistics of anyon in the 3d TQFT \Eq{eq:3dTQFT-DS},
we can insert two line operators as two 1-circles $S^1$ linked in a 3-sphere $S^3$ \cite{Wang2019diz1901.11537},
by using the fact that $S^3 = (D^2 \times S^1)\cup_{T^2} (S^1 \times D^2)$ with the gluing $\cup_{T^2}$ along ${T^2}$
and the homology group $H_1(D^2 \times S^1, \Z)=\Z$.
We show that
the braiding statistics of these anyon line operators are encoded by the $\cS$ (for mutual-statistics) 
and $\cT$ (for exchange or self-statistics) matrices in \Sec{sec:Z2t-1}.
The semion and $\overline{\text{semion}}$ have exchange or self statistics, $\ii$ and $-\ii$, by half-braiding.
The anyon condensation on 2d boundary $(\partial M)^2$ of this 3d TQFT is achieved by setting 
$\cA  \bigg\rvert_{\partial \mathcal{M}} =a_1 + a_2  \bigg\rvert_{\partial \mathcal{M}} = 0$, when
semion-$\overline{\text{semion}}$ can condense on ${\partial \mathcal{M}}$.\\[4mm]
Similarly, for the 5d TQFT \Eq{eq:5dTQFT-DS},
we have the 2-worldsheet of anyonic strings given by the following 2-surface operator
on a closed 2-manifold ${\Sigma^2}$:
$$
\text{semionic string}:\; \exp(\ii \oint_{\Sigma^2}  (\mathscr{C}+ \cB)), \quad
\overline{\text{semionic}}\text{ string}:\; \exp(\ii \oint_{\Sigma^2}  \mathscr{C}), \quad
\text{semion-$\overline{\text{semionic}}$ string}: \;\exp(\ii \oint_{\Sigma^2} \cB).
$$
We abbreviate the {semionic string} as $s$-string,
the $\overline{\text{semionic}}$ string as $\bar{s}$-string,
and the {semionic-$\overline{\text{semionic}}$ string} as $s\bar{s}$-string.
We can insert two $S^2$ 2-surface operators linked in a 5-sphere $S^5$ \cite{Wang2019diz1901.11537, Wan2019oyr1904.00994},
by using the fact that $S^5 = (D^3 \times S^2)\cup_{S^2 \times S^2} (S^2 \times D^3)$ with the gluing $\cup_{S^2 \times S^2}$ along ${S^2 \times S^2}$
and the homology group $H_2(D^3 \times S^2, \Z)=\Z$.
The braiding statistics of  $s$-string, $\bar{s}$-string and  $s\bar{s}$-string are the same as that of 
$s$, $\bar{s}$ and $s\bar{s}$ anyons, up to a dimensional reduction of spacetime from 5d to 3d, and a reduction of operators from 2d to 1d.  
Thus, by analogy, 
a gapped 4d boundary $(\partial M)^4$ of this 5d TQFT can be achieved by setting 
$\cB  \bigg\rvert_{\partial \mathcal{M}} = 0$, when
we have the {\bf{\emph{anyon-string condensation}}} (here only $s\bar{s}$-string) on ${\partial \mathcal{M}}$.
\label{footnote:semion-string}
}
\newpage

This path integral definition of partition function gives rise to long-range entangled states of a 2-form gauge theory.
The 5d gauge theory has dynamical objects of electric and magnetic 1-strings, whose electric and magnetic 2-worldsheets
are gauge invariant 2-surface operators from $B$ and $b$ gauge fields respectively.

We particularly consider the 5d topological term (see footnote \ref{footnote:Steenrod} for clarification on math notations)
\bea
\omega^{5}_{j} =(-1)^{\int_{M^5} B_j \Sq^1 B_j} = (-1)^{\int_{M^5} B_j (\frac{1}{2} \delta B_j )},
\eea
where $B \in \cH^2(M^5,\Z_2)$ is a discrete 2-cocycle assigning a 2-simplex (a triangle as a 2-face) on $M^5$
to a $\Z_2$-valued coefficient. 
The $\delta$ is a coboundary operator, which sends $B \in \cH^2(M^5 ,\Z_2)$ of a $\Z_2$-valued 2nd cohomology class to 
$\Sq^1 B =  \frac{1}{2} \delta B  \in \cH^3(M^5,\Z_2)$ of a $\Z_2$-valued 3rd cohomology class.
Next, for $j=1,\dots.n$, as an interface intersected by totally $n$-layers,
we have the associated higher classifying space,
\bea
\prod_{j=1}^{n} (\B\mathbb{G}_j)=\B^2 (\prod_{j=1}^{n} (\Z_2)_j )= \B^2 ( (\Z_2)_1 \times (\Z_2)_2 \times \dots \times (\Z_2)_n ),
\eea
and an associated topological term contributed from each $j$-th layer:
\bea \label{eq:BdB}
\omega^{5} =(-1)^{\int_{M^5} \sum_{j=1}^n B_j \Sq^1 B_j }.
\eea
As we stated in \Sec{sec:trivialization-cohomology/cobordism-topological-term},
the trivialization of the topological term can be used to construct a one-lower-dimensional (here 3+1D, 4d) interface sitting at the junction of $n$ layers of 5d TQFTs.

To trivialize the topological term ${\int_{M^5} \sum_{j=1}^n B_j\cup \Sq^1 B_j}$, we notice that
\bea
\sum_{j=1}^n  B_j \Sq^1 B_j  &=& 
(\sum_{j=1}^n  B_j) \Sq^1 (\sum_{k=1}^n  B_k)  
- \Sq^1 (\sum_{j <  k}^{j,k \in \{ 1, \dots, n\}}  B_j B_k)\nn
\\
&=& 
(\sum_{j=1}^n  B_j) \Sq^1 (\sum_{k=1}^n  B_k)   \mod 2.
\label{eq:j1nBSq1B}
\eea
Here we use the face that if two cohomology classes $\cX \in \cH^2(M^5,\Z_2)$ and $\cY \in \cH^2(M^5,\Z_2)$,
 we have $\Sq^1(\cX\cY)=w_1(TM)\cX\cY$ on an $M^5$.
 Notice that $M$ with a SO or Spin structure,  
on an orientable manifold we have $w_1(TM)=0$, thus $\Sq^1(\cX\cY)=0$, similarly,
$\Sq^1 (\sum_{j <  k}^{j,k \in \{ 1, \dots, n\}}  B_j B_k)=0$ on SO or Spin manifolds.

Furthermore, for  $\cX \in \cH^2(M^5,\Z_2)$, we can show that  \cite{Wan2018zql1812.11968, Wan2019oyr1904.00994}
$$
\cX \Sq^1 \cX + \Sq^2 \Sq^1  \cX = \frac{1}{2} \tilde{w}_1(TM) P(\cX),
$$
where $\tilde{w}_1$ is a twisted Stiefel-Whitney class of a mod 4 class, and the $P(\cX)$ is the Pontryagin square of $\cX$ sending $\cX \in \cH^2(M,\Z_2)$ to 
$P(\cX) \in \cH^4(M,\Z_4)$.
Again on SO, Spin or orientable manifolds, we have $\tilde{w}_1(TM)=0$, thus we have
$$
\Big(\cX \Sq^1 \cX =  \Sq^2 \Sq^1  \cX  \mod 2 \Big)  \bigg\rvert_{\text{orientable }\mathcal{M}};
$$
so
\bea \label{eq:j1nBSq1B-2}
\Big( (\sum_{j=1}^n  B_j) \Sq^1 (\sum_{k=1}^n  B_k) =  \Sq^2 \Sq^1 (\sum_{k=1}^n  B_k) \mod 2 \Big)  \bigg\rvert_{\text{orientable }\mathcal{M}}.
\eea
Below let us enumerate possible approaches to trivializing the topological term based on breaking or extension:
\begin{enumerate}[label=\textcolor{blue}{\arabic*.}, ref={\arabic*},leftmargin=*]
\item \label{1st-} 
Breaking the electric sector completely $B_j=0$:\\
We may call the background field $B_j$ associated to 
the electric sector of 1-symmetry $\Z_{2,[1]}^e$.
In terms of the first expression in \Eq{eq:breaking-intro-summary},
we can consider the breaking which breaks all electric sectors completely
 in terms of the groups  or in terms of the higher classifying space: 
$$0 \stackrel{\iota}{\longrightarrow} (\prod_{j=1}^{n} (\Z_{2,[1]}^e)_j ), \quad\quad\quad
pt \stackrel{\iota}{\longrightarrow} \B^2 (\prod_{j=1}^{n} (\Z_2)_j^e ),$$ breaks to a point $pt$.
In this case, we thus set all the electric $$B_j=0 \mod 2$$ for all $j=1,\dots.n$,
thus \Eq{eq:BdB} is trivialized to a trivial term as 1 (no topological term).
Such a breaking process can define a gapped interface.
We can also comprehend this gapped interface by the 
{\bf{\emph{anyonic string condensation}}}
on the interface.
By setting the 
$B_j  \bigg\rvert_{\text{interface}} = 0$ on the interface, 
we indeed have
\Eq{eq:j1nBSq1B} = 0 mod 2,
thus $\omega^{5}$ becomes a trivial topological term;
see the footnote \ref{footnote:semion-string},
actually we have the {semionic-$\overline{\text{semionic}}$ string}
($s\bar{s}$-string) condensed on the interface.

\item  \label{2nd-}
Breaking the electric sector along the $\sum_{j=1}^n B_j =0$:\\
We can trivialize the topological term by breaking to set
\bea \label{eq:sum-B=0}
\sum_{j=1}^n B_j = B_1+B_2+\dots + B_n=0 \mod 2.
\eea
By setting  
$\sum_{j=1}^n B_j  \bigg\rvert_{\text{interface}} = 0$ on the interface, 
we indeed have
\Eq{eq:j1nBSq1B} = 0 mod 2,
thus $\omega^{5}$ becomes a trivial topological term.
In terms of the second expression of \Eq{eq:breaking-intro-summary},
what we do is breaking one single linear combinatory $\Z_{2,[1]}$ out of all $(\prod_{j=1}^{n} (\Z_{2,[1]}^e)_j )$.

\item  \label{3rd-}
Extension 
$(\B^2\Z_2)_j \hookrightarrow (\B^2\Z_4)_j \stackrel{r}{\longrightarrow}  (\B^2\Z_2)_j$:\\
As shown in \Refe{Wan2018djl1812.11955, Wan2019oyr1904.00994},
the  $\Sq^1\cX$ for $\cX \in \cH^2(M,\Z_2)$ 
can be trivialized if we lift it to $\tilde \cX \in \cH^2(M,\Z_4)$
such that ${\cX}= \tilde{\cX}\mod 2$.
For any 5-cocycle 
$\omega^{5}=\cY\cup \Sq^1\cX$ with $\cX, \cY \in \cH^2(M,\Z_2)$, \Refe{Wan2018djl1812.11955, Wan2019oyr1904.00994} proposes a way for trivialization
via the extension \Eq{eq:Z2Z4Z2Bj} or fibration \Eq{eq:Z2Z4Z2Bjfiber}, by lifting the $\Z_{2,[1]}$-symmetry to a $\Z_{4,[1]}$-symmetry for each $j=1, \dots, n$ layer:
\bea \label{eq:Z2Z4Z2Bj}
(\Z_{2,[1]})_j \to (\Z_{4,[1]})_j   &\stackrel{r}{\longrightarrow}& (\Z_{2,[1]})_j  .\\
 \label{eq:Z2Z4Z2Bjfiber}
(\B^2\Z_2)_j \hookrightarrow (\B^2\Z_4)_j &\stackrel{r}{\longrightarrow}&  (\B^2\Z_2)_j.
\eea
The solution of trivialization requires solving the equation $r^* \omega^{5}= \delta ( \beta^4)$ in the higher and enlarged classifying space $\B^2\Z_4$
after the pullback (here the $r^*$ means the pullback from $\B^2\Z_2$ to $\B^2\Z_4$) :
\bea
\label{eq:r^*omega^5}
r^*\omega^{5}&=&r^*(\cY \cup \Sq^1 {\cX}) = \delta ( \beta^4).\\
\text{solution:  } \beta^4&=&\cY \cup \ga^2 \quad 
\text{ with } \quad
 \ga^2( \tilde{\cX})=\frac{ \tilde{\cX}^2 -  \tilde{\cX}}{2} \mod 2.
 \eea
The $ \tilde{\cX}  \in \cH^2(M,\Z_4)$ is a $\Z_4$-valued 2-cochain satisfying ${\cX}= \tilde{\cX}\mod 2$, and $\gamma^2: \Z_4\to \Z_2$ is a function 
$
	\gamma^2( \tilde{\cX})_{i,j,k}= \frac{( \tilde{\cX}_{i,j,k})^2- ( \tilde{\cX}_{i,j,k})}{2},
$
which maps the $\Z_4$-valued 2-cochain to a $\Z_2$-valued 2-cochain on a 2-simplex with its three vertices labeling $(i,j,k)$ written in the subscript.

$\Sq^1 \cX = \cX \hcup{1} \cX$
on a 3-simplex with vertices (0,1,2,3) can be \emph{split} into 2-cochains $ \gamma^2$ in the following way:\footnote{We use the fact: 
$(\Sq^1 \cX)_{0,1,2,3}= (\cX \hcup{1} \cX)_{0,1,2,3}=  \cX_{0,1,2}\cX_{0,2,3}-\cX_{0,1,3}\cX_{1,2,3}=\cX_{0,1,2}\cX_{0,2,3}-\cX_{0,1,3}(\cX_{0,1,2}- \cX_{0,1,3} + \cX_{0,2,3} ) $
$=({\cX}_{0,1,2} + {\cX}_{0,1,3})({\cX}_{0,1,3} + {\cX}_{0,2,3})- 2 (\cX_{0,1,2}+  \cX_{0,2,3})\cX_{0,1,3}
=({\cX}_{0,1,2} + {\cX}_{0,1,3})({\cX}_{0,1,3} + {\cX}_{0,2,3})$ mod 2, in the last line of \Eq{eq:gamma^2}.}
\bea
&&(\delta \gamma^2(\tilde{\cX}))_{0,1,2,3} = - \gamma^2(\tilde{\cX}_{0,1,2})+\gamma^2(\tilde{\cX}_{0,1,3})-\gamma^2(\tilde{\cX}_{0,2,3})+\gamma^2(\tilde{\cX}_{1,2,3})\nn\\
&&= - \gamma^2(\tilde{\cX}_{0,1,2})+\gamma^2(\tilde{\cX}_{0,1,3})-\gamma^2(\tilde{\cX}_{0,2,3})+\gamma^2({\tilde{\cX}_{0,1,2}-\tilde{\cX}_{0,1,3}+\tilde{\cX}_{0,2,3}})\nn\\
&&= (\tilde{\cX}_{0,1,2} + \tilde{\cX}_{0,1,3})(\tilde{\cX}_{0,1,3} + \tilde{\cX}_{0,2,3}) 
=  (r(\tilde{\cX}))\hcup{1} (r(\tilde{\cX})) =  \Sq^1  (r(\tilde{\cX})) = \Sq^1{\cX} \mod 2, \quad\quad\quad
\label{eq:gamma^2}
\eea
with the reduction map $r$ in \Eq{eq:Z2Z4Z2Bj}.
Via \Eq{eq:gamma^2},  then we can show \Eq{eq:r^*omega^5} is true:
$$
\delta ( \beta^4 (r(\tilde{\cX}) ) )=
\delta (\cY \cup \ga^2 (r(\tilde{\cX}) )  )
=(\delta \cY) \cup
\ga^2 (r(\tilde{\cX}) ) 
+
\cY \cup \delta\ga^2 (r(\tilde{\cX}) ) 
=\cY \cup \Sq^1  (r(\tilde{\cX}))=
\cY \cup  \Sq^1{\cX} =
\omega^{5}.
$$
Above we use the fact
$(\delta \cY)$ is a 3-coboundary (i.e., a trivial 3-cocycle) because $\cY$ is a 2-cocycle,
thus $(\delta \cY) \cup
\ga^2 (r(\tilde{\cX}) )$ is also a 5-coboundary (i.e., a trivial 5-cocycle).
To trivialize
\Eq{eq:j1nBSq1B},
we take ${\cX}=\cY=B_j$ for each layer, 
so each layer $ B_j \Sq^1 B_j$ can be trivialized via the above extension.

\item \label{4th-}
Extension 
$(\B^2\Z_2)_j \hookrightarrow (\B^2\Z_4)_j \stackrel{r}{\longrightarrow}  (\B^2\Z_2)_j$:\\
Instead of trivializing each layer as the previous Approach \ref{3rd-}, 
which \emph{only} constructs a gapped interface of decoupled $n$ layers,
we can construct a gapped interface of joint $n$ layers. 
To this end, we trivialize \Eq{eq:j1nBSq1B} jointly together, 
we take ${\cX}=\cY=(\sum_{j=1}^n  B_j)$, 
so
\Eq{eq:j1nBSq1B} $(\sum_{j=1}^n  B_j) \Sq^1 (\sum_{k=1}^n  B_k)=\cY \cup  \Sq^1{\cX}$ can be trivialized via the extension:
\bea \label{eq:Z2Z4Z2Bj}
(\Z_{2,[1]})_{\text{sum}} \to (\Z_{4,[1]})_{\text{sum}}   &\stackrel{r}{\longrightarrow}& (\Z_{2,[1]})_{\text{sum}}  .\\
 \label{eq:Z2Z4Z2Bjfiber}
(\B^2\Z_2)_{\text{sum}} \hookrightarrow (\B^2\Z_4)_{\text{sum}} &\stackrel{r}{\longrightarrow}&  (\B^2\Z_2)_{\text{sum}}.
\eea
Here this ${\Z_{2,[1]}}_{\text{sum}}$ is a joint 1-symmetry whose 2-cochain background field is
the sum $(\sum_{j=1}^n  B_j)$.

\item \label{5th-}
Extension 
$\B\Z_2 \hookrightarrow \B\Spin \stackrel{r}{\longrightarrow}  \B\SO$:\\
As shown in \Refe{Wan2018djl1812.11955, Wan2019oyr1904.00994},
the  $\Sq^1\cX$ for $\cX \in \cH^2(M,\Z_2)$ on manifolds with SO structure
can be trivialized if we lift it to $\tilde \cX \in \cH^2(M,\Z_4)$
on manifolds with Spin structure, via the fibration
\bea  \label{eq:Z2SOSpinjfiber}
\B\Z_2 \hookrightarrow \B\Spin \stackrel{r}{\longrightarrow}  \B\SO.
\eea
Namely, for
any $\Sq^2 \Sq^1 {\cX}$ with $\cX \in \cH^2(M,\Z_2)$,
the Wu formula shows
$$
\Sq^2 \Sq^1 {\cX}= (w_2(TM)+w_1(TM)^2 )\;\Sq^1 {\cX},
$$
while the Spin structure requires the first and the second Stiefel-Whitney class $w_1(TM)=w_2(TM)=0$. So take
${\cX}=(\sum_{j=1}^n  B_j)$,
we have \Eq{eq:j1nBSq1B} = \Eq{eq:j1nBSq1B-2} is trivialized by lifting to a Spin manifold.
The extended Spin structure implies that if we treat the 5d bulk $ B \Sq^1 B$ as a higher-symmetry SPTs (as an iTQFT, i.e., $B$ is ungauged),
then there exists possible 4d $\Z_2$ gauge theory whose gauge charge carries \emph{emergent fermion}.

\end{enumerate}
The above equalities in Approaches \ref{1st-}-\ref{5th-}
all are only mod 2 relations. 

Let us summarize the properties of the interfaces for the given 4+1D higher-gauge TQFT bulk layers $\sum_{j=1}^n  B_j \Sq^1 B_j $
constructed via the above  Approaches \ref{1st-}-\ref{5th-}.
\begin{itemize}
\item Approach \ref{1st-} \emph{decoupled} cell layers: After gauging the $B_j$ fields on bulk cells,
each 4+1D TQFT layer has its own $s_j\bar{s}_j$-string ({semion-$\overline{\text{semionic}}$ string}) 
condensed on the 3+1D interface, but each 4+1D layer stays \emph{decoupled} from other 4+1D layers.

\item Approach \ref{2nd-} \emph{coupled} cell layers: After gauging the $B_j$ fields on bulk cells,
all 4+1D TQFT layers have a joint $s\bar{s}$ string condensed on the 3+1D interface, while each 4+1D layer \emph{couples} with other 4+1D layers.
The closed 2-worldsheet of the joint $s\bar{s}$ string corresponds to the 2-surface operator:
$\exp(\ii \pi \oint_{\Sigma^2} (\sum_{j=1}^n  B_j))$. 
(However, there could be additional other types of compatible objects also condensing on the boundary, 
compatible with the joint $s\bar{s}$ string.)

\item Approach \ref{3rd-} \emph{decoupled} cell layers: After gauging the $B_j$ fields on bulk cells,
we can construct a 3+1D gapped gauge theory with $\Z_4$-valued 2-cochain (or 2-form) gauge fields on the boundary of each layer.
These 3+1D $\Z_4$-valued 2-cochain gauge theories are \emph{decoupled} from each other if we do not identify their fractional excitations of each layer 
as the same excitations 

\item Approach \ref{4th-}  \emph{coupled} cell layers:
After gauging the $B_j$ fields on bulk cells,
we can construct a 3+1D joint gapped gauge theory with $\Z_4$-valued 2-cochain (or 2-form) gauge fields on the interface.
There is a 3+1D $\Z_4$-valued 2-cochain gauge theory \emph{coupled} to all layers.
The joint $\Z_4$-valued 2-cochain is obtained from extending the
$\cX= (\sum_{j=1}^n  B_j) \in  \cH^2(M,\Z_2)$
to $\tilde \cX= (\sum_{j=1}^n  \tilde{B}_j)  \in \cH^2(M,\Z_4)$,
where $\tilde{B}_j$ is  a $\Z_4$-valued 2-cochain gauge field extended from the $\Z_2$-valued 2-cochain gauge field $B_j$.
The joint $\Z_4$-valued 2-cochain $\tilde \cX$ gauge field lives on the 3+1D gapped interface.

\item Approach \ref{5th-}  \emph{decoupled} cell layers:
If we extend SO to Spin on each cell layer, we can construct a 3+1D emergent fermion $\Z_2$ gauge theory on the boundary of each layer.
These 3+1D emergent fermion $\Z_2$ gauge theories are \emph{decoupled} from each other if we do not identify their fractional excitations of each layer 
as the same excitations (e.g. if we do not identify their emergent fermions as the same fermion).

\end{itemize}

In the above decoupled cell layers,  we do not identify their fractional excitations as the same excitations.
The construction of decoupled cell layers imply the {\bf {\emph{non-fracton}}} behavior.
However, if we do identify their fractional excitations as the same excitations, then which means that the excitations can move \emph{freely} across the interface to each 
cell layer  --- this also indicates {\bf {\emph{non-fracton}}} behavior.
In any case, we expect to construct the {\bf {\emph{liquid cellular states}}} from Approach \ref{1st-}, \ref{3rd-}, and \ref{5th-}.

We should remind ourselves that although we have glued the 4+1D TQFTs along with a 3+1D interface, it is possible
to use them to construct 4+1D cellular states as well as 5+1D cellular states, depending on
whether we use the 4+1D layers as the codimension-0 or codimension-1 defects in the defect network.
Since Approach \ref{2nd-} and \ref{4th-}
have \emph{coupled} layers,
they can be used to construct  non-trivial {\bf {\emph{cellular states}}}, i.e., non-tensor product states between the cell layers. 


\subsection{More examples of higher symmetry and time-reversal symmetry}
\label{sec:More-examples}

In this subsection, we summarize some possible useful facts to construct cellular states via exotic gapped interfaces with 
ordinary symmetry, higher symmetry, or time-reversal symmetry. We leave the full explorations of the 
constructed cellular states for future work.

\begin{enumerate}[label=\textcolor{blue}{\arabic*.}, ref={\arabic*},leftmargin=*]
\item
For quantum systems with  the internal 1-form $\Z_2$-symmetry 
(also on manifolds with special orthogonal group SO structure),
the following 3+1D (4d) and 4+1D (5d) $\Z_4$-class topological terms must have 1-form $\Z_2$-symmetry-enforced gapless boundary states 
\cite{Wan2018djl1812.11955, Cordova2019bsd1910.04962, 
Thorngren2020aph2001.11938}:
\bea
P(B),\\
A P(B), 
\eea
with the Pontryagin square $P(B) \in \cH^4(M,\Z_4)$, $A \in \cH^1(M,\Z_4)$, $B \in \cH^2(M,\Z_2)$, and $A P(B) \in \cH^4(M,\Z_4)$. 

\item
For quantum systems with time-reversal $\Z_2^T$ symmetry 
(for manifolds with orthogonal group O structure) and the internal 1-form $\Z_2$-symmetry,
there exists a $\Z_2$-class topological term, whose 
symmetry-enforced boundary states must also be gapless \cite{Wan2019oyr1904.00994, Cordova2019bsd1910.04962}:
\bea
B\Sq^1 B+  \Sq^2 \Sq^1 B= \frac{1}{2} \tilde{w}_1(TM) P(B).
\eea
The above symmetry-enforced gapless boundary states mean that if the full symmetry is preserved,
the boundary states cannot be fully gapped, not possible even for a gapped long-range entangled TQFT in order to saturate the 't Hooft anomaly 
of a global symmetry.\footnote{A 
  theorem on anomaly obstructions to symmetry preserving gapped phases
 is  proven  in Cordova-Ohmori \cite{Cordova2019bsd1910.04962}.}
However, as long as some of the full symmetry is broken, spontaneously or explicitly,
the 't Hooft anomalies of the above topological terms can be saturated by a (symmetry-breaking) gapped TQFT.

\item 
{\bf {\emph{Composite string v.s. p-string condensations}}}:
\Refe{Wang2018edf1801.05416} gives an example of gapped boundary conditions which can be relevant to
 the p-string condensation in fracton physics \cite{Ma2017aogpstring1701.00747}. For 
 $3+1$D $\mathbb{Z}_2^{4}$ gauge theory whose partition function is
$(-1)^{\int A_1\cup A_2\cup A_3\cup A_4}, 
$ with $A_j \in \cH^1(M,(\Z_2)_j)$
there exists a boundary condition
\begin{equation} 	\label{eq:boundaryZ24}
	A_1\cup A_2 =0, \quad A_3\cup A_4=0,
\end{equation}
obtained from the extension construction:
$$
1\to (\mathbb{Z}_2)_{\text{interface}} \to (D_4\times \mathbb{Z}_2^2)_{\text{interface}}\to (\mathbb{Z}_2^4)_{\text{bulk}}\to 1.
$$
The $A_1\cup A_2 =0$ and $\quad A_3\cup A_4=0$ boundary 
condition means that two types of composite strings can condense on the gapped boundary: 
The worldsheet of the first composite string is formed by the intertwining particle worldlines of $A_1$ and $A_2$.
The worldsheet of the second composite string is formed by the intertwining particle worldlines of $A_3$ and $A_4$. (See more details in Section 7 of \Refe{Wang2018edf1801.05416}.)
For another example, given a boundary condition with a constraint:
\bea
	A_i\cup A_j \cup A_j =0,
\eea
the worldvolume of a composite membrane (formed by the intertwining particle worldlines of $A_i$, $A_j$ and $A_k$) can end on the boundary.
This is analogous to {\bf{\emph{p-membrane condensation}}} \cite{Ma2017aogpstring1701.00747}.

More generally, given a boundary condition with a constraint:
\bea
	A_i\cup B_j =0,
\eea
the worldvolume of a {\bf{\emph{composite membrane}}} (formed by the intertwining particle worldline of $A_i$ and string worldsheet of $B_j$) can end on the boundary.

Composite string v.s. p-string condensations:
Indeed the composite condensation means that a composite object form by the cup product of 
the lower-dimensional objects condensing on the interface. 
In our case, a composite object can be a composite string,
the lower-dimensional objects can be a set of particles (that form a string-like object).
The particles of a composite string are located on the 0-simplex vertices of different 1-simplex links 
(say link variables $A_1$ and $A_2$), such that the link variables $A_1$ and $A_2$ further form a 2-simplex. 
The link variables $A_1$ and $A_2$ have the codimension-2 dual plaquettes located at different codimension-2 spacetime layers
(or the codimension-3 dual plaquettes located at different codimension-3 spatial layers). 
This suggests a possible relation of the composite string on a simplicial complex, versus that of the p-string on a cubic lattice \cite{Ma2017aogpstring1701.00747}.

To construct {\bf {\emph{cellular states}}} with a long-range entangled bulk, we should gauge the bulk internal symmetries 
(e.g. gauge the 1-form $\Z_2$-symmetry with dynamical 2-cochain $B$ fields).
We can put the cell layers with the above gauged topological terms. It will be interesting to study more consistent
{cellular states} by gluing the above interfaces in the future.

\end{enumerate}


\noindent
{\bf Note added}: During the preparation of this manuscript on the construction of general phases of matter via
gluing gauge-(higher)-symmetry-breaking or gauge-(higher)-symmetry-extension interfacial defects, the author becomes aware
that two recent inspiring works \cite{Wen2020pri2002.02433, Aasen2020zru2002.05166} are somehow related to the author's construction.
In particular, some of \Refe{Wen2020pri2002.02433}'s results can be considered as the
gauge-breaking gapped interface construction of our case.
In comparison, our constructions of gluing 
gauge-higher-symmetry-breaking or gauge-higher-symmetry-extension interfaces remain original and new to the literature.

\section{Acknowledgements}

JW warmly thanks the organizers and participants of the meeting at Banff International Research Station on ``Fractons and Beyond''
from January 26 to January 31, 2020, during which part of the present work is done. 
JW thanks Zheyan Wan and Yunqin Zheng 
for previous collaborations and inspiring discussions on \cite{Wan2018djl1812.11955, Wan2018bns1812.11967HAHSI, Wan2018zql1812.11968, Wan2019oyr1904.00994}.
This work is also supported by NSF Grant DMS-1607871
``Analysis, Geometry and Mathematical Physics'' and Center for Mathematical Sciences and Applications at
Harvard University.\\

\bibliography{NonLiquidCellular.bib}

\end{document}